\documentclass[12pt]{article}
\usepackage{txfonts}

\usepackage[none]{hyphenat}
\usepackage{graphicx}
\usepackage{color}
\usepackage[rotateright]{rotating}

\title{The tectonic cause of mass extinctions and the genomic contribution to biodiversification}
\author
{\small Dirson Jian Li\\
\small Department of Applied Physics, Xi'an Jiaotong
University, Xi'an 710049, China}

\date{}
\begin{document}

\baselineskip22pt

\setlength{\parskip}{12pt}

\maketitle
\sloppy

{\bf
\begin{center}{Abstract}\end{center}

Despite numerous mass extinctions in the Phanerozoic eon, the overall trend in biodiversity evolution was not blocked and the life has never  been wiped out. Almost all possible catastrophic events (large igneous province, asteroid impact, climate change, regression and transgression, anoxia, acidification, sudden release of methane clathrate, multi-cause etc.) have been proposed to explain the mass extinctions. However, we should, above all, clarify at what timescale and at what possible levels should we explain the mass extinction? Even though the mass extinctions occurred at short-timescale and at the species level, we reveal that their cause should be explained in a broader context at tectonic timescale and at both the molecular level and the species level. The main result in this paper is that the Phanerozoic biodiversity evolution has been explained by reconstructing the Sepkoski curve based on climatic, eustatic and genomic data. Consequently, we point out that the P-Tr extinction was caused by the tectonically originated climate instability. We also clarify that the overall trend of biodiversification originated from the underlying genome size evolution, and that the fluctuation of biodiversity originated from the interactions among the earth's spheres. The evolution at molecular level had played a significant role for the survival of life from environmental disasters.

}

\clearpage
\begin{center}{\bf \large RESULTS}\end{center}

Let us go back to the early history of our planet, and gaze at these just originated lives. They seemed so delicate, however they were indeed persistent and dauntless. They had a lofty aspiration to live on until the end of the earth; otherwise the rare opportunity of this habitable planet in the wildness of space may be wasted. Their story continued and was recorded in the big book of stratum. This story was so magnificent that we were moved to tears time and again. Was the life just lucky to survive from all the disasters, or innately able to contend with any possible challenges in the environment? Before answering this question, we should explain the evolution of biodiversity by appropriate driving forces.

Again, let us go back to mid nineteenth century, and size up the situations for the founders of evolutionism. They were completely unaware of the molecular evolution; they knew little about the marine regression or transgression and paleoclimate; and they possessed poor fossil records. However, they still pointed out the right direction to understand the evolution of life by their keen insight. What is the mission then for contemporary evolutionists in floods of genomic and stratum data? Can we go a little further than endless debates?

The Sepkoski curve based on fossil records indicates the Phanerozoic biodiversity evolution \cite{Sepkoski 2002} \cite{Bambach et al. 2004} \cite{Rohde and Muller 2005}, where we can observe five mass extinctions, the background extinction, and its increasing overall trend. The main purpose of this paper is to explain the Sepkoski curve by a tectono-genomic curve based on climatic, eustatic (sea level) and genomic data. We propose a split scenario to study the biodiversity evolution at the species level and at the molecular level separately. We construct a tectonic curve based on climatic and eustatic data to explain the fluctuations in the Sepkoski curve. And we also construct a genomic curve based on genomic data to explain the overall trend of the Sepkoski curve. Thus, we obtain a tectono-genomic curve by synthesizing the tectonic curve and the genomic curve, which agrees with the Sepkoski curve not only in overall trend but also in detailed fluctuations (Fig 1):
$$Curve\_Sepkoski\approx Curve\_TectonoGenomic.$$
We observe that both the tectono-genomic curve and the Sepkoski curve decline at each time of the five mass extinctions (O-S, F-F, P-Tr, Tr-J and K-Pg). The growth rates of the tectono-genomic curve and the Sepkoski curve also coincide with each other. Hence, we show that the biodiversity evolution is driven by both the tectonic movement and the genome size evolution. The main steps in constructing the tectono-genomic curve are as follows.

(1) We obtained the consensus climate curve ($Curve\_CC$), the consensus sea level curve ($Curve\_SL$) and the biodiversification curve ($Curve\_BD$) to describe the Phanerozoic climate change, sea level fluctuation and biodiversity variation respectively (Fig 2a). (i) We obtained $Curve\_CC$ by synthesizing the following three independent results on Phanerozoic climate change in a pragmatic approach (Fig S1a): Berner's atmosphere $CO_2$ curve \cite{Berner 1998}, the Phanerozoic global climatic gradients revealed by climatically sensitive sediments \cite{Boucot and Gray 2001} \cite{Boucot et al. 2009}, and the Phanerozoic $^{87}Sr/^{86}Sr$ curve \cite{Raymo 1991}; (ii) We obtained $Curve\_SL$ by synthesizing the result in ref. \cite{Hallam 1992} and the results in ref. \cite{Haq et al 1987} \cite{Haq and Schutter 2008} (Fig. S1c); and (iii) We obtained $Curve\_BD$ based on fossil record (Fig. 2d).

(2) We calculated the correlation coefficients $r^{\rho}_{\mu\nu}$ among $Curve\_CC$, $Curve\_SL$ and $Curve\_BD$ (Table 1). The correlation coefficient between $Curve\_BD$ and $Curve\_SL$ in the Phanerozoic eon is $r^{PMC}_{SB}=0.564$, which generally indicates a same phase between $Curve\_BD$ and $Curve\_SL$. The correlation coefficients between $Curve\_BD$ and $Curve\_CC$, and between $Curve\_SL$ and $Curve\_CC$ in the Paleozoic era are $r^P_{BC}=0.114>0$ and $r^P_{CS}=0.494>0$ respectively, which generally indicate the same variation pattern (or the same phase) of $Curve\_CC$ with $Curve\_BD$ and $Curve\_SL$ in the Paleozoic era. While the correlation coefficients between $Curve\_BD$ and $Curve\_CC$, and between $Curve\_SL$ and $Curve\_CC$ in the Mesozoic era are $r^M_{BC}=-0.431<0$ and $r^M_{CS}=-0.617<0$ respectively, which indicate a ``climate phase reverse event'' from same phase to opposite phase in P-Tr boundary. In the supplementary methods, we confirm the reality of such a ``climate phase reverse event'' by verifications for $10$ group curves based on candidate climate, biodiversity and sea level data. Therefore, when constructing the tectonic curve based on $Curve\_SL$ and $Curve\_CC$, we chose a positive sign for $Curve\_SL$ throughout the Phanerozoic eon; and we chose a positive sign for $Curve\_CC$ only in the Paleozoic era, but a negative sign for $Curve\_CC$ in the Mesozoic and Cenozoic eras (Fig S1e).

(3) The overall trend in biodiversity evolution is about an exponential function \cite{Hewzulla et al. 1999}: $N_{genus}=N^0_{genus}\ \exp(-t/\tau_{BD}).$  Based on the relationship between certain average genome sizes in taxa and their origin time, we found that the overall trend in genome size evolution is also an exponential function \cite{Sharov 2007} \cite{Li and Zhang 2010} (Fig 3a): $N_{genome}=N^0_{genome}\ \exp(-t/\tau_{GS}).$ The log-normal genome size distributions (Fig S2a, 3b) and the exponential asymptotes of the accumulation origination and extinction number of genera (Fig 2d) also indicate the exponential growth trend in genome size evolution. We found that the ``e-folding'' time of the biodiversity evolution $\tau_{BD}=259.08$ Million years (Myr) is approximately equal to the ``e-folding'' time of the genome size evolution $\tau_{GS}=256.56$ Myr (Fig 3d): $$\tau_{BD}\approx \tau_{GS}.$$
Hence, we can explain the overall trend in biodiversity evolution by constructing the genomic curve based on $\tau_{GS}$.

In the split scenario, we can explain the declining Phanerozoic background extinction rates \cite{Raup and Sepkoski 1982} \cite{Newman and Eble 1999} according to the equation:
$$rate_{o+e}=exp(-k_{GS}\cdot(-t+542.0)) \cdot rate\_essential,$$
where the declining factor $exp(-k_{GS}\cdot(-t+542.0))$ is due to the increasing overall trend in genome size evolution (Fig 2c). The underlying genomic contribution to the biodiversity evolution prevents the life from being completely wiped out by uncertain disasters.

So far, we have explained the declining background extinction rates and the increasing overall trend of the Sepkoski curve. The remaining problem is to explain the mass extinctions. Since we have successfully fulfilled the tectono-genomic curve to explain the Sepkoski curve, the reasons that caused the fluctuations in the tectono-genomic curve are just what caused the mass extinctions. We should emphasize here that the fluctuations in the tectono-genomic curve have nothing to do with the fossil data. According to the methods in constructing the tectono-genomic curve, we conclude that the mass extinctions were caused by both the sea level fluctuations and the climate changes. We refer it as the tectonic cause of the mass extinctions, which rules out any celestial explanations.

Furthermore, we point out that the greatest P-Tr extinction uniquely involved the climate phase reverse event, which occurred not just coincidentally with the formation of Pangaea and the atmosphere composition variation \cite{Berner 1998} \cite{Berner and Kothavala 2001} \cite{Berner et al. 2003}. The fossil record indicates a two-stage pattern at the Guadalupian-Lopingian boundary (GLB) \cite{Jin 1991} \cite{Jin 1993} \cite{Stanley and Yang 1994} and at the Permian-Triassic Boundary (PTB) \cite{Shen et al. 2011} \cite{Jin 2000}. In detail, it also indicates a multi-episode pattern in the PTB stage \cite{Xie et al. 2005} \cite{Chen and Benton 2012}. The P-Tr mass extinction was by no means just one single event. The multi-stage/episode pattern can hardly be explained by the large igneous province event \cite{Renne and Basu 1991} \cite{Campbell et al. 1992}. We can explain the above two stages by two sharp peaks observed in $d\_CC$ (the variation rate curve of $Curve\_CC$) at GLB and PTB respectively, which show that the temperature increased extremely rapidly at GLB and decreased extremely rapidly at PTB (Fig 2b). The different climate at GLB and at PTB resulted in different extinction time for Fusulinina (at GLB) and Endothyrina (at PTB).

At last, we will focus on the genomic contribution to the biodiversity evolution. We can obtain both the phylogenetic tree of species (Fig S3a, 4c by $M_{ci}$) and the evolutionary tree of $64$ codons (Fig 4a, S3b by $M_{codon}$) based on the same codon interval correlation matrix $\Delta$. This is a direct evidence to show the close relationship between the molecular evolution and the biodiversity evolution. On one hand, the result is reasonable in obtaining the tree of species. This universal phylogenetic method based on $M_{ci}$ applies for Bacteria, Archaea, Eukarya and virus. On the other hand, the result is valid in understanding the genetic code evolution \cite{Trifonov et al. 2006} \cite{Trifonov et al. 2001} \cite{Wong and Lazcano 2009}. And an average codon distance curve $Barrier$ based on $M_{codon}$ reveals a midway ``barrier'' in the genetic code evolution (Fig 4b, S3c). Moreover, we can testify the three-stage pattern (Basal metazoa, Protostomia and Deuterostomia) in Metazoan origination \cite{Shu 2008} according to the genome size evolution. Favorable phylogenetic trees can also be obtained by the correlation matrices $M_{gs}$ based on genome size data (Fig 3c, S2c, S2d).

\clearpage

\begin{center}{\bf \large METHODS}\end{center}

\section{Data resources and notations}

\subsection{Data resources}

\noindent (1) Phanerozoic climate change data: ref. [4], [5], [6], [7];

\noindent (2) Phanerozoic sea level fluctuation data: ref. [8], [9], [10];

\noindent (3) Phanerozoic biodiversity variation based on fossil records: ref. [1], [2], [3];

\noindent (4) Genome size databases: Animal Genome Size Database \cite{Animal Genome Size Database}, Plant DNA C-values Database \cite{Plant DNA C-values database};

\noindent (5) Whole genome database: GenBank.

\subsection{Notations}

\begin{equation}
\begin{array}{rcl}
\mbox{Sepkoski curve}&:&\ Curve\_Sepkoski\\
\mbox{teconto-genomic curve}&:&\ Curve\_TectonoGenomic\\
\mbox{time}&:&\ t,T\\
\mbox{biodiversity curves}&:&\ Curve\_BD, BD, Total\mbox{-}BD\\
\mbox{sea level curves}&:&\ Curve\_SL, S^1, S^2, S_w\\
\mbox{climate curves}&:&\ Curve\_CC, C^1, C^2, C^3, C_{w1}, C_{w2}, C_w\\
\mbox{correlation coefficients}&:&\ r^\rho_{\mu\nu}, R^+, R^-, \Delta R, Q, Q',\Delta Q\\
\mbox{climate phases}&:&\ CP I, CP II, CP III\\
\mbox{genome sizes}&:&\ G, G_{sp}, G_{mean\_log}, G_{sd\_log},G^*\\
\mbox{biodiversity variation rates}&:&\ rate\_ori, rate\_ext, rate\_essential\\
\mbox{derivative curves}&:&\ d\_CC, d\_SL, d\_BD\\
\mbox{overall trends}&:&\ OT\mbox{-}BD, OT\mbox{-}GS\\
\mbox{e-folding times and growth rates}&:&\ \tau_{BD}, k_{BD}, \tau_{GS},k_{GS}\\
\mbox{genomes size distributions and matrices}&:&\ D_{gs}, M_{gs}\\
\mbox{codon interval distributions and matrices}&:&\ D_{ci}, \Delta,  M_{ci},  M_{codon}\\
\mbox{genetic code evolutionary curves}&:&\ Barrier, Hurdle.
\end{array}
\end{equation}

\subsection{Math notations}

Let $sum(V)$, $mean(V)$, $std(V)$, $log(V)$ and $exp(V)$ denote respectively the summation, mean, stand deviation, logarithm and exponent of a vector $V(i)$, $i=1,2,...,i_m$:
\begin{eqnarray}
sum(V)&=&\sum_{i=1}^{i_m}V(i)\\
mean(V)&=&\frac{1}{i_m}\ sum(V)\\
std(V)&=&\sqrt{mean((V-mean(V))^2)}\\
log(V)&=&[\log_e(V(1)),\log_e(V(2)),...,\log_e(V(i_m))]\\
exp(V)&=&[\exp(V(1)),\exp(V(2)),...,\exp(V(i_m))].
\end{eqnarray}

Especially, let $nondim(V)$ denote the operation of nondimensionalization for a dimensional vector $V$,
\begin{equation} nondim(V)=(V-mean(V))/std(V).\end{equation}
In this paper, we obtain respectively the dimensionless vectors $Curve\_BD$, $Curve\_CC$, $Curve\_SL$, etc. after nondimensionalization based on the dimensional raw data of biodiversity curve, climate curve and sea level curve in the Phanerozoic eon.

Let $corrcoef(V,U)$, $max(V,U)$, $min(V,U)$ and $[V,U]$ denote respectively the correlation coefficient, maximum and minimum of a pair of vectors $V(i)$ and $U(i)$ ($i=1,2,...,i_m$):
\begin{eqnarray} corrcoef(V,U)&=&\frac{\sum_{i=1}^{i_m}(V(i)-mean(V))(U(i)-mean(U))}{\sqrt{\sum_{i=1}^{i_m}(V(i)-mean(V))^2}\sqrt{\sum_{i=1}^{i_m}(U(i)-mean(U))^2}}\\
max(V,U)&=&[max(V(1),U(1)),max(V(2),U(2)),...,max(V(i_m),U(i_m))]\\
min(V,U)&=&[min(V(1),U(1)),min(V(2),U(2)),...,min(V(i_m),U(i_m))].
\end{eqnarray}

Let $\frac{\mathrm{d}}{\mathrm{d}t}(V)$ denote the discrete derivative of $V(t)$ with respect to time $t$:
\begin{equation}
\frac{\mathrm{d}}{\mathrm{d}t}(V)=[\frac{\mathrm{d}V}{\mathrm{d}t}|_{t=t(1)}, ..., \frac{\mathrm{d}V}{\mathrm{d}t}|_{t=t(i_m)}],
\end{equation}
where $V(t)=[V(1),V(2),...,V(i_m)]$ is an $i_m$-element discrete function of time $t=[t(1), t(2),..., t(i_m)]$.
The linear interpolation of $V$ is denoted by:
\begin{equation}
[V(1),V(2),...,V(i_m')]=interp([t(1),...,t(i_m)],[V(1),...,V(i_m)],[t(1),...,t(i_m')]).
\end{equation}
The concatenation of function $V(t)$ between period $t([P_1])=[t(i_1), t(i_1+1), ..., t(i_2)]$ and period $t([P_2])=[t(t_2+1), t(i_2+2), ..., t(i_3)]$ is denoted by:
\begin{equation}
[V([P_1]),V([P_2])]=[V(i_1),...,V(i_2),V(i_2+1),...,V(i_3)],
\end{equation}
where $P_1=[i_1,i_1+1,...,i_2]$ and  $P_2=[i_2+1,i_2+2,...,i_3]$ are parts of the indices. For a $i_m-$by$-j_m$ array $M(i,j)$, let $M(i,:)$ denote
\begin{equation}
M(i,:)=[M(i,1),M(i,2),...,M(i,j_m)].
\end{equation}

\section{Understanding the Sepkoski curve through the tectono-genomic curve}

The Phanerozoic biodiversity curve has been explained in this paper. We propose a split scenario for the biodiversity evolution:
\begin{equation}
\mbox{Biodiversity evolution}=\mbox{Tectonic contribution}+\mbox{Genomic contribution}.
\end{equation}
We construct a tectono-genomic curve based on climatic, eustatic (sea level) and genomic data, which agrees with the Phanerozoic biodiversity curve based on fossil records very well. We explain the P-Tr extinction by a climate phase reverse event.
And we point out that the biodiversity evolution was driven independently at the species level as well as at the molecular level.

\section{The overall trend of biodiversity evolution}

\subsection{Motivation}

A split scenario is propose to separate the Phanerozoic biodiversity evolution curve into its exponential growth part and its variation part.

\subsection{The exponential outline of the Sepkoski curve}

The Phanerozoic biodiversity curve (namely the Sepkoski curve) can be obtained based on fossil records. We denote the Phanerozoic genus number biodiversity curve in ref. [2] after linear interpolation by (Fig 1):
\begin{equation} Curve\_Sepkoski(t):\ \mbox{ref. [2]}, \end{equation}
which is a $5421$-element function of time $t$, from $542$ million years ago (Ma) to $0$ Ma in step of $0.1$ million of years (Myr):
\begin{equation}
\begin{array}{rcl}
t&=&[t(1),t(2),t(3),...,t(5419),t(5420),t(5421)]\\
&=&[542.0, 541.9, 541.8,...,0.2, 0.1, 0].
\end{array}
\end{equation}

The outline of $Curve\_Sepkoski(t)$ is an exponential function:
\begin{equation} N_{genus}(t)=N^0_{genus}\ \exp (-t/\tau_{BD}), \end{equation}
where the genera number constant is $N^0_{genus}=2690$ genera, and the ``e-folding time'' of the biodiversity evolution is $\tau_{BD}=259.08$ Myr.

\subsection{The split scenario of the Sepkoski curve}

We define the total biodiversity curve $Total\mbox{-}BD$ in the Phanerozoic eon by the logarithm of $Curve\_Sepkoski$:
\begin{equation} Total\mbox{-}BD=log(Curve\_Sepkoski(t)),\end{equation}
which is also a $5421$-element function of time $t$.
According to the linear regression analysis, the regression line of $Total\mbox{-}BD$ on $t$ is defined as the overall trend of total biodiversity curve:
\begin{equation}
\begin{array}{rcl}
OT\mbox{-}BD &=& log(N_{genus}(t)) \\
&=& k_{BD}\cdot(-t)+\log(N^0_{genus}),
\end{array}
\end{equation}
where the growth rate of biodiversity evolution, namely the slope of this regression line, is $k_{BD}=1/\tau_{BD}=0.0038598$ $\mbox{Myr}^{-1}$.

We propose a ``split scenario'' in observing the Phanerozoic biodiversity evolution by separating the Sepkoski curve into its exponential growth part and its variation part. In this scenario, the total biodiversity curve $Total\mbox{-}BD$ can be written as the summation of its linear part $OT\mbox{-}BD$ and its net variation part $BD$ (Fig. 2d):
\begin{equation}
Total\mbox{-}BD=OT\mbox{-}BD+BD.
\end{equation}

Hence, we obtain the biodiversity curve $Curve\_BD$ after nondimensionalization of $BD$:
\begin{equation} Curve\_BD=nondim(BD).\end{equation}

\section{The tectonic cause of mass extinctions}

\subsection{Motivation}

We construct the tectonic curve based on the climatic and eustatic data in consideration of the phase relationships among $Curve\_BD$, $Curve\_CC$ and $Curve\_SL$.

\subsection{The consensus climate curve}

We denote the three independent results on Phanerozoic global climate in ref. [5] [6], [7], [4] as $C^1_0$, $C^2_0$, $C^3_0$ respectively after linear interpolation:
\begin{equation} C^1_0(t):\ \mbox{ref. [5] [6]},\end{equation}
\begin{equation} C^2_0(t):\ \mbox{ref. [7]},\end{equation}
\begin{equation} C^3_0(t):\ \mbox{ref. [4]}.\end{equation}
The missing $^{87}Sr/^{86}Sr$ in ref. [7] in lower Cambrian are obtained from ref. \cite{Veizer et al. 1999} for $C^2_0$. We obtain three dimensionless global climate curves after nondimensionalization:
\begin{equation} C^1(t)=nondim(C^1_0(t)), \end{equation}
\begin{equation} C^2(t)=nondim(C^2_0(t)), \end{equation}
\begin{equation} C^3(t)=nondim(C^3_0(t)). \end{equation}

Hence, we obtain the consensus climate curve $Curve\_CC$ by synthesizing the above three results $C^1$, $C^2$ and $C^3$ (Fig. S1a):
\begin{equation} Curve\_CC=nondim((C^1+C^2+C^3)/3). \end{equation}

\subsection{The consensus sea level curve}

We denote the Phanerozoic sea level curves in ref. [8] and in ref. [9] [10] as $S^1_0$ and $S^2_0$ (via linear interpolation) respectively:
\begin{equation} S^1_0(t):\ \mbox{ref. [8]}, \end{equation}
\begin{equation} S^2_0(t):\ \mbox{ref. [9] [10]}. \end{equation}
And we obtain the dimensionless sea level curves after nondimensionalization:
\begin{equation}S^1(t)=nondim(S^1_0(t)),\end{equation}
\begin{equation}S^2(t)=nondim(S^2_0(t)). \end{equation}

Hence we obtain the consensus sea level curve $Curve\_SL$ by synthesizing the two results $S^1$ and $S^2$ (Fig. S1c):
\begin{equation} Curve\_SL=nondim((S^1+S^2)/2).\end{equation}

We can obtain the derivative curves $d\_CC$, $d\_SL$ and $d\_BD$ respectively as follows (Fig. 2b):
\begin{eqnarray}
d\_CC&=&\frac{\mathrm{d}}{\mathrm{d} t}(Curve\_CC)\\
d\_SL&=&\frac{\mathrm{d}}{\mathrm{d} t}(Curve\_SL)\\
d\_BD&=&\frac{\mathrm{d}}{\mathrm{d} t}(Curve\_BD).
\end{eqnarray}

\subsection{Correlation coefficients among $Curve\_CC$, $Curve\_SL$ and $Curve\_BD$}

So far, we have obtained the first group ($n=1$) of curves $Curve\_CC$, $Curve\_SL$ and $Curve\_BD$ to describe the Phanerozoic climate, sea level and biodiversity. They are all $5421$-element functions of time $t$.

There are three eras (Paleozoic, Mesozoic and Cenozoic) in the Phanerozoic eon, the time $t$ in the Phanerozoic eon can be concatenated as follow:
\begin{equation}t=[t([P]),t([M]),t([C])],\end{equation}
where the indices for the Paleozoic, Mesozoic and Cenozoic are as follows respectively:
\begin{eqnarray}
P&=&[(5421-5420),...,(5421-2510)],\ \mbox{for Paleozoic from 542.0 Ma to 251.0 Ma}, \\
M&=&[(5421-2510+1),...,(5421-655)],\ \mbox{for Mesozoic from 251.0 Ma to 65.5 Ma}, \\
C&=&[(5421-655+1),...,5421],\ \mbox{for Cenozoic from 65.5 Ma to today}.
\end{eqnarray}
Similarly, we define the indices for the other periods as follows:
\begin{eqnarray}
PMC&:&\ \mbox{for Phanerozoic from 542.0 Ma to 0 Ma}, \\
PM&:&\ \mbox{for Paleozoic and Mesozoic from 542.0 Ma to 65.5 Ma}, \\
MC&:&\ \mbox{for Mesozoic and Cenozoic from 251.0 Ma to 0 Ma}, \\
P \backslash L&:&\ \mbox{for Paleozoic except for Lopingian from 542.0 Ma to 260.4 Ma}, \\
L&:&\ \mbox{for Lopingian from 260.4 Ma to 251.0 Ma}, \\
L.M.Tr&:&\ \mbox{for Lower and Middle Triassic from 251.0 Ma to 228.7 Ma}, \\
M \backslash L.M.Tr&:&\ \mbox{\small for Mesozoic except for Lower and Middle Triassic from 228.7 Ma to 65.5 Ma}.
\end{eqnarray}

We can calculate the correlation coefficients $r^\rho_{\mu\nu}$ among $Curve\_CC$, $Curve\_SL$ and $Curve\_BD$ in certain periods respectively (Data\_2):
\begin{equation}
r^\rho_{\mu\nu}=corrcoef(\mbox{curve }\mu([\rho]),\mbox{curve }\nu([\rho]))
\end{equation}
where the subscripts
\begin{equation} \mu, \nu= C, S, B \end{equation}
for the curves $Curve\_CC$, $Curve\_SL$ and $Curve\_BD$ respectively, and the superscript
\begin{equation} \rho=P, M, C, PMC, PM, MC, P \backslash L, L, L.M.Tr, M \backslash L.M.Tr \end{equation}
for the corresponding periods respectively.

{\bf Note:} The correlation coefficients generally agree with one other in the calculations between $Curve\_BD$ and any of $Curve\_SL$, $S^1$, $S^2$, or between $Curve\_BD$ and any of $Curve\_CC$, $C^1$, $C^2$, $C^3$, i.e. in general:
\begin{equation}
r^\rho_{\mu\nu}(n)\sim r^\rho_{\mu\nu}(n'),\ n,n'=1,2,...,10.
\end{equation}
Therefore, the phase relationship of $Curve\_CC$, $Curve\_SL$ and $Curve\_BD$ is generally irrelevant with the weights in obtaining $Curve\_CC$ and $Curve\_SL$. The correlation coefficients are also irrelevant whether we nondimensionalize the curves, for instance:
\begin{equation}
\begin{array}{rcl}
&\ &corrcoef((S^1([P])+S^2([P]))/2, BD([P]))\\
&=&corrcoef(nondim((S^1([P])+S^2([P]))/2), nondim(BD([P])))\\
&=&corrcoef(Curve\_SL([P]), Curve\_BD([P]))\\
&=&r^P_{SB}.
\end{array}
\end{equation}

{\bf Note:} The first group ($n=1$) of curves $Curve\_CC$, $Curve\_SL$ and $Curve\_BD$ is the best among the $10$ similar groups of curves to describe the Phanerozoic climate, sea level and biodiversity.

\subsection{Three climate phases}

We propose three climate patterns CP I, CP II and CP III in the Phanerozoic eon based on the positive or negative correlations among $Curve\_CC$, $Curve\_SL$ and $Curve\_BD$. Interestingly, the time between the positive correlation periods and the negative correlation periods agree with the Paleozoic-Mesozoic boundary and the Mesozoic-Cenozoic boundary.

(1) We have
\begin{eqnarray}
r^P_{SB}&=&0.5929>0\\
r^P_{BC}&=&0.1136>0\\
r^P_{CS}&=&0.4942>0
\end{eqnarray}
which indicate the positive correlations among $Curve\_CC$, $Curve\_SL$ and $Curve\_BD$ in the Paleozoic era. This is called the first climate pattern (CP I);

(2) We have
\begin{eqnarray}
r^M_{SB}&=&0.9054>0\\
r^M_{BC}&=&-0.4308<0\\
r^M_{CS}&=&-0.6171<0
\end{eqnarray}
which indicate the negative correlations between $Curve\_CC$ and $Curve\_SL$ and between $Curve\_CC$ and $Curve\_BD$, and the positive correlation between $Curve\_SL$ and $Curve\_BD$ in the Mesozoic era. This is called the second climate pattern (CP II);

(3) We have
\begin{eqnarray}
r^C_{SB}&=&-0.8314<0\\
r^C_{BC}&=&-0.8814<0\\
r^C_{CS}&=&0.9501>0
\end{eqnarray}
which indicate the negative correlations between $Curve\_CC$ and $Curve\_BD$ and between $Curve\_SL$ and $Curve\_BD$, and the positive correlation between $Curve\_SL$ and $Curve\_CC$ in the Cenozoic era. This is called the third climate pattern (CP III).

We define the average correlation coefficient $R^+$ in the positive correlation periods:
\begin{equation}
R^+=\frac{w^P\cdot r^P_{SB}+w^P\cdot r^P_{BC}+w^P\cdot r^P_{CS}+ w^M\cdot r^M_{SB}+ w^C\cdot r^C_{CS}} {w^P +w^P +w^P + w^M + w^C },
\end{equation}
and the average correlation coefficient $R^-$ in the negative correlation periods:
\begin{equation}
R^-=\frac{ w^M\cdot r^M_{BC}+ w^M\cdot r^M_{CS}+ w^C\cdot r^C_{SB}+ w^C\cdot r^C_{BC}} { w^M + w^M + w^C + w^C },
\end{equation}
where the weights $w^{\rho}$ are the durations of Paleozoic, Mesozoic and Cenozoic respectively:
\begin{eqnarray}
w^P&=&542.0-251.0=291.0\ \mbox{Myr}\\
w^M&=&251.0-65.5=185.5\ \mbox{Myr}\\
w^C&=&65.5\ \mbox{Myr}.
\end{eqnarray}
And we denote the difference between $R^+$ and $R^-$ as
\begin{equation}
\Delta R= R^+-R^-.
\end{equation}

We define the average abstract correlation coefficient $Q$ for the positive as well as the negative correlation periods as:
\begin{equation}
Q=\frac{1}{w^{P} + w^{M} + w^{C}}\sum_{\rho=P, M, C} w^{\rho}\cdot (|r^\rho_{SB}| + |r^\rho_{BC}| + |r^\rho_{CS}|),
\end{equation}
and the average abstract correlation coefficient $Q'$ for the mixtures of positive and negative correlation periods as:
\begin{equation}
Q'=\frac{1}{w^{PMC} + w^{PM} + w^{MC}}\sum_{\rho=PMC, PM, MC} w^{\rho}\cdot (|r^\rho_{SB}| + |r^\rho_{BC}| + |r^\rho_{CS}|),
\end{equation}
where the remaining weights $w^{\rho}$ are:
\begin{eqnarray}
w^{PMC}&=&542.0 \ \mbox{Myr}\\
w^{PM}&=&542.0-65.5=476.5\ \mbox{Myr}\\
w^{MC}&=&251.0\ \mbox{Myr}.
\end{eqnarray}
And we denote the difference between $Q$ and $Q'$ as
\begin{equation}
\Delta Q= Q-Q'.
\end{equation}

We found that the abstract correlation coefficients $|r^{PMC}_{\mu\nu}|$, $|r^{PM}_{\mu\nu}|$ and $|r^{MC}_{\mu\nu}|$ in the mixtures of positive and negative periods $\rho=PMC, PM, MC$ are obviously less than the abstract values $|r^{P}_{\mu\nu}|$, $|r^{M}_{\mu\nu}|$ and $|r^{C}_{\mu\nu}|$ in the positive or negative periods, namely in the Paleozoic, Mesozoic and Cenozoic eras. Therefore, the three climate patterns naturally correspond to the Paleozoic, Mesozoic and Cenozoic eras respectively. Based on the data of the first group (n=1) of curves $Curve\_CC$, $Curve\_SL$ and $Curve\_BD$, we have:
\begin{eqnarray}
R^+=R^+(1)&>&0\ (\mbox{tend to be equal to }1)\\
R^-=R^-(1)&<&0\ (\mbox{tend to be equal to }-1)\\
\Delta R=\Delta R(1)&\gg&0\\
Q=Q(1)&\sim&1\ (\mbox{tend to be equal to }1)\\
Q'=Q'(1)&\sim&0\ (\mbox{tend to be equal to }0)\\
\Delta Q=\Delta Q(1)&>&0
\end{eqnarray}
which furthermore shows that the division of three climate patterns CP I, CP II and CP III is essential property of the evolutionary earth's spheres.

{\bf Note:} These relations are still valid for the other groups of curves ($n=2,3,...,10$).

\subsection{The P-Tr extinction was caused by the climate phase reverse between CP I and CP II}

We summarize the reasons to explain the P-Tr extinction by the climate phase reverse event as follows.
\begin{itemize}
  \item Successful explanation of the Sepkoski curve by the tectono-genomic curve based on the climate phase reverse event (Fig 1)
  \item The climate phase reverse event between CP I and CP II happened at P-Tr boundary (Fig 2a)
  \item The sharp peaks of $d\_CC$ at the Guadalupian-Lopingian boundary and at the P-Tr boundary (Fig 2b)
  \item Abnormal climate trend in the Lopingian epoch
  \item Different animal extinction patterns at the Guadalupian-Lopingian boundary and at the P-Tr boundary.
\end{itemize}

\subsection{The tectonic curve and the tectonic contribution to the biodiversity variation}

The phase of $Curve\_SL$ is about the same with the phase of $Curve\_BD$ in the Phanerozoic eon. And the phase of $Curve\_CC$ is about the same with the phase of $Curve\_BD$ in the Paleozoic era (CP I), while it is about the opposite in the Mesozoic era (CP II) and in the Cenozoic era (CP III). Accordingly, we define the associate tectonic curve $Curve\_Tectonic\_0$ by combining the consensus sea level curve and the consensus climate curve as follow (Fig S1e):
\begin{equation}
\begin{array}{rcl}
Curve\_Tectonic\_0 = [(Curve\_SL([P])&+&Curve\_CC([P]))/2,\\
(Curve\_SL([MC])&-&Curve\_CC([MC]))/2].
\end{array}
\end{equation}
We define the tectonic curve $Curve\_Tectonic$ with the same standard deviation of the net variation biodiversity curve $BD$:
\begin{equation}
Curve\_Tectonic=(Curve\_Tectonic\_0 - mean(Curve\_Tectonic\_0)) \cdot a_{std},
\end{equation}
where
\begin{equation}
a_{std}=\frac{std(BD)}{std(Curve\_Tectonic\_0 - mean(Curve\_Tectonic\_0))}.
\end{equation}

The tectonic curve $Curve\_Tectonic$ represents the tectonic (sea level and climate) contribution to the biodiversity evolution. We can calculate the correlation coefficient between the tectonic curve and the biodiversity curve in the Paleozoic era or in the Mesozoic and Cenozoic eras:
\begin{equation}
\begin{array}{rcl}
r^P_{B+}&=& corrcoef(Curve\_Tectonic([P]), Curve\_BD([P]))\\
&=&0.421,
\end{array}
\end{equation}
\begin{equation}
\begin{array}{rcl}
r^{MC}_{B-}&=&corrcoef(Curve\_Tectonic([MC]), Curve\_BD([MC]))\\
&=&0.878.
\end{array}
\end{equation}
Accordingly, we found that the  tectonic curve $Curve\_Tectonic$ is positively correlated with the biodiversity curve $Curve\_BD$ either in the Paleozoic era or in the Mesozoic and Cenozoic eras.

\section{The genomic contribution to the biodiversity evolution}

\subsection{Motivation}

We construct the genomic curve based on the observation of equality between the growth rate $k_{GS}$ in genome size evolution and the growth rate $k_{BD}$ in biodiversity evolution.

\subsection{The overall trend of genome size evolution}

\subsubsection{The log-normal distribution of genome size}

We found that the genome sizes of species in a taxon are log-normally distributed in general, which were verified in the following $7$ taxa (Fig. S2a):
\begin{equation}
log(G(\lambda,sp(\lambda))) \mbox{ are normally distributed},
\end{equation}
where $G(\lambda,sp(\lambda))$ are the genome sizes of all the species $sp(\lambda)$ ($sp(\lambda)=1,2,...,s_m(\lambda)$) in the taxon $\lambda$ in the genome size databases, and
\begin{equation}
\begin{array}{rcl}
\lambda=1 &:& \mbox{Diploblostica}\\
\lambda=2 &:& \mbox{Protostomia}\\
\lambda=3 &:& \mbox{Deuterostomia}\\
\lambda=4 &:& \mbox{Bryophyte}\\
\lambda=5 &:& \mbox{Pteridophyte}\\
\lambda=6 &:& \mbox{Gymnosperm}\\
\lambda=7 &:& \mbox{Angiosperm}.
\end{array}
\end{equation}
Due to the additivity of normal distribution, the genome sizes of animals, plants, or eukaryotes are also log-normal distributed. We obtain the means of logarithm of genome sizes and the standard deviations of logarithm of genome sizes as follows:
\begin{equation} G^P_{mean\_log}(\lambda)=mean(log(G(\lambda,sp(\lambda)))), \end{equation}
and
\begin{equation} G^P_{sd\_log}(\lambda)=std(log(G(\lambda,sp(\lambda)))), \end{equation}
where $sp(\lambda)=1,2,...,s_m(\lambda)$.
Denote $G^*$ as the mean logarithm of genome sizes of all the contemporary eukaryotes:
\begin{equation} G^*=mean(log(G(sp))),\end{equation}
where $sp$ is all the contemporary eukaryotes in the genome size databases.

{\bf Note:} The log-normal distribution of genome size can be demonstrated by the common intersection point $\Omega$ for the following regression lines (Fig 3b):
\begin{eqnarray}
&\mbox{regression line of}& G_{mean\_log}(\lambda')  \mbox{ on } G_{sp}(\lambda')\\
&\mbox{regression line of}& G_{mean\_log}(\lambda') \pm \chi \cdot G_{sd\_log}(\lambda') \mbox{ on } G_{sp}(\lambda')\\
&\mbox{regression line of}& G_{mean\_log}(\lambda') \pm \chi' \cdot G_{sd\_log}(\lambda') \mbox{ on } G_{sp}(\lambda')\\
&\mbox{regression line of}& max(G(\lambda',sp(\lambda'))) \mbox{ on } G_{sp}(\lambda')\\
&\mbox{regression line of}& min(G(\lambda',sp(\lambda'))) \mbox{ on } G_{sp}(\lambda')\\
&\mbox{regression line of}& G_{mean\_log}^P(\lambda)  \mbox{ on } G_{sp}^P(\lambda)\\
&\mbox{regression line of}& G_{mean\_log}^P(\lambda) \pm \chi \cdot G_{sd\_log}^P(\lambda) \mbox{ on } G_{sp}^P(\lambda)\\
&\mbox{regression line of}& G_{mean\_log}^P(\lambda) \pm \chi' \cdot G_{sd\_log}^P(\lambda) \mbox{ on } G_{sp}^P(\lambda)\\
&\mbox{regression line of}& max(G(\lambda,sp(\lambda))) \mbox{ on } G_{sp}^P(\lambda)\\
&\mbox{regression line of}& min(G(\lambda,sp(\lambda))) \mbox{ on } G_{sp}^P(\lambda)
\end{eqnarray}
where $\lambda=1,2,...,7$ for the above $7$ taxa, $\lambda'=1,2,...,19+53$ for $19$ animal taxa and $53$ angiosperm taxa, $\chi=1.5677$ and $\chi_1=3.1867$. The values of $G_{sd\_log}$ tend to decline with respect to $G_{sp}$ that is proportional to the origin time of taxa (Fig S2b).

\subsubsection{The exponential overall trend of genome size evolution}

We assume the approximate origin times $T(\lambda)$ for the taxa $\lambda=1,2,...,7$ as follows:
\begin{equation}
\begin{array}{rcl}
T(1)&=&560.0\ \mbox{Ma}\\
T(2)&=&542.0\ \mbox{Ma, PreCm-Cm}\\
T(3)&=&525.0\ \mbox{Ma}\\
T(4)&=&488.3\ \mbox{Ma, Cm-O}\\
T(5)&=&416.0\ \mbox{Ma, S-D}\\
T(6)&=&359.2\ \mbox{Ma, D-C}\\
T(7)&=&145.5\ \mbox{Ma, J-K}
\end{array}
\end{equation}
We observed a rough proportional relationship between $G^P_{mean\_log}(\lambda)$ and $T(\lambda)$. Because $G^P_{mean\_log}(\lambda)$ is the mean genome size of the ``contemporary species'', we should introduce a new notion (the specific genome size) to indicate the mean genome sizes of the ``ancient species'' in taxa $\lambda=1,2,...,7$ at its origin time $T(\lambda)$. Here, we define the specific genome size $G^P_{sp}$ as:
\begin{equation} G^P_{sp}(\lambda)=G^P_{mean\_log}(\lambda)-\chi \cdot G^P_{sd\_log}(\lambda), \end{equation}
where we let $\chi=1.5677$ such that the intercept of the regression line of $G^P_{sp}(\lambda)$ on $T(\lambda)$ is equal to $G^*$. We found that $G^P_{sp}(\lambda)$ is generally proportional to $T(\lambda)$ (Fig. 3a). We define the regression line of $G^P_{sp}(\lambda)$ on  $T(\lambda)$ as overall trend of genome size curve:
\begin{equation}
OT\mbox{-}GS=k_{GS}(-t)+log(N^0_{genome}).
\end{equation}
This equation is equivalent to the exponential overall trend of genome size evolution:
\begin{equation} N_{genome}(t) = N^0_{genome}\ \exp(-t/\tau_{GS}),
\end{equation}
where the genome size constant is $N^0_{genome}=2.16\times10^9$ base pairs (bp) and the ``e-folding time'' in genome size evolution is $\tau_{GS}=256.56$ (Myr). The growth rate (namely the slope) of $OT\mbox{-}GS$ is $k_{GS} =1/\tau_{GS} =0.0038977$ $\mbox{Myr}^{-1}$.

{\bf Note:} The exponential overall trend of genome size evolution obtained in the Phanerozoic eon can be extrapolated to the Precambrian period. This extrapolation result according to the value of $k_{GS}$ is reasonable to show that the least genome size at $3800$ Ma (about the beginning of life) is about several hundreds of base pairs (Fig 3d).

\subsection{The agreement between the overall trend of genome size evolution and the overall trend of biodiversity evolution}

We found the closely relationship between the genome size evolution and the biodiversity evolution (Fig 3d). Both the overall trend of genome size evolution and the overall trend of biodiversity evolution are exponential; and the exponential growth rate in the genome size evolution ($k_{GS}=0.0038977$ $\mbox{Myr}^{-1}$) (Fig 3a, 3d) is approximately equal to the exponential growth rate in the biodiversity evolution ($k_{BD}=0.0038598$ $\mbox{Myr}^{-1}$) (Fig 2d, 3d):
\begin{equation}
k_{GS} \approx k_{BD},
\end{equation}
which is equivalent to that the e-folding time in the genome size evolution ($\tau_{GS}=256.56$ Myr) is approximately equal to the e-folding time in the biodiversity evolution ($\tau_{BD}=259.08$ Myr):
\begin{equation} \tau_{GS} \approx \tau_{BD}. \end{equation}

\subsection{Explanation of the declining Phanerozoic background extinction rates}

Let $rate\_ori$ and $rate\_ext$ denote the Phanerozoic biodiversity origination rate and extinction rate respectively:
\begin{equation}
rate\_ori:\ \mbox{ref. [2]},
\end{equation}
\begin{equation}
rate\_ext:\ \mbox{ref. [2]},
\end{equation}
which agree with each other in general.
The difference and the average of them are as follows respectively:
\begin{equation}
rate_{o-e}=(rate\_ori-rate\_ext)/2,
\end{equation}
\begin{equation}
rate_{o+e}=(rate\_ori+rate\_ext)/2,
\end{equation}
where $rate_{o-e}$ should agree with $d\_BD$ according to their definitions, and $rate_{o+e}$ represents the variation of biodiversity in the Phanerozoic eon. The outline of $rate_{o+e}$ indicates the declining Phanerozoic background extinction rates \cite{Flessa and Jablonski 1985} \cite{Van Valen 1985} \cite{Sepkoski 1991} \cite{Gilinsky 1994} \cite{Alroy 1998}.

We define an essential biodiversity background variation rate by:
\begin{equation}
rate\_essential=[amp(1) \cdot rate_{o+e}(1),amp(2) \cdot rate_{o+e}(2),...,amp(5421)\cdot rate_{o+e}(5421)],
\end{equation}
where
\begin{equation}
amp=exp(k_{GS}\cdot(-t+542.0)).
\end{equation}
The outline of $rate\_essential$ is generally horizontal (NOT declining). Especially, the peaks of the curve $rate\_essential$ at P-Tr boundary and at K-Pg boundary are very high, which naturally divide the Phanerozoic eon into three climate phases (Fig 2c).

In the split scenario of biodiversification, we can explain the ``declining'' background extinction rates in the Phanerozoic eon. Firstly, there does not exist a tendency in the essential biodiversity background rate curve $rate\_essential$. This essential rate was caused by the random tectonic contribution (no tendency) to the biodiversity evolution:
\begin{equation}
rate\_essential=\frac{\mbox{variation of biodiversity}}{\mbox{tectonic contribution to biodiversity}}.
\end{equation}
Then, the declining tendency in the observed background extinction or origination rates was caused by the genomic contribution to the biodiversity evolution:
\begin{equation}
rate_{o+e}=\frac{\mbox{variation of biodiversity}}{\mbox{tectonic contribution+genomic contribution to biodiversity}}.
\end{equation}
It follows that (Fig 2c):
\begin{equation}
rate_{o+e}=exp(-k_{GS}\cdot(-t+542.0)) \cdot rate\_essential,
\end{equation}
where $rate_{o+e}$ is declining due to the factor $exp(-k_{GS}\cdot(-t+542.0))$.

The genomic contribution to the biodiversity plays a significant role in the robustness of biodiversity evolution: the random tectonic contribution can hardly wipe out all the life on the earth thanks for the exponential growth genomic contribution to the biodiversity evolution.

\subsection{Calculating the origin time of taxa based on the overall trend of genome size evolution}

\subsubsection{The three-stage pattern in Metazoan origination}

We can calculate the origin time of animal taxa according to the linear relationship between the origin time and the specific genome size. We obtained the specific genome sizes of the $19$ taxa in the Animal Genome Size Database (Nematodes, Chordates, Sponges, Ctenophores, Tardigrades, Miscellaneous Inverts, Arthropod, Annelid, Myriapods, Flatworms, Rotifers, Cnidarians, Fish, Echinoderm, Molluscs, Bird, Reptile, Amphibian, Mammal):
\begin{equation} G_{sp}^{animal}(\lambda^{animal})= G_{mean\_log}^{animal}(\lambda^{animal})- \chi\cdot G_{sd\_log}^{animal}(\lambda^{animal}),
\end{equation}
where $\lambda^{animal}=1,2,...,19$.
We can obtain the origin order of these $19$ taxa by comparing their specific genome sizes. Hence, we can classify these $19$ taxa into Basal metazoa, Protostomia and Deuterostomia according to cluster analysis of their specific genome sizes (Data\_3). Our result supports the three-stage pattern in Metazoan origination based on fossil records \cite{Conway-Morris 1993} \cite{Conway-Morris 1989} \cite{Budd and Jensen 2000} \cite{Shu 2005} \cite{Shu et al. 2004} \cite{Valentine 2001} \cite{Shu et al. 2009}.

\subsubsection{On angiosperm origination}

Similarly, we can calculate the origin time of angiosperm taxa according to the linear relationship between the origin time and the specific genome size. We obtained the specific genome sizes of the $53$ taxa of angiosperms in the Plant DNA C-value Database (we chose the taxa whose number of species is greater than $20$ in the calculations):
\begin{equation}
G_{sp}^{angiosperm}(\lambda^{angiosperm})= G_{mean\_log}^{angiosperm}(\lambda^{angiosperm})- \chi \cdot G_{sd\_log}^{angiosperm}(\lambda^{angiosperm}),
\end{equation}
where $\lambda^{angiosperm}=1,2,...,53$.
We can obtain the origin order of these $53$ taxa by comparing their specific genome sizes. Hence, we can classify these $53$ taxa into Dicotyledoneae and Monocotyledoneae (Data\_3).

{\bf Note:} The validity of our theory on genome size evolution is supported by its reasonable explanation of metazoan origination and angiosperm origination.

{\bf Notation:} We denote the mean logarithm genome size, the standard deviation genome size and the specific genome sizes by concatenations for all the $19$ animal taxa and the $53$ plant taxa:
\begin{eqnarray}
G_{mean\_log}&=&[\ G_{mean\_log}^{animal},\ G_{mean\_log}^{angiosperm}\ ]\\
G_{sd\_log}&=&[\ G_{sd\_log}^{animal},\ G_{sd\_log}^{angiosperm}\ ]\\
G_{sp}&=&[\ G_{sp}^{animal},\ G_{sp}^{angiosperm}\ ].
\end{eqnarray}

\subsection{The phylogenetic tree based on the correlation among genome size distributions}

We found that the phylogenetic tree for taxa can be easily obtained based on the correlation coefficients among their genome size distributions. We denote the genome size distribution for a taxon $\lambda$ by:
\begin{equation}
D_{gs}(\lambda,:)=[D_{gs}(\lambda,1), D_{gs}(\lambda,2),..., D_{gs}(\lambda,k),...,D_{gs}(\lambda,cutoff_{gs})],
\end{equation}
where there are $D_{gs}(\lambda,k)$ species in taxon $\lambda$ whose genome size is between $(k-1)\cdot step_{gs}$ and $k\cdot step_{gs}$, the genome size step $step_{gs}=0.01$ picogram (pg) and the genome size cutoff is $cutoff_{gs}=2000$.
Hence, we define the genome size distribution distance matrix $M_{gs}(\lambda_1,\lambda_2)$ among taxa by:
\begin{equation}
M_{gs}(\lambda_1,\lambda_2)=1-corrcoef(D_{gs}(\lambda_1,:),D_{gs}(\lambda_2,:)),
\end{equation}
by which, we can draw the phylogenetic tree of the taxa.

We can obtain the genome size distributions $D^P_{gs}(\lambda,:)$ and consequently obtain the genome size distribution distance matrix $M^P_{gs}(\lambda_1,\lambda_2)$ among the above $7$ taxa as follows:
\begin{equation}
M^P_{gs}(\lambda_1,\lambda_2)=1-corrcoef(D^P_{gs}(\lambda_1,:),D^P_{gs}(\lambda_2,:)),
\end{equation}
where $\lambda_1, \lambda_2 = 1,2,...,7$. Hence, we can draw the phylogenetic tree of the $7$ taxa based on $M^P_{gs}$ (Fig S2c).

We can obtain the genome size distributions $D^{animal}_{gs}(\lambda,:)$ and consequently obtain the genome size distribution distance matrix $M^{animal}_{gs}(\lambda_1,\lambda_2)$ among the above $19$ animal taxa as follows:
\begin{equation}
M^{animal}_{gs}(\lambda_1^{animal},\lambda_2^{animal}) = 1-corrcoef(D^{animal}_{gs}(\lambda_1^{animal},:),D^{animal}_{gs}(\lambda_2^{animal},:)),
\end{equation}
where $\lambda_1^{animal}, \lambda_2^{animal} = 1,2,...,19$. Hence, we can draw the phylogenetic tree of the $19$ taxa based on $M^{animal}_{gs}$ (Fig 3c).

We can obtain the genome size distributions $D^{angiosperm}_{gs}(\lambda,:)$ and consequently obtain the genome size distribution distance matrix $M^{angiosperm}_{gs}(\lambda_1,\lambda_2)$ among the $25$ angiosperm taxa (we chose $25$ angiosperm taxa whose number of species is greater than $50$ in the Plant DNA C-value database in order to obtain nontrivial distributions) as follows:
\begin{equation}
M^{angiosperm}_{gs}(\lambda_1^{angiosperm},\lambda_2^{angiosperm}) = 1-corrcoef(D^{angiosperm}_{gs}(\lambda_1^{angiosperm},:),D^{angiosperm}_{gs}(\lambda_2^{angiosperm},:)),
\end{equation}
where $\lambda_1^{angiosperm}, \lambda_2^{angiosperm} = 1,2,...,25$. Hence, we can draw the phylogenetic tree of the $25$ taxa based on $M^{angiosperm}_{gs}$ (Fig S2d).

These phylogenetic trees based on genome size distribution distance matrices generally agree with the traditional phylogenetic trees respectively, which is an evidence to show the close relationship between the genome evolution and the biodiversity evolution.

{\bf Software:} PHYLIP to draw the phylogenetic trees (Neighbor-Joining) in this paper \cite{Felsenstein 1981}.

\subsection{The varying velocity of molecular clock among taxa}

The growth rates $k_{GS}(\lambda)$ of overall genome size evolution $OT_{taxa}(\lambda)$ for taxa $\lambda$ are not constant, though we have an average growth rate $k_{GS}$ for $OT\mbox{-}GS$. We have an approximate relationship that the earlier the origin time $T_{ori}(\lambda)$ is, the slower the growth rate $k_{GS}(\lambda)$ is:
\begin{equation}
(k_{GS}(\lambda)-k_{GS})\cdot T_{ori}(\lambda) \doteq \hat{G},
\end{equation}
where the constant $\hat{G}$ is the difference between the intercept of the overall trend of mean logarithm genome size $OT_{mean\_log}$ and the intercept of $OT\mbox{-}GS$.

\subsection{The genomic curve and the genomic contribution to the biodiversity evolution}

We define the genomic curve by a straight line with slope $k_{GS}$ and the undetermined intercept $b_{today}$:
\begin{equation}
Curve\_Genomic=k_{GS}\cdot (-t)+b_{today},
\end{equation}
which represents the exponential contribution to the biodiversity evolution.

\section{Construction of the tectono-genomic curve}

\subsection{The synthesis scheme for the tectono-genomic curve}

The above undetermined intercept of the genomic curve can be defined as:
\begin{equation}
b_{today}=Curve\_Sepkoski(today)-Curve\_Tectonic(today)
\end{equation}
such that $Curve\_TectonoGenomic(5421)=Curve\_Sepkoski(5421)$.

We define the tectono-genomic curve by synthesizing the tectonic curve $Curve\_Tectonic$ and the genomic curve $Curve\_Genomic$ (Fig 1):
\begin{equation}
Curve\_TectonoGenomic=\exp(Curve\_Tectonic+Curve\_Genomic),
\end{equation}
which agrees very well with the Phanerozoic biodiversity curve $Curve\_Sepkoski$:
\begin{equation}
Curve\_TectonoGenomic \approx Curve\_Sepkoski.
\end{equation}
Thus, the Sepkoski curve based on fossil records can be explained by the tectono-genomic curve based on climatic, eustatic and genomic data.

\subsection{The driving forces of biodiversity evolution at the molecular level and at the species level}

Thus, we have explained the Sepkoski curve in the split scenario. The exponential growth part in the Phanerozoic biodiversity evolution was driven by the genome size evolution on one hand, and the variation of the the Phanerozoic biodiversity evolution was caused by the Phanerozoic sea level fluctuation and climate change on the other hand.

The successful explanation of the Phanerozoic biodiversity curve $Curve\_Sepkoski$ shows that the driving force of the biodiversity evolution is the tectono-genomic driving force. There are two independent tectonic and genomic driving forces in the biodiversity evolution. The first driving force originated from the plate tectonics movement at the species level; while the second driving force originated from the genome evolution at the molecular level.

\section{The error analysis and reasonability analysis}

\subsection{The agreement between the Sepkoski curve and the tectono-genomic curve}

\subsubsection{The error analysis of the consensus climate curve}

We obtain the first weighted average climate curve $C_{w1}$ by choosing the corresponding $\Delta R(n)$, $n=2,3,4$ as the weights $w1$ for $C^1$, $C^2$ and $C^3$ as follows:
\begin{equation}
\begin{array}{rcl}
w1&=&[\Delta R(2),\Delta R(3),\Delta R(4)]/(\Delta R(2)+\Delta R(3)+\Delta R(4))\\
&=&[0.3454,\ 0.1611,\ 0.4935],
\end{array}
\end{equation}
hence,
\begin{equation}
C_{w1}=nondim(w1(1) \cdot C^1+w1(2) \cdot C^2+w1(3) \cdot C^3).
\end{equation}

We obtain the second weighted average climate curve $C_{w2}$ by choosing the corresponding correlation coefficients as the weights $w2$ for $C^1$, $C^2$ and $C^3$ as follows:
\begin{equation}
\begin{array}{rcl}
w2&=&[corrcoef(Curve\_CC,C^1), corrcoef(Curve\_CC,C^2),\\
 & & corrcoef(Curve\_CC,C^3)]/\\
& &(corrcoef(Curve\_CC,C^1)+ corrcoef(Curve\_CC,C^2)+\\
 & &+corrcoef(Curve\_CC,C^3))\\
&=&[0.4865,\ 0.2796,\ 0.2339],
\end{array}
\end{equation}
hence,
\begin{equation}
C_{w2}=nondim(w2(1) \cdot C^1+w2(2) \cdot C^2+w2(3) \cdot C^3).
\end{equation}

We can obtain a weighted average climate curve $C_{w}$ by choosing the average of $w1$ and $w2$ as the weights $w$ for $C^1$, $C^2$ and $C^3$ as follows:
\begin{equation}
\begin{array}{rcl}
w&=&(w1+w2)/2\\
&=&[0.4159,\ 0.2204,\ 0.3637],
\end{array}
\end{equation}
hence,
\begin{equation}
C_{w}=nondim(w(1) \cdot C^1+w(2) \cdot C^2+w(3) \cdot C^3),
\end{equation}
which agrees with $Curve\_CC$.

The weights $w1$ or $w2$ can be referred to as credibilities for the independent curves $C^1$, $C^2$ and $C^3$. Both of $C_{w1}$ and $C_{w2}$ are reasonable estimations of the Phanerozoic climate. So, we can consider the zone between $C_{w1}$ and $C_{w2}$ as the error range of $Curve\_CC$, whose upper range $C_{upper}$ and lower range $C_{lower}$ are about as follows (Fig S1b):
\begin{equation}
C_{upper}=max(C_{w1},C_{w2}),
\end{equation}
\begin{equation}
C_{lower}=min(C_{w1},C_{w2}).
\end{equation}

\subsubsection{The error analysis of the consensus sea level curve}

We obtain the weighted average sea level curve $S_{w}$ by choosing the corresponding $\Delta R(n)$, $n=10,11$ as the weights $w'$ for $S^1$ and $C^2$ as follows:
\begin{equation}
\begin{array}{rcl}
w'&=&[\Delta R(10), \Delta R(11)]/(\Delta R(10)+\Delta R(11))\\
&=&[0.4872,\ 0.5128],
\end{array}
\end{equation}
hence,
\begin{equation}
S_{w}=nondim(w'(1) \cdot S^1+w'(2) \cdot S^2),
\end{equation}
which agrees with $Curve\_SL$.

We can consider the zone between $S^1$ and $S^2$ as the error range of $Curve\_SL$, whose upper range $S_{upper}$ and lower range $S_{lower}$ are about as follows (Fig S1c):
\begin{equation}
S_{upper}=max(S^1,S^2),
\end{equation}
\begin{equation}
S_{lower}=min(S^1,S^2).
\end{equation}

\subsubsection{The error analysis of the Sepkoski curve}

We can consider the zone between $Curve\_S\_AllGenera$ and $Curve\_S\_WellResolvedGenera$ as the error range of $Curve\_Sepkoski$ (Fig 1):
\begin{equation} Curve\_S\_AllGenera:\ \mbox{ref. [3]},
\end{equation}
\begin{equation} Curve\_S\_WellResolvedGenera:\ \mbox{ref. [3]}, \end{equation}
where $Curve\_S\_AllGenera$ is the Phanerozoic biodiversity curve based on all the genera in Sepkoski's data and $Curve\_S\_WellResolvedGenera$ is the Phanerozoic biodiversity curve based on well resolved genera in Sepkoski's data.

\subsubsection{The error analysis of the tectono-genomic curve}

In consideration of the error ranges of $Curve\_CC$ and $Curve\_SL$ as well as their  phase relationships, we define the associate upper tectono-genomic curve $Curve\_TG\_upper\_0$ and the associate lower tectono-genomic curve $Curve\_TG\_lower\_0$ as follow:
\begin{equation}
\begin{array}{l}
Curve\_TG\_upper\_0 = [(S_{upper}([P])+C_{upper}([P]))/2, (S_{upper}([MC])-C_{lower}([MC]))/2],
\end{array}
\end{equation}
\begin{equation}
\begin{array}{l}
Curve\_TG\_lower\_0 = [(S_{lower}([P])+C_{lower}([P]))/2, (S_{lower}([MC])-C_{upper}([MC]))/2].
\end{array}
\end{equation}
Furthermore, in the similar process and with the same parameters in construction of the tectono-genomic curve, we can obtain the upper range and the lower range of the tectono-genomic curve as follows (Fig 1):
\begin{equation}
\begin{array}{l}
Curve\_TG\_upper = \exp(Curve\_Genomic + \\
\hspace{1cm} + a_{std} \cdot (Curve\_TG\_upper\_0  - mean(Curve\_TG\_upper\_0))),
\end{array}
\end{equation}
\begin{equation}
\begin{array}{l}
Curve\_TG\_lower = \exp(Curve\_Genomic + \\
\hspace{1cm} + a_{std} \cdot (Curve\_TG\_lower\_0 - mean(Curve\_TG\_lower\_0))).
\end{array}
\end{equation}

\subsection{The reasonability of the principal conjectures}

\subsubsection{Reasonability of the climate phase reverse based on $r_{\mu\nu}^{\rho}(n)$}

We can obtain the following $10$ groups of curves to describe the Phanerozoic climate, sea level and biodiversity:
\begin{equation}
\begin{array}{lcclclc}
n=1\ &:&\  Curve\_SL&,&\  Curve\_BD&,&\ Curve\_CC\\
n=2\ &:&\  Curve\_SL&,&\  Curve\_BD&,&\ C^1\\
n=3\ &:&\  Curve\_SL&,&\  Curve\_BD&,&\ C^2\\
n=4\ &:&\  Curve\_SL&,&\  Curve\_BD&,&\ C^3\\
n=5\ &:&\  Curve\_SL&,&\  Curve\_BD&,&\ C_{w1}\\
n=6\ &:&\  Curve\_SL&,&\  Curve\_BD&,&\ C_{w2}\\
n=7\ &:&\  Curve\_SL&,&\  Curve\_BD&,&\ C_{w}\\
n=8\ &:&\  S^1&,&\  Curve\_BD&,&\ Curve\_CC\\
n=9\ &:&\  S^2&,&\  Curve\_BD&,&\ Curve\_CC\\
n=10\ &:&\  S_w&,&\  Curve\_BD&,&\ Curve\_CC\\
\end{array}
\end{equation}
And we can obtain the correlation coefficients $r^\rho_{\mu \nu}(n)$ among these groups of curves (Data\_2), where
\begin{equation}
\mu,\nu=S,B,C,C^1,C^2,C^3,C_{w1},C_{w2},C_{w},S^1,S^2,S_w
\end{equation}
for the curves $Curv\_SL$, $Curve\_BD$, $Curve\_CC$, $C^1$, $C^2$, $C^3$, $C_{w1}$, $C_{w2}$, $C_w$, $S^1$, $S^2$ and $S_w$ respectively.

We can define the corresponding average correlation coefficients for all the $10$ groups of curves ($n=1,2,...,10$) as follows:
\begin{equation}
R^+(n), R^-(n), \Delta R(n), Q(n), Q'(n), \Delta Q(n).
\end{equation}

The conclusions on the climate phases CP I, CP II and CP III based on the first group of curves ($n=1$) still hold for the cases of the other groups of curves ($n=2,3,...,10$). Namely, the following equations holds in general:
\begin{eqnarray}
r^P_{SB}(n)&>&0\\
r^P_{BC}(n)&>&0\\
r^P_{CS}(n)&>&0
\end{eqnarray}
for CP I,
\begin{eqnarray}
r^M_{SB}(n)&>&0\\
r^M_{BC}(n)&<&0\\
r^M_{CS}(n)&<&0
\end{eqnarray}
for CP II, and
\begin{eqnarray}
r^C_{SB}(n)&<&0\\
r^C_{BC}(n)&<&0\\
r^C_{CS}(n)&>&0
\end{eqnarray}
for CP III.

Furthermore, we have
\begin{eqnarray}
R^+(n)&>&0\ (\mbox{tend to be equal to }1)\\
R^-(n)&<&0\ (\mbox{tend to be equal to }-1)\\
\Delta R(n)&\gg&0\\
Q(n)&\sim&1\ (\mbox{tend to be equal to }1)\\
Q'(n)&\sim&0\ (\mbox{tend to be equal to }0)\\
\Delta Q(n)&>&0
\end{eqnarray}
which shows that the division of three climate phases is an essential property in the evolution rather than just random phenomenon in math games.

The explanation of the P-Tr extinction based on the phase reverse at P-Tr boundary is therefore valid regardless the disagreement in the raw data of the Phanerozoic climate and sea level. Especially, $\Delta R(1)$ and $\Delta Q(1)$ are relatively the maximum among these $10$ groups of curves, hence we chose the optimal first group of curves to describe the Phanerozoic climate, sea level and biodiversity throughout this paper.

The climate system was not stationary when coupling with the other earth's spheres around P-Tr boundary. We calculate the correlation coefficients $r_{\mu \nu}^{\rho}$, where $\rho= P\backslash L, L, L.M.Tr, M\backslash L.M.Tr$ in detail around P-Tr boundary. The curve $Curve\_CC$ varies instead in the opposite phase with $Curve\_SL$ and $Curve\_BD$ in Lopingian yet; and it varies instead in the same phase with $Curve\_SL$ and $Curve\_BD$ in Lower and Middle Triassic.

\subsubsection{Reasonability of the split scenario}

We summarize the reasons to propose the split scenario in observing the biodiversity evolution as follows.

(1) Evidences to support the close relationship between the genome evolution and the biodiversity evolution:
\begin{itemize}
  \item Exponential growth in both the genome size evolution and the biodiversity evolution
  \item Agreement between genome size growth rate $k_{GS}$ and biodiversity growth rate $k_{BD}$, namely $\tau_{GS}\approx \tau_{BD}$
  \item Favorable phylogenetic trees based on $M_{gs}^P$, $M_{gs}^{animal}$, $M_{gs}^{angiosperm}$, $M_{ci}^{all}$, $M_{ci}^{euk}$
  \item Verification of the three-stage pattern in Metazoan origination and the classification of dicotyledoneae and monocotyledoneae in angiosperm origination based on the overall trend in genome size evolution
  \item Reasonable extrapolation of the overall trend in genome size evolution obtained in Phanerozoic eon to the Precambrian period
  \item The relationship between phylogenetic trees of species by $M_{ci}$ and the evolutionary tree of codons by $M_{codon}$ based on the same matrix $\Delta$.
\end{itemize}

(2) Successful applications of the split scenario:
\begin{itemize}
  \item Explanation of the Sepkoski curve by the tectono-genomic curve in the split scenario
  \item Error analysis agreement between $Curve\_Sepkoski$ and $Curve\_TectonoGenomic$
  \item Explanation of the declining Phanerozoic background extinction rates
  \item Explanation of the robustness of biosphere in the tremendously changing environment.
\end{itemize}

\section{The genetic code evolution as the initial driving force in the biodiversity evolution}

\subsection{The evolutionary relationship between the tree of life and the tree of codon}

\subsubsection{The codon interval distribution $D_{ci}$}

We can obtain both the phylogenetic tree of species and the evolutionary tree of $64$ codons based on the codon interval distributions in the whole genomes. For a certain species $\alpha$ and a certain codon $n_c$ ($n_c=1,2,...,64$ for $64$ codons), we define the ``codon interval'' $I(n_c,\alpha,p)$ as the distance between a pair ($p$) of neighboring codon $n_c$'s in the whole genome sequence. We define the codon interval distribution
\begin{equation}
D_{ci}(n_c,\alpha,:)=[D_{ci}(n_c,\alpha,1), D_{ci}(n_c,\alpha,2),...,D_{ci}(n_c,\alpha,cutoff_{ci})].
\end{equation}
as the distribution of all the codon intervals $I(n_c,\alpha,p)$ in the whole genome sequence (reading in only one direction), where there are $D_{ci}(i)$ pairs of codon $n_c$'s with the distance $i$ (the cutoff of distance in the calculations is set as $cutoff_{ci}=1000$ bases). For a group of $N$ species, there are $64 \times N$ $cutoff_{ci}$-dim vectors $D_{ci}(n_c,\alpha,:)$.

{\bf Example:} The ``GGC'' codon interval distribution of the following ``genome $\alpha_0$'' is $D_{ci}(``GGC\mbox{''},\alpha_0,:)=[0,0,1,3,5,1,0,0,0,0]$, where $cutoff_0=10$.
$$
\mbox{{\color{red}GGC}AU{\color{red}GGC}UU{\color{red}GGC}AUC{\color{red}GGC}A{\color{red}GGC}AU{\color{red}GGC}A{\color{red}GGC}{\color{red}GGC}AU{\color{red}GGC}A{\color{red}GGC}UU{\color{red}GGC}AGCA}
$$
And the ``GCA'' codon interval distribution of the same ``genome $\alpha_0$'' is $D_{ci}(``GCA\mbox{''},\alpha_0,:)=[0,0,1,1,2,1,1,0,1,1]$.
$$
\mbox{G{\color{green}GCA}UGGCUUG{\color{green}GCA}UCG{\color{green}GCA}G{\color{green}GCA}UG{\color{green}GCA}GGCG{\color{green}GCA}UG{\color{green}GCA}GGCUUG{\color{green}GCA}{\color{green}GCA}}
$$
Hence, the correlation coefficient between $D_{ci}(``GGC\mbox{''},\alpha_0,:)$ and $D_{ci}(``GCA\mbox{''},\alpha_0,:)$ is
$$
corrcoef(D_{ci}(``GGC\mbox{''},\alpha_0,:),D_{ci}(``GCA\mbox{''},\alpha_0,:))=0.7235.
$$

\subsubsection{The codon interval correlation matrix $\Delta$}

The codon interval correlation matrix $\Delta(n_c,\alpha,\beta)$ for a group of $N$ species is defined as the $64 \times N \times N$ matrix of the correlation coefficients between pairs of vectors $D_{ci}(n_c,\alpha)$ and $D_{ci}(n_c,\beta)$:
\begin{equation} \Delta(n_c,\alpha,\beta)=corrcoef(D_{ci}(n_c,\alpha,:),D_{ci}(n_c,\beta,:)). \end{equation}

\subsubsection{Calculating the codon interval distance matrix of species $M_{ci}$ according to $\Delta$}

We can obtain the $N \times N$ codon interval distance matrix $M_{ci}(\alpha,\beta)$ of the $N$ species by averaging the $64 \times N \times N$ correlation coefficients with respect to the $64$ codons: \begin{equation}
M_{ci}(\alpha,\beta) = 1- \frac{1}{64} \sum_{n_c=1}^{64} \Delta(n_c,\alpha,\beta).
\end{equation}
Hence, we can draw the phylogenetic tree of $N$ species based on $M_{ci}$.

The method to obtain phylogenetic trees of species based on the codon interval distance matrices is valid not only for eukarya but also for bacteria, archaea and virus. The phylogenetic trees of species based on the codon interval distance matrices generally agree with the traditional phylogenetic trees respectively, which is also an evidence to show the close relationship between the genome evolution and the biodiversity evolution.

\subsubsection{Calculating the distance matrix of codons $M_{codon}$ according to $\Delta$}

We can obtain the $64 \times 64$ distance matrix of codons $M_{codon}$ by averaging the $64 \times N \times N$ correlation coefficients with respect to the $N$ species:
\begin{equation}
M_{codon}(n_c,n_c') = 1- corrcoef(\Delta(n_c,:,:),\Delta(n_c',:,:)).
\end{equation}
Hence, we can draw the evolutionary tree of $64$ codons based on $M_{codon}$.

The evolutionary tree of codons based on $M_{codon}$ agrees with the traditional understanding of the genetic code evolution. Thus, we can obtain both the phylogenetic tree of species and the evolutionary tree of $64$ codons based on the same codon interval correlation matrix $\Delta$. This is an evidence to show the close relationship between the genetic code evolution and the biodiversity evolution. The principal rules in the biodiversity evolution may concern the primordial molecular evolution.

\subsection{The tree of life and the tree of codon (example 1)}

Based on the genomes of $748$ bacteria, $55$ archaea, $16$ eukaryotes and $133$ viruses (GeneBank, up to 2009), we can obtain the codon interval correlation matrices $\Delta^{all}$. For the eukaryotes with several chromosomes, the codon interval distributions are obtained by averaging the codon interval distributions with respect to the chromosomes of the certain species. Consequently, we can obtain the reasonable phylogenetic tree of these speces (Fig S3a) and the reasonable tree of $64$ codons (Fig 4a) by calculating $M_{ci}^{all}(\alpha,\beta)$ and $M_{codon}^{all}(n_c,n_c')$ from $\Delta^{all}$:
\begin{equation}
M_{ci}^{all}(\alpha,\beta) = 1- \frac{1}{64} \sum_{n_c=1}^{64} \Delta^{all}(n_c,\alpha,\beta),
\end{equation}
and
\begin{equation}
M_{codon}^{all}(n_c,n_c') = 1- corrcoef(\Delta^{all}(n_c,:,:),\Delta^{all}(n_c',:,:)).
\end{equation}

\subsection{The tree of life and the tree of codon (example 2)}

Based on the genomes of $16$ eukaryotes, we can obtain the codon interval correlation matrices $\Delta^{euk}$. Consequently, we can obtain the reasonable phylogenetic tree of these $16$ eukaryotes (Fig 4c) and the reasonable tree of $64$ codons (Fig S3b) by calculating $M_{ci}^{euk}(\alpha,\beta)$ and $M_{codon}^{euk}(n_c,n_c')$ from $\Delta^{euk}$.
If there are several chromosomes ($chr(\alpha)=1,2,...,c_m(\alpha)$) in the genome of eukaryote $\alpha$, the codon interval distributions of the chromosomes of species $\alpha$ are $D^{euk}_{ci}(n_c,\alpha,chr(\alpha),:)$. The codon interval correlation matrix is:
\begin{equation}
\Delta^{euk}(n_c,\alpha,chr(\alpha),\beta,chr(\beta)) = corrcoef(D^{euk}_{ci}(n_c,\alpha,chr(\alpha),:), D^{euk}_{ci}(n_c,\beta,chr(\beta),:)). \end{equation}
Consequently, we can calculating the codon interval distance matrix of species: \begin{equation}
M^{euk}_{ci}(\alpha,\beta) = 1- \frac{1}{64} \sum_{n_c=1}^{64} (\frac{1}{c_m(\alpha)}\frac{1}{c_m(\beta)} \sum_{chr(\alpha)=1}^{c_m(\alpha)} \sum_{chr(\beta)=1}^{c_m(\beta)}  \Delta^{euk}(n_c, \alpha,chr(\alpha), \beta,chr(\beta)))
\end{equation}
and the distance matrix of codons:
\begin{equation}
M^{euk}_{codon}(n_c,n_c') = 1- corrcoef(\Delta^{euk}(n_c, :,:,:,:), \Delta^{euk}(n_c', :,:,:,:))
\end{equation}
The phylogenetic tree of eukaryotes by this chromosome average method (for $M_{ci}^{euk}$) generally agrees with the tree by the chromosome average method (for $M_{ci}^{all}$).

\subsection{Three periods in genetic code evolution}

We arrange the $64$ codons in the ``codon\_aa'' order by considering the codon chronology order firstly and considering the amino acid chronology order secondly according to the results in [27]:
\begin{equation}
\begin{array}{cccc}
\mbox{codon chronology:}\\
(1)GGC,GCC,&(2)GUC,GAC,&(3)GGG,CCC,&(4)GGA,UCC,\\
(5)GAG,CUC,&(6)GGU,ACC,&(7)GCG,CGC,&(8)GCU,AGC,\\
(9)GCA,UGC,&(10)CCG,CGG,&(11)CCU,AGG,&(12)CCA,UGG,\\
(13)UCG,CGA,&(14)UCU,AGA,&(15)UCA,UGA,&(16)ACG,CGU,\\
(17)ACU,AGU,&(18)ACA,UGU,&(19)GAU,AUC,&(20)GUG,CAC,\\
(21)CUG,CAG,&(22)AUG,CAU,&(23)GAA,UUC,&(24)GUA,UAC,\\
(25)CUA,UAG,&(26)GUU,AAC,&(27)CUU,AAG,&(28)CAA,UUG,\\
(29)AUA,UAU,&(30)AUU,AAU,&(31)UUA,UAA,&(32)UUU,AAA,
\end{array}
\end{equation}
\begin{equation}
\begin{array}{lcr}
\mbox{amino acid chronology:}\\
(1)\ G,\ (2)\ A,\ (3)\ V,\ (4)\ D,\ (5)\ P,\ (6)\ S,\ (7)\ E,\ (8)\ L,\ (9)\ T,\ (10)\ R,\\
(11)I,(12)Q,(13)N,(14)K,(15)H,(16)F,(17)C,(18)M,(19)Y,(20)W.
\end{array}
\end{equation}

We define the average correlation curves Hurdle curve and Barrier curve as follows:
\begin{equation}
Hurdle(\alpha)=mean(M_{codon}(\alpha,:)),
\end{equation}
\begin{equation}
Barrier(\alpha)=mean(\{M_{codon}(\beta,\beta'):|\beta-\alpha| \leq n_{barr}\ \mbox{and}\ |\beta'-\alpha| \leq n_{barr}\}),
\end{equation}
where $n_{barr}=8$.

According to the observations of the certain positions of the three terminal codons in the evolutionary tree of codons (Fig 4a, S3b) and the certain shapes of the Hurdle curve and the Barrier curve (Fig 4b, S3c), we propose three periods in the genetic code evolution:
\begin{equation}
\mbox{(1) initial period, (2) transition period, and (3) fulfillment period,}
\end{equation}
which are separated by the three terminal codons and correspond to the origination of three terminal codons respectively. We observe that the curve $Barrier$ begins at a level of $Barrier\sim0.4$, then overcome a ``barrier'' of level $Barrier\sim0.5$, and at last reach a low place of level $Barrier\sim0.3$ (Fig 4b). Between the initial period and the fulfillment period, we can observe some considerably higher values in the curves $Barrier$ and $Hurdle$, which indicates a ``barrier'' in the middle period of the genetic code evolution. The overall trend of the curve $barrier$ is declining. This ``barrier'' in the curve $Barrier$ corresponds to the narrow palace in the middle of the tree of $64$ codons based on $M_{codon}$.

\section{A heuristic model on the coupled earth spheres}

\subsection{The strategy of biodiversification}

The robustness of biodiversification was ensured by the genomic contributions, without which the biodiversity on the earth can hardly survive the tremendous environmental changes. The mechanism of genome evolution is independent from the rapid environmental change during mass extinctions, which ensures the continuity of the evolution of life: all the phyla survived from the Five Big mass extinctions; more families (in ratio) survived from the mass extinctions than genera. The mass extinctions had only influenced some non-fatal aspects of the living system (e.g. wipeout of some genera or families), whose influence for the vital or more essential aspects of living system (e.g. the advancement aspect) was limited. The living system seems to be able to respond freely to any possible environmental changes on the earth. The sustainable development of the living system in the high risk earth environment was ensured at the molecular level rather than at the species level.

\subsection{The tectonic timescale coupling of earth's spheres}

The three patterns CP I, CP II and CP III in the Phanerozoic eon indicate the tectonic timescale coupling of earth's spheres. The driving force in the biodiversity evolution should be explained in a tectonic timescale dynamical mechanism. Although the P-Tr mass extinction happened rapidly within several $10^4$ years, its cause should be explained in a broader context at the tectonic timescale. Overemphasis of the impacts of occasional events did not quite touch the core of the biodiversity evolution.

\subsection{A triple pendulum model to explain the climate phase reverse event}

The phase relationship among $Curve\_BD$, $Curve\_SL$ and $Curve\_CC$ can be simulated by a triple pendulum model (Fig S1d) with the coupling constants $k_1$, $k_2$ and a varying coupling $k_3(t)=(1-\epsilon\arctan(t/t_0)/(\pi/2)) \cdot k_3$:
\begin{equation}
\left\{
\begin{array}{l}
\frac{\mathrm{d}^2}{\mathrm{d}t^2} \xi= -\xi - k_1(\xi-\eta)-k_3(t)(\xi-\zeta)\\
\frac{\mathrm{d}^2}{\mathrm{d}t^2} \eta= -\eta - k_2(\eta-\zeta)-k_1(\eta-\xi)\\
\frac{\mathrm{d}^2}{\mathrm{d}t^2} \zeta= -\zeta - k_3(t)(\zeta-\xi)-k_2(\zeta-\eta).\\
\end{array}
\right.
\end{equation}
This model shows that the climate phase reverse can achieve by just varying the coupling $k_3(t)$ from $k_3(1+\epsilon)$ to $k_3(1-\epsilon)$, $\epsilon \ll 1$.

\noindent{\bf Acknowledgements} My warm thanks to Jinyi Li for valuable discussions. Supported by the Fundamental Research Funds for the Central Universities.

\clearpage
\begin{figure}
  \centering
  \includegraphics[width=18cm]{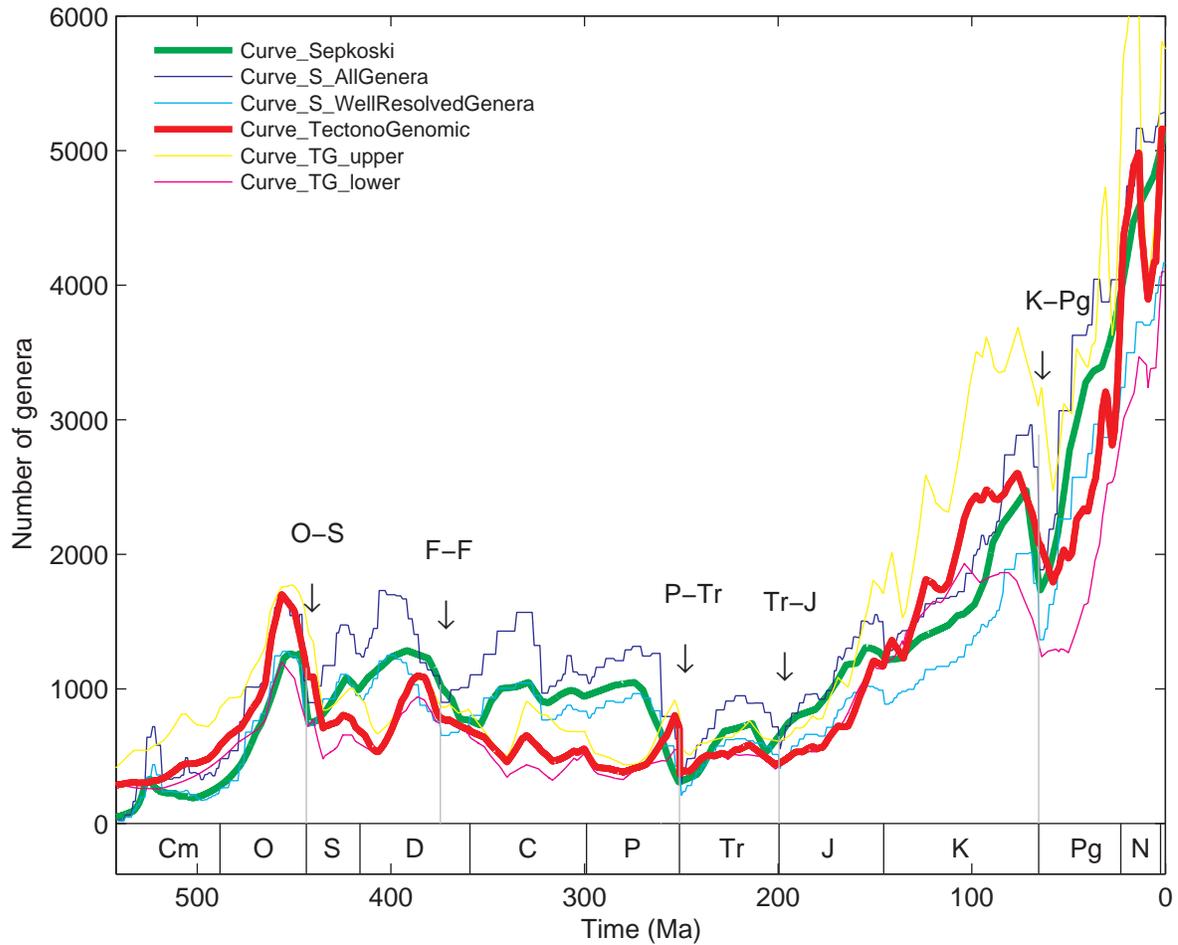}
  \caption{{\bf Explanation of the Sepkoski curve by a tectono-genomic curve.} $Curve\_TectonoGenomic$ generally agrees with $Curve\_Sepkoski$ not only in overall trends but also in detailed fluctuations (including some very detailed fluctuation agreement with $Curve\_S\_AllGenera$). The error range of the Sepkoski curve is about between $Curve\_S\_AllGenera$ and $Curve\_S\_WellResolvedGenera$. The error range of the tectono-genomic curve is about between $Curve\_TG\_upper$ and $Curve\_TG\_lower$. }\label{1}
\end{figure}

\clearpage
\begin{figure}
  \centering
  \includegraphics[width=8cm]{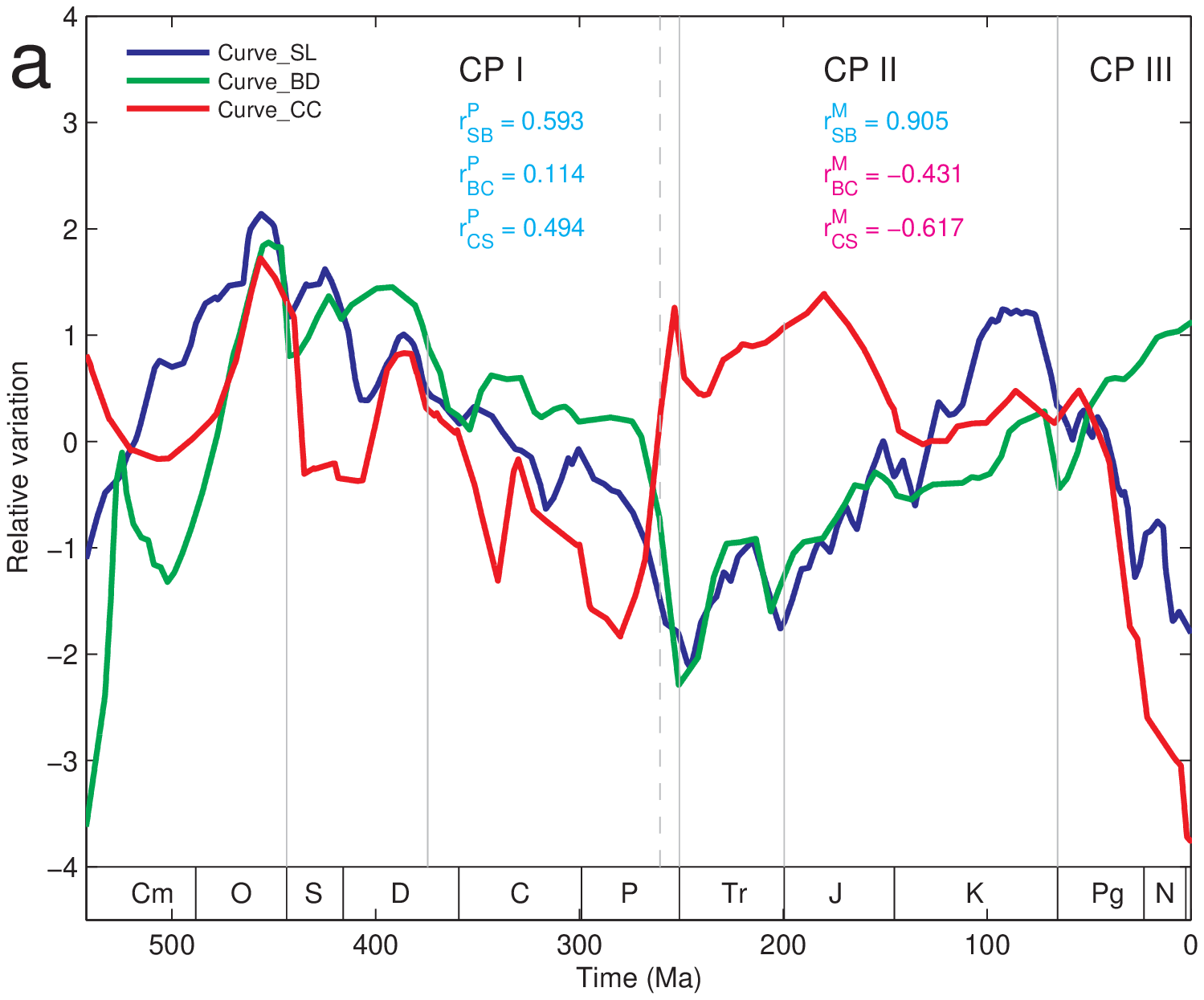}
  \includegraphics[width=8cm]{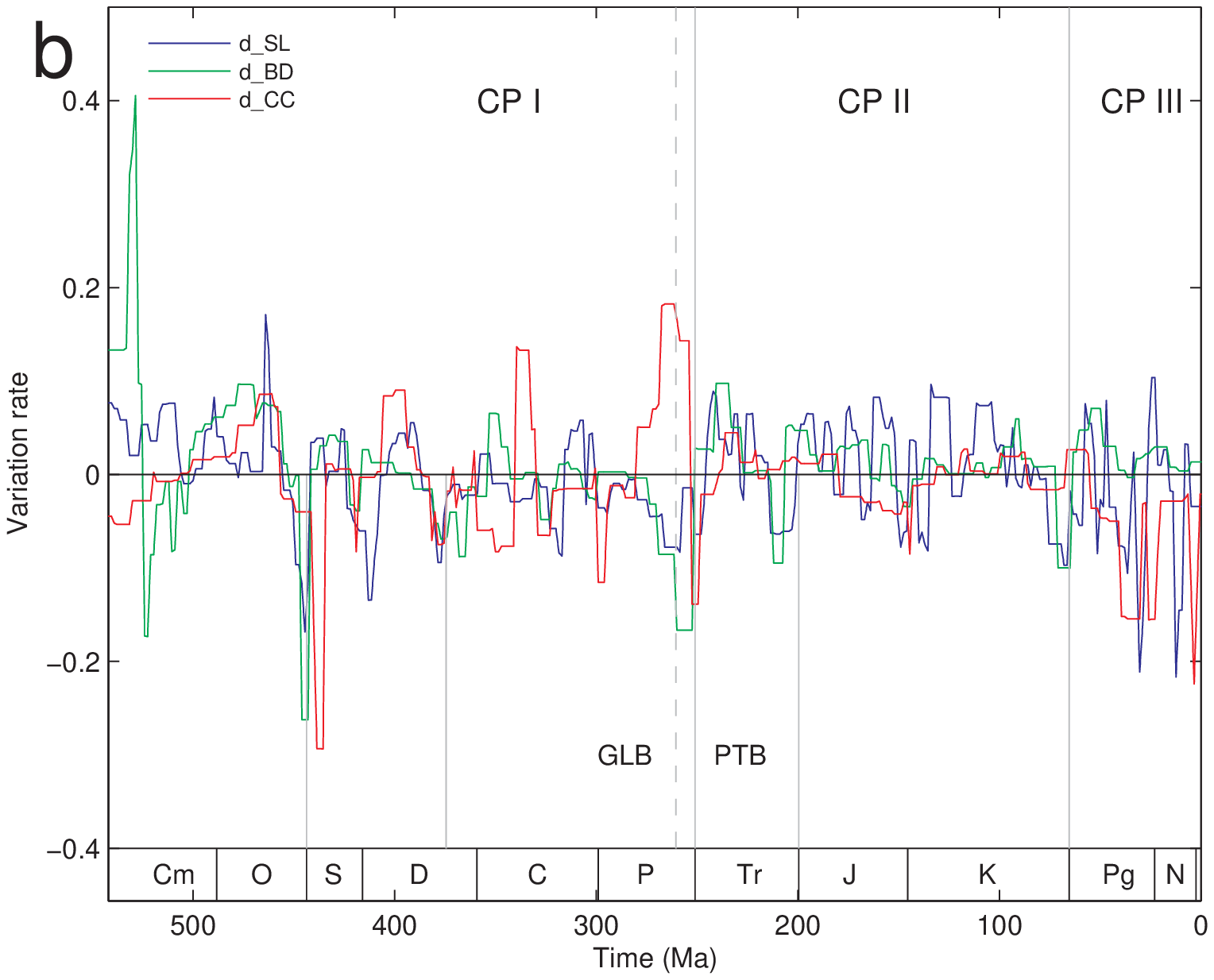}\\
  \includegraphics[width=8cm]{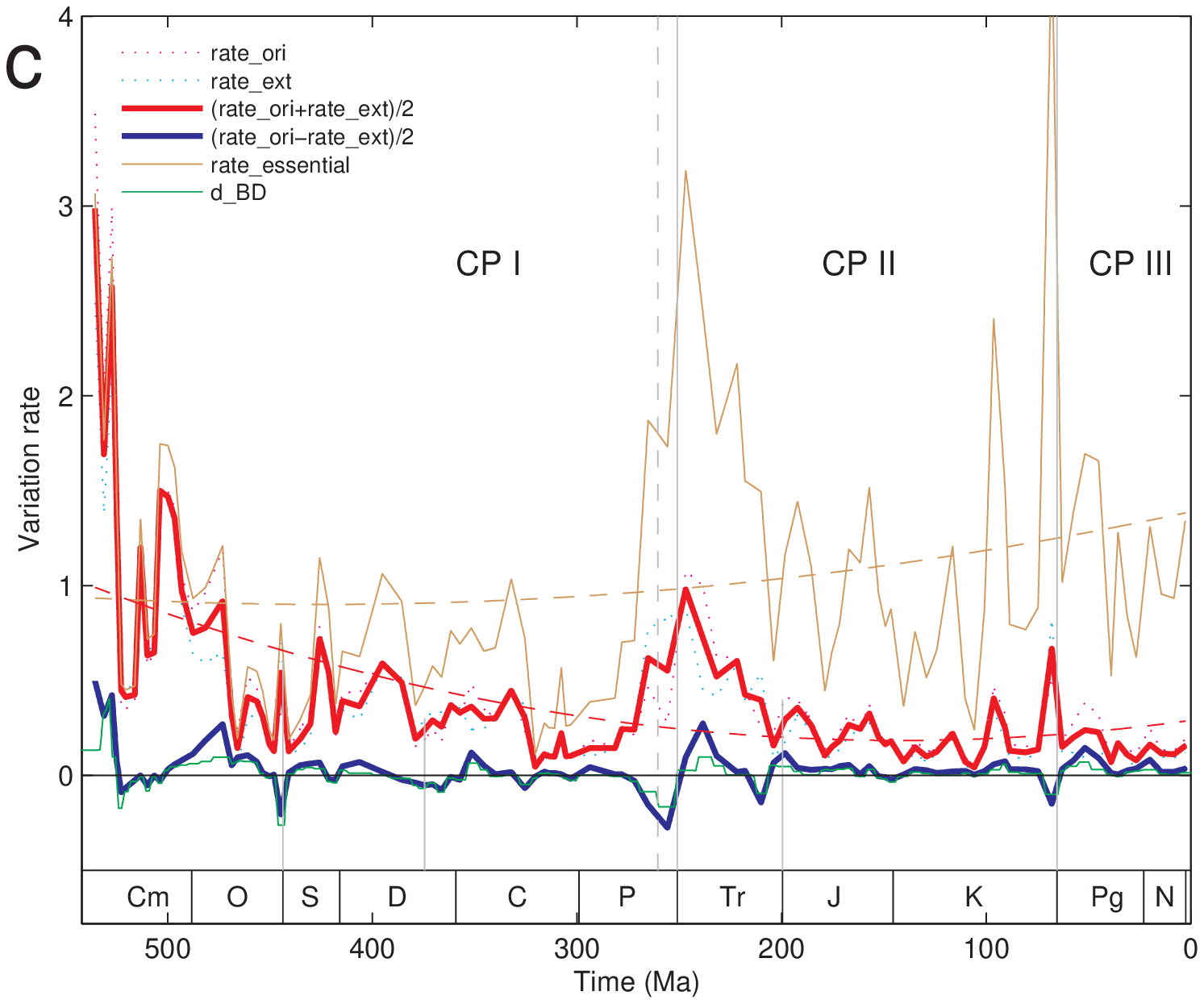}
  \includegraphics[width=8cm]{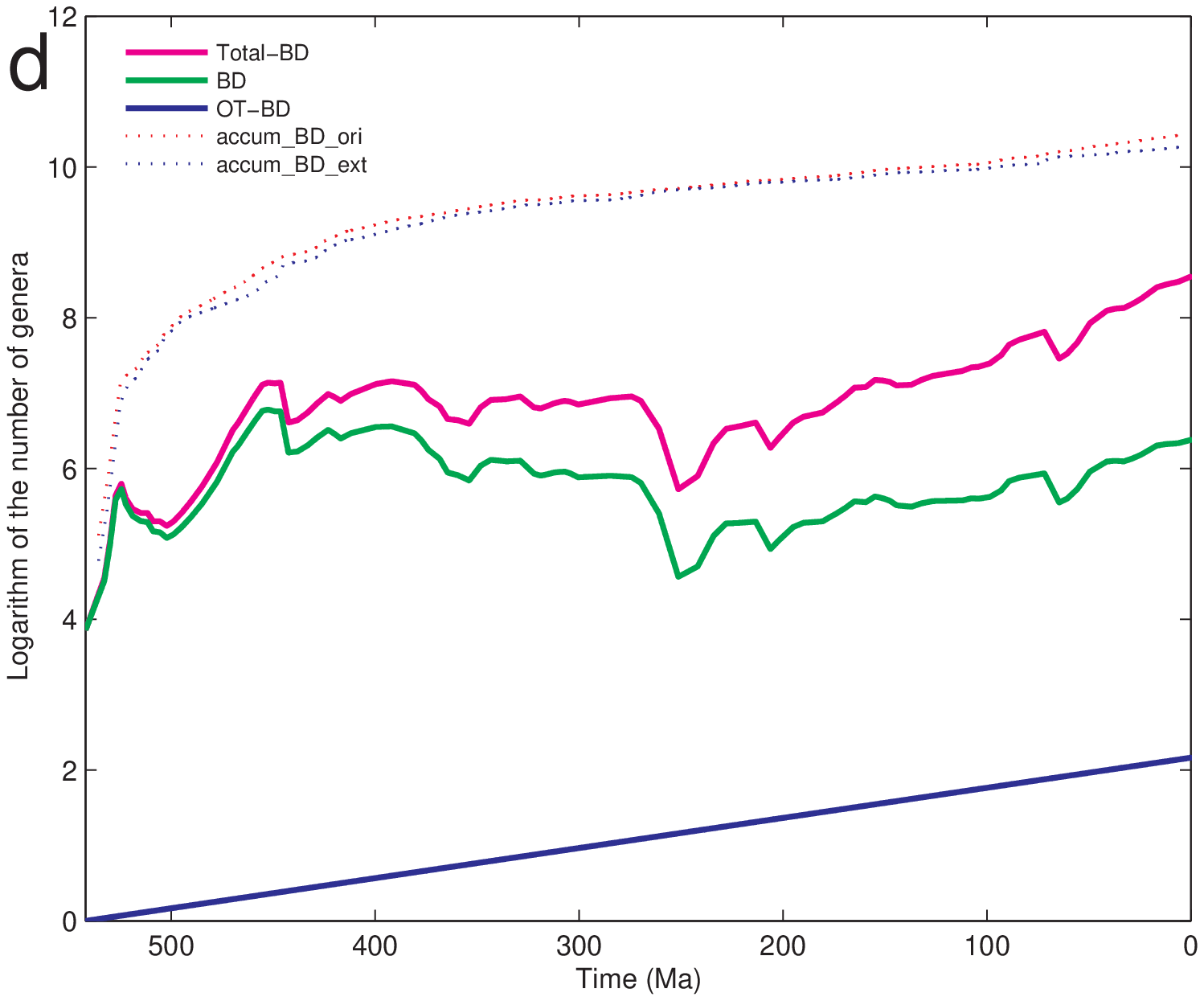}
  \caption{{\bf The tectonic contribution to the fluctuations in the biodiversity evolution.} {\bf a} The consensus climate curve, the consensus sea level curve and the biodiversification curve. There are three climate phases CP I, II, III naturally corresponds to Paleozoic, Mesozoic and Cenozoic respectively. $Curve\_BD$ generally agrees with $Curve\_SL$. $Curve\_BD$ only agrees with $Curve\_CC$ in the Paleozoic era, but varies oppositely with $Curve\_CC$ in the Mesozoic and Cenozoic eras in general. {\bf b} Climate, sea level and biodiversification variation rate curves. We can observe a sharp upward peak at GLB and a sharp downward peak at PTB on the curve $d\_CC$. {\bf c} Explanation of the declining Phanerozoic background extinction rates. The overall trend of the essential biodiversity background variation rate $rate\_essential$ is about horizontal, while the overall trend of the origination and extinction rate curves $rate\_ori$, $rate\_ext$ and their average decline due to the increasing genomic contribution. {\bf d} The total biodiversity curve $Total\mbox{-}BD$ is equal to its net variation $BD$ plus its overall trend $OT\mbox{-}BD$. Also, the overall trends of accumulation origination and extinction biodiversity curves are exponential.}\label{1}
\end{figure}

\clearpage
\begin{figure}
  \centering
  \includegraphics[width=8cm]{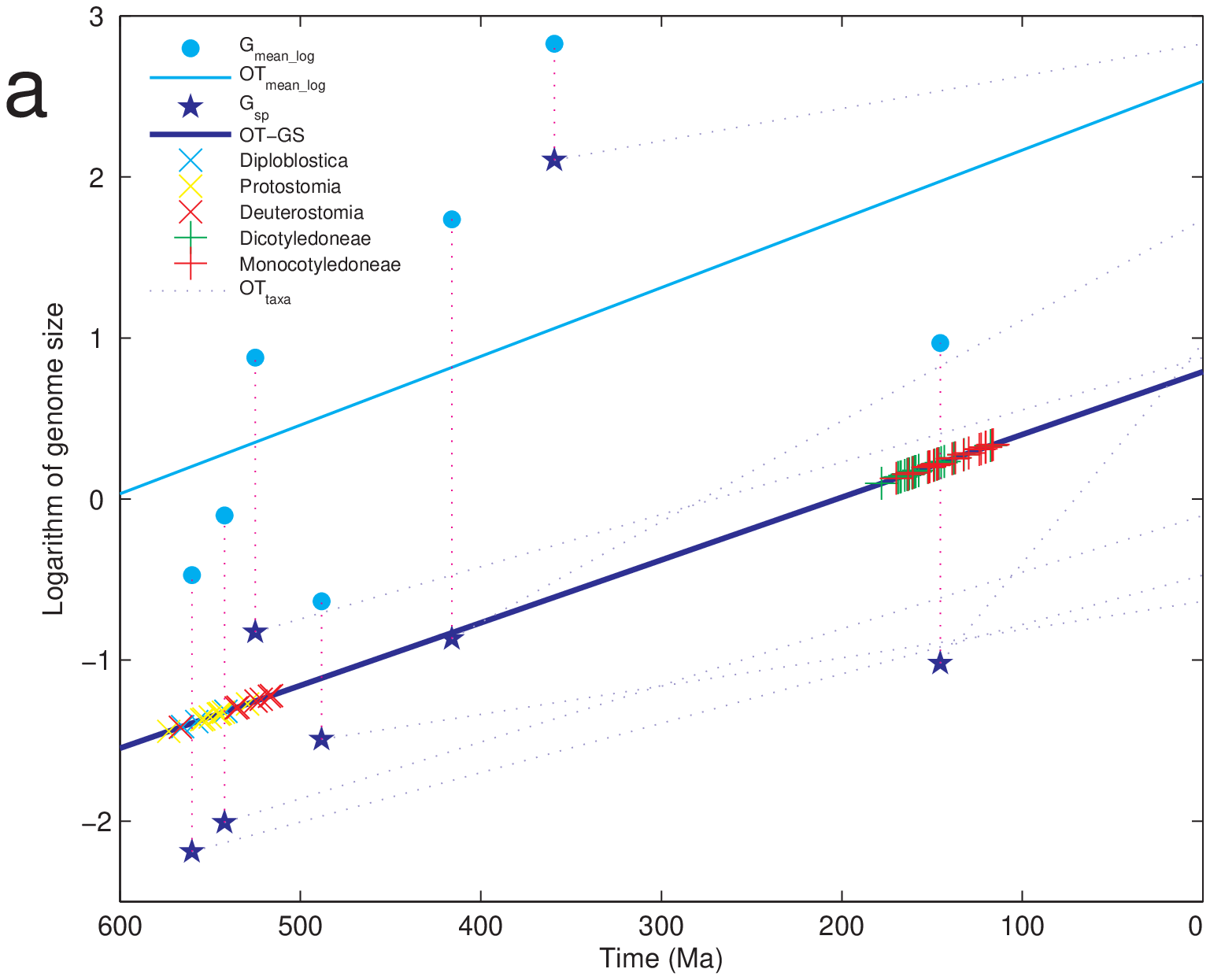}
  \includegraphics[width=8cm]{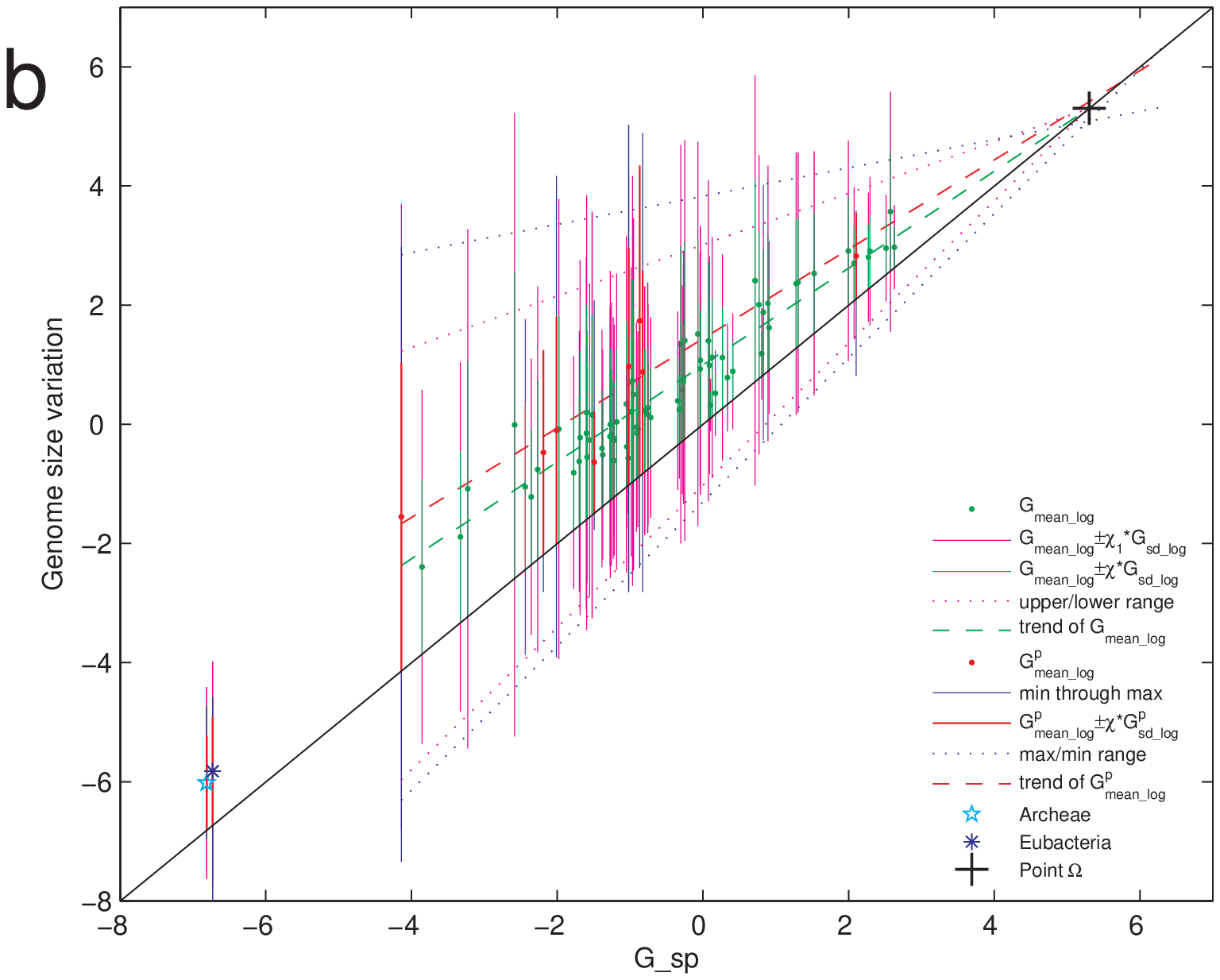}\\
  \includegraphics[width=7cm]{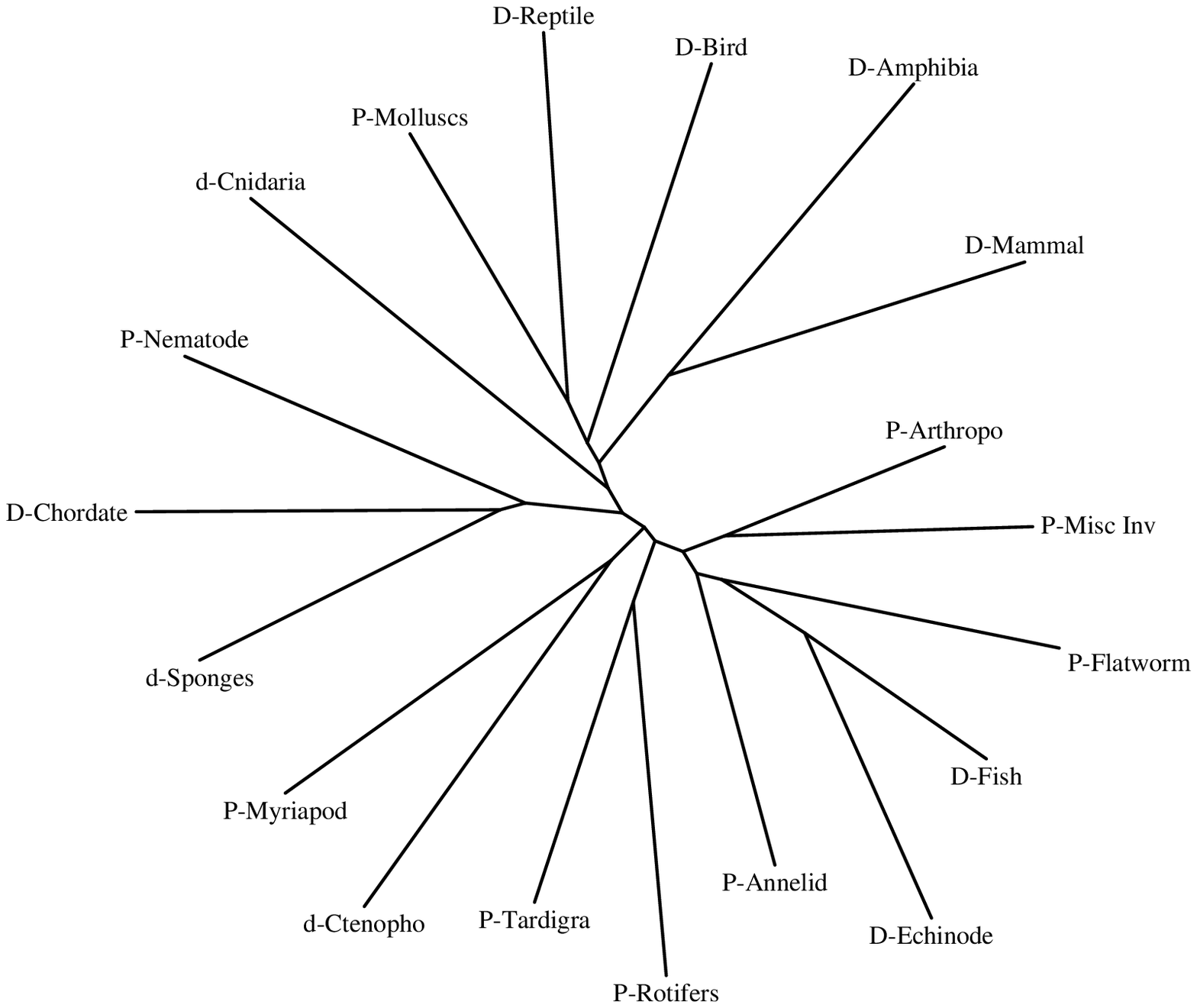}\hspace{0.7cm}
  \includegraphics[width=8cm]{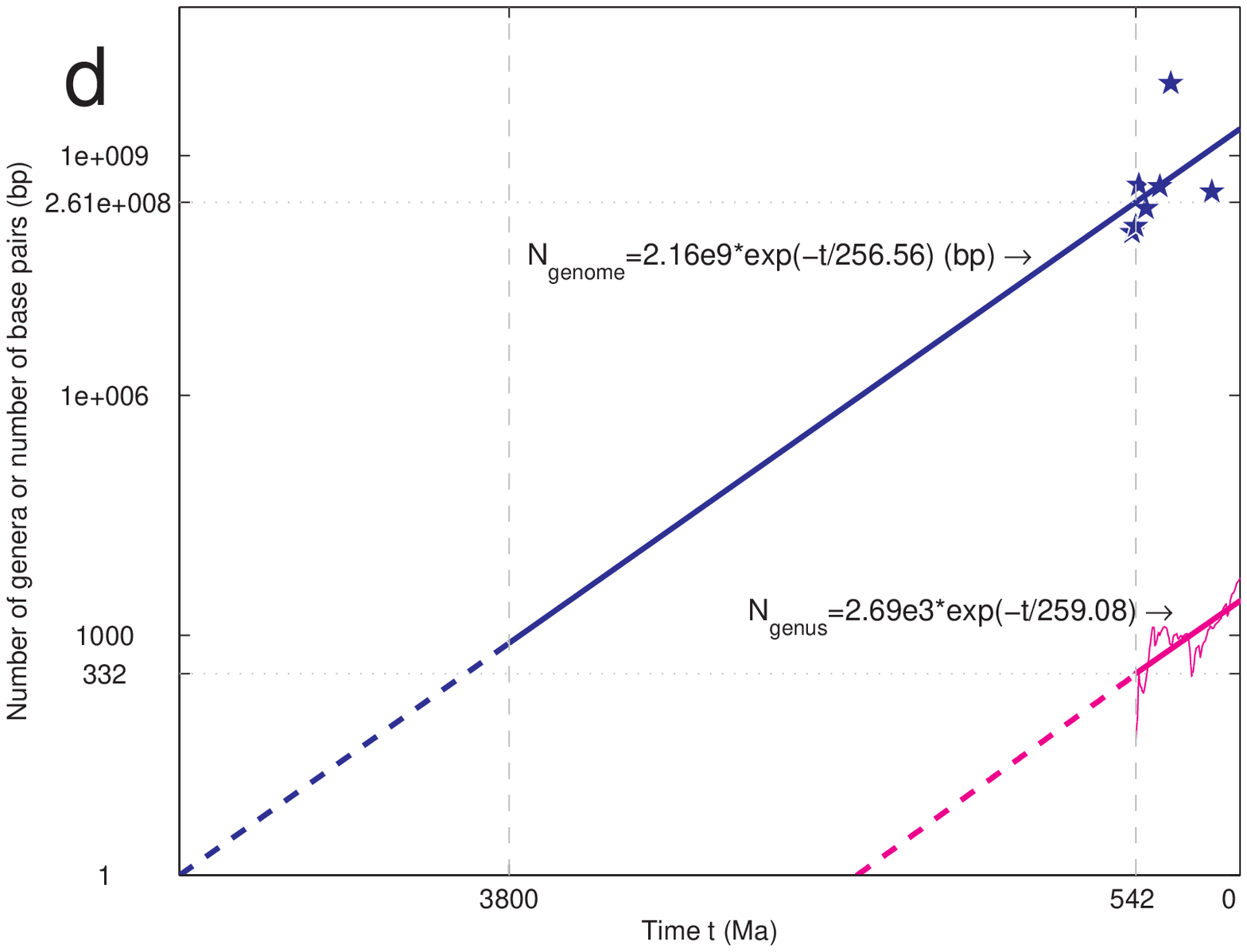}
  \caption{{\bf The genomic contribution to the overall trend of the biodiversity evolution.} {\bf a} The overall trend in the genome size evolution and its applications: (i) Prediction of origin time of taxa in Diploblostica, Protostomia and Deuterostomia indicates three-stage pattern in the metazoan origination; (ii) Prediction of origin time of angiosperm taxa differ between Dicotyledoneae and Monocotyledoneae. {\bf b} Proof of the log-normal distribution of genome size in taxa by the common intersection point $\Omega$. {\bf c} The phylogenetic tree of animal taxa obtained by $M_{gs}^P$. {\bf d} Agreement between the ``e-folding'' time $\tau_{BD}$ in biodiversity evolution and the ``e-folding'' time $\tau_{GS}$ in genome size evolution. Also, reasonable extrapolation of the overall trend of the genome size evolution obtained in the Phanerozoic eon to the Precambrian periods.}\label{1}
\end{figure}

\clearpage
\begin{figure}
  \centering
  \includegraphics[width=5cm]{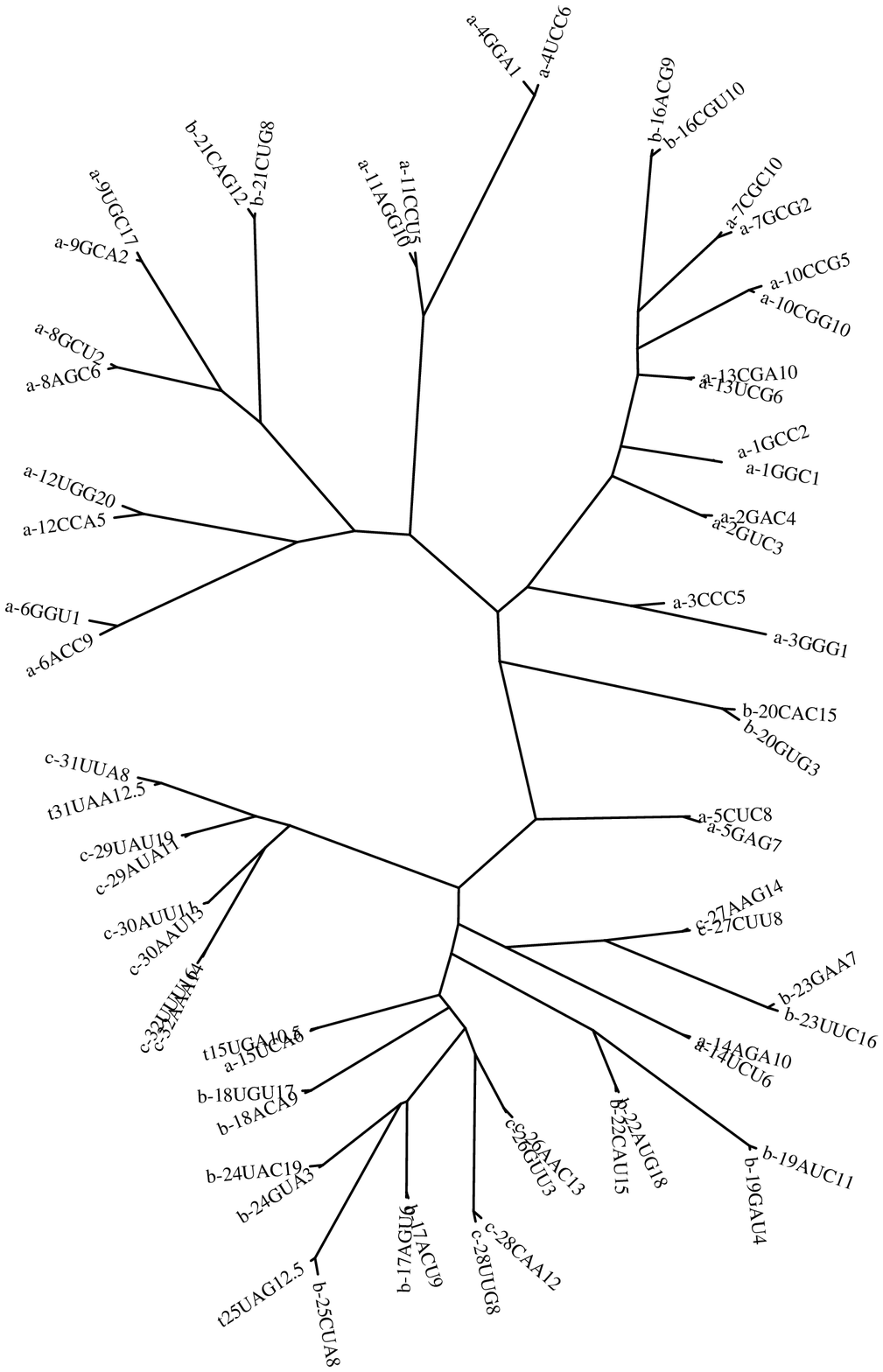}
  \includegraphics[width=5.4cm]{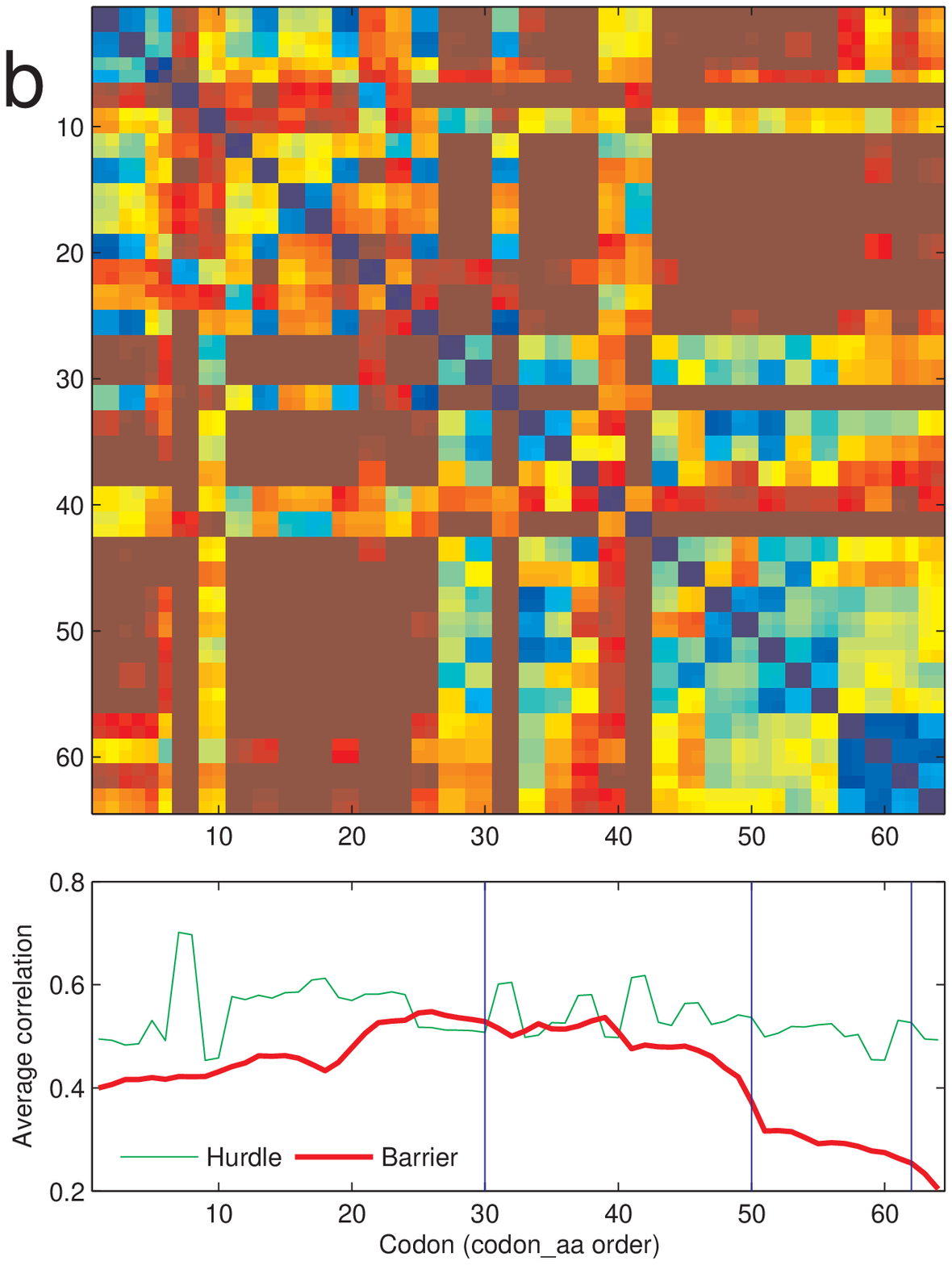}
  \includegraphics[width=5cm]{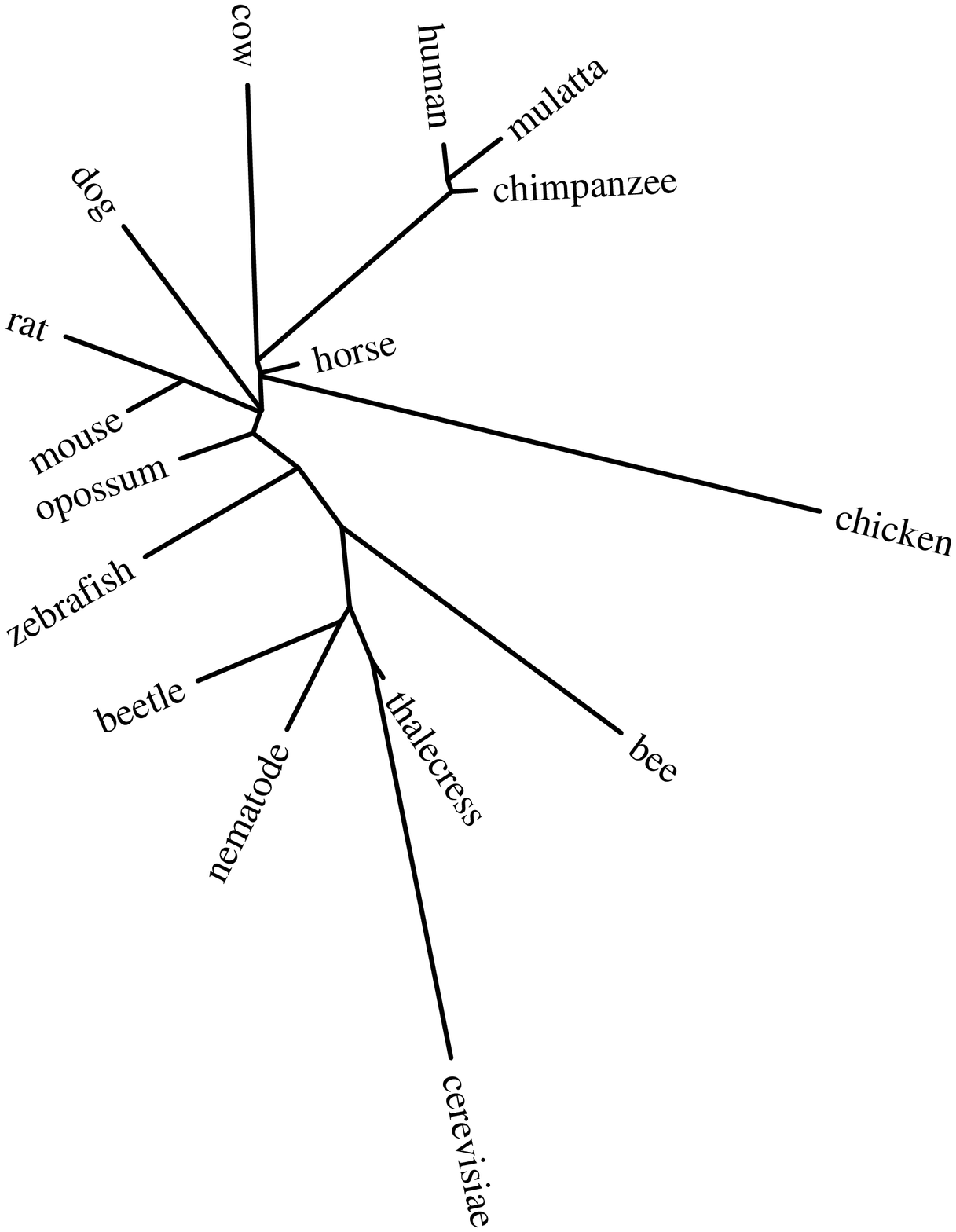}
  \caption{{\bf Relationship between the molecular evolution and the biodiversity evolution.} {\bf a} The evolutionary tree of codons obtained by $M_{codon}^{all}$, which agrees with the codon chronolgy. The codons in (a) initial period, (b) transition period and (c) fulfilment period are in green, blue and red respectively. {\bf b} The codon distance matrix $M_{codon}^{all}$ and its averaging curve $Barrier$. There was a midway high ``barrier'' ($Barrier\approx 0.5$) in the genetic code evolution between the initial period and the fulfillment period. {\bf c} The phylogenetic tree of eukaryotes obtained by their codon interval distance matrix $M_{ci}^{euk}$.}\label{1}
\end{figure}

\clearpage
\begin{figure}
  \centering
  \includegraphics[width=18cm]{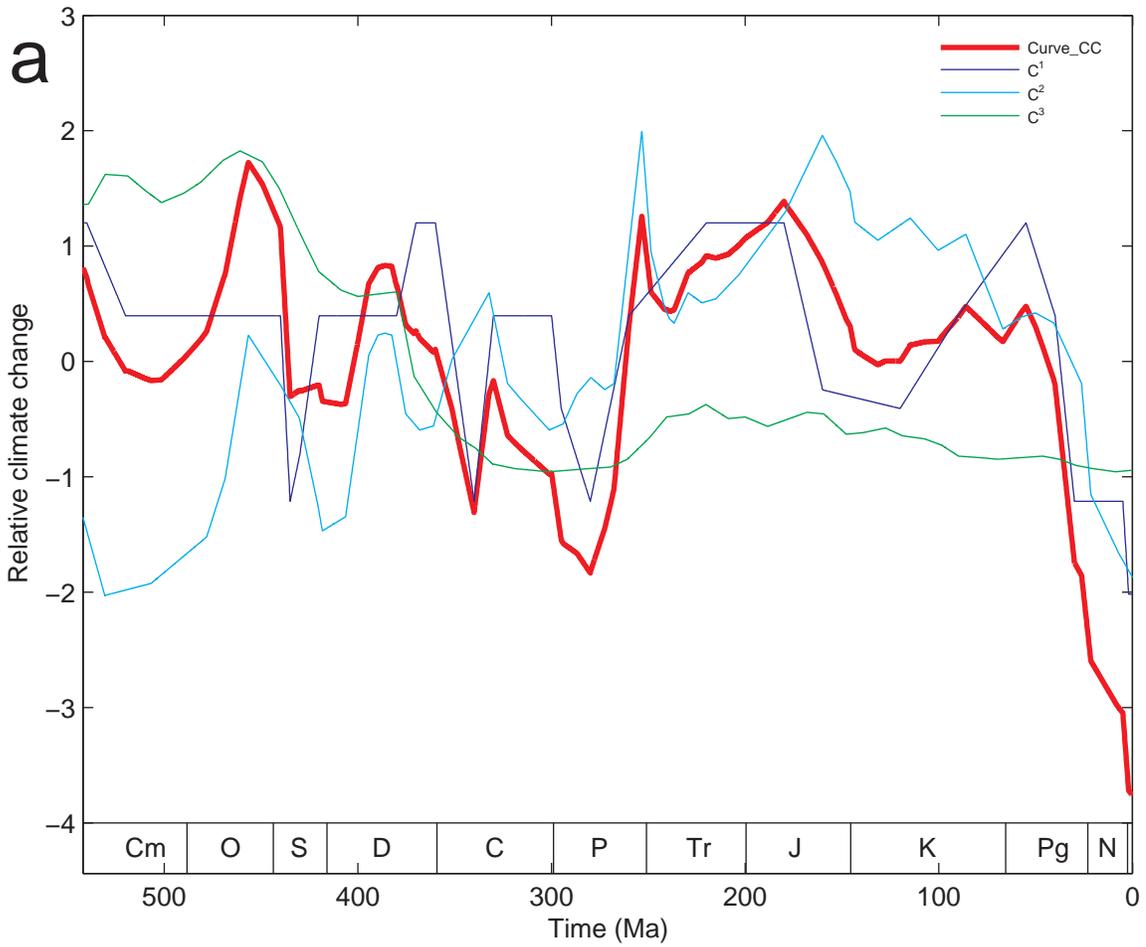}
  \caption{{\bf Fig. S1a}\hspace{0.3cm} The climate curves.}\label{1}
\end{figure}

\clearpage
\begin{figure}
  \centering
  \includegraphics[width=18cm]{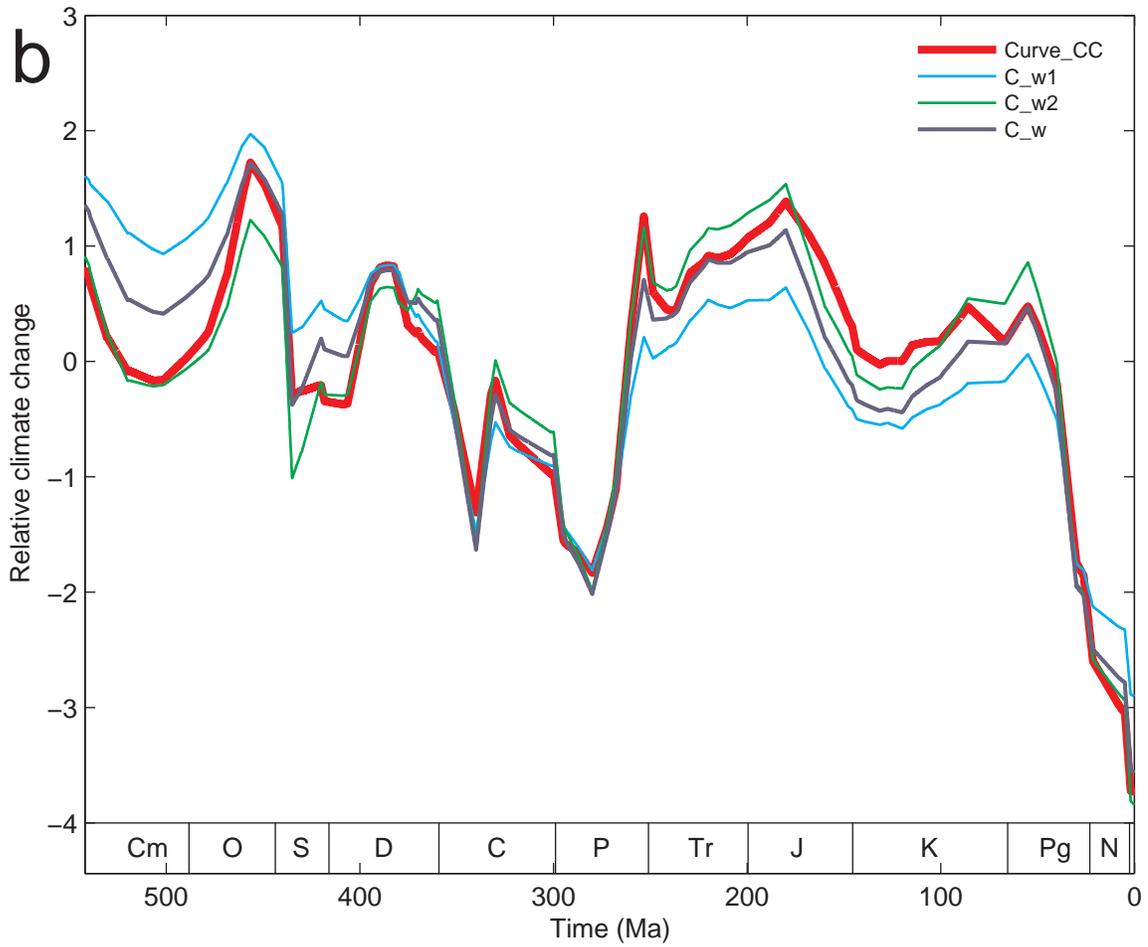}
  \caption{{\bf Fig. S1b}\hspace{0.3cm} The error range of $Curve\_CC$.}\label{1}
\end{figure}

\clearpage
\begin{figure}
  \centering
  \includegraphics[width=18cm]{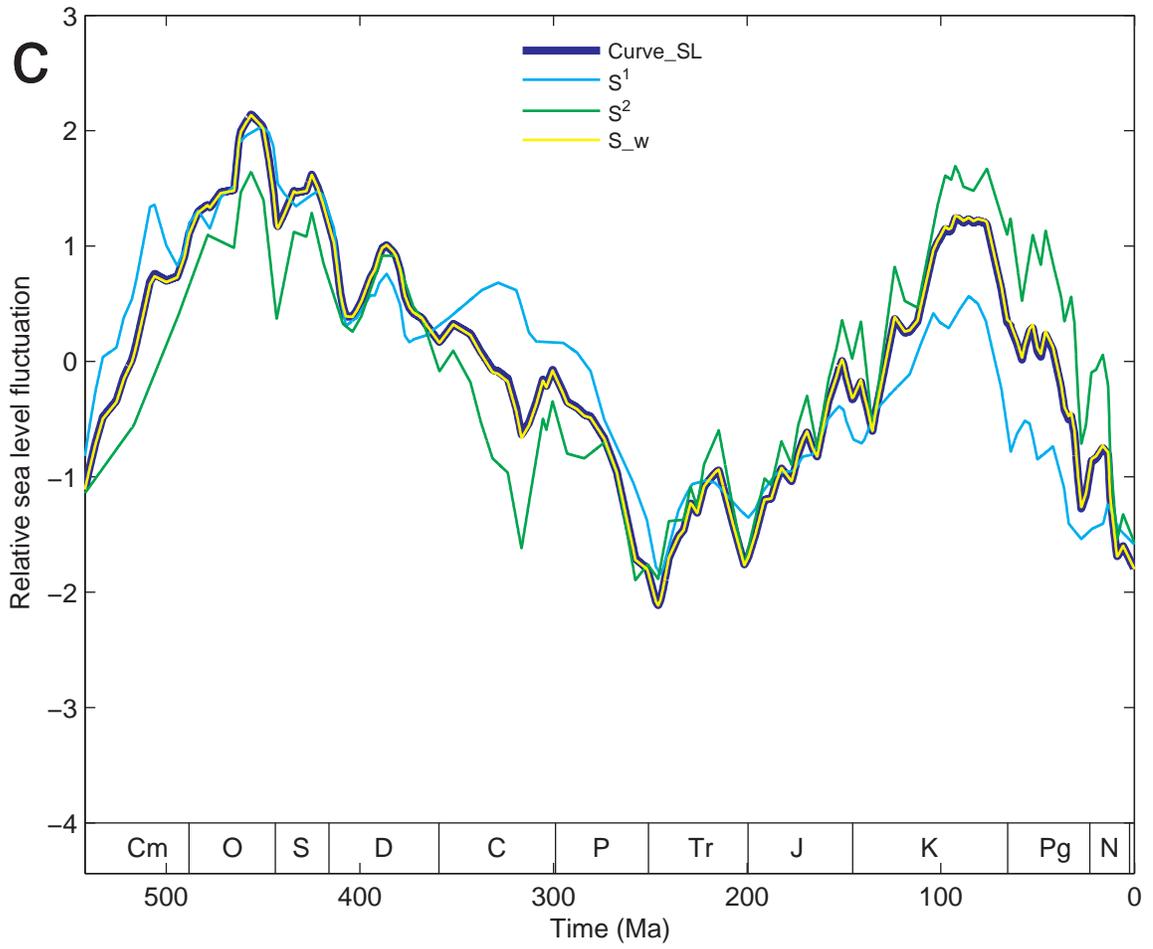}
  \caption{{\bf Fig. S1c}\hspace{0.3cm} The error range of $Curve\_SL$.}\label{1}
\end{figure}

\clearpage
\begin{figure}
  \centering
  \includegraphics[width=18cm]{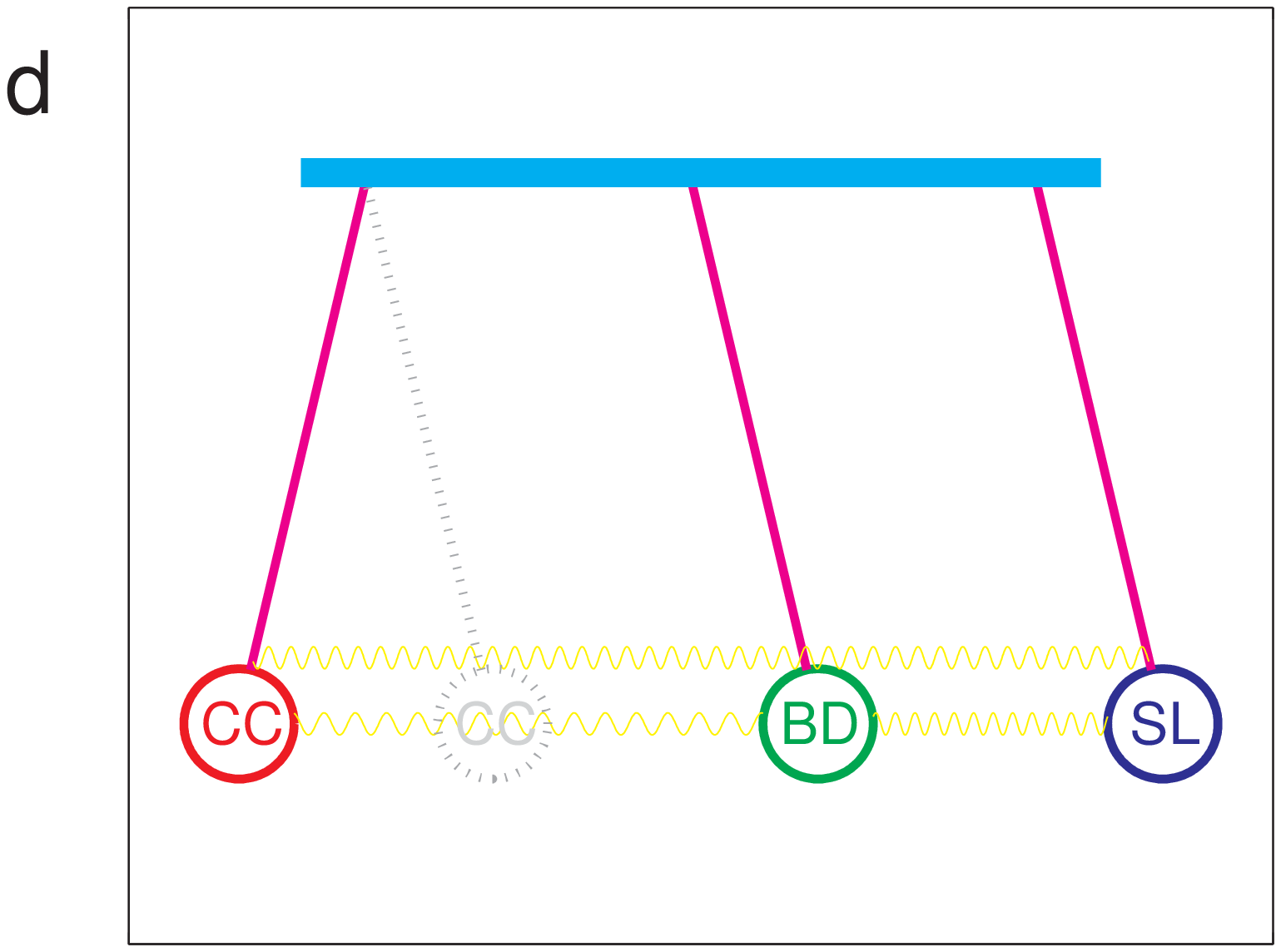}
  \caption{{\bf Fig. S1d}\hspace{0.3cm} Simulating climate phase reverse by a triple pendulum model.} \label{1}
\end{figure}

\clearpage
\begin{figure}
  \centering
  \includegraphics[width=18cm]{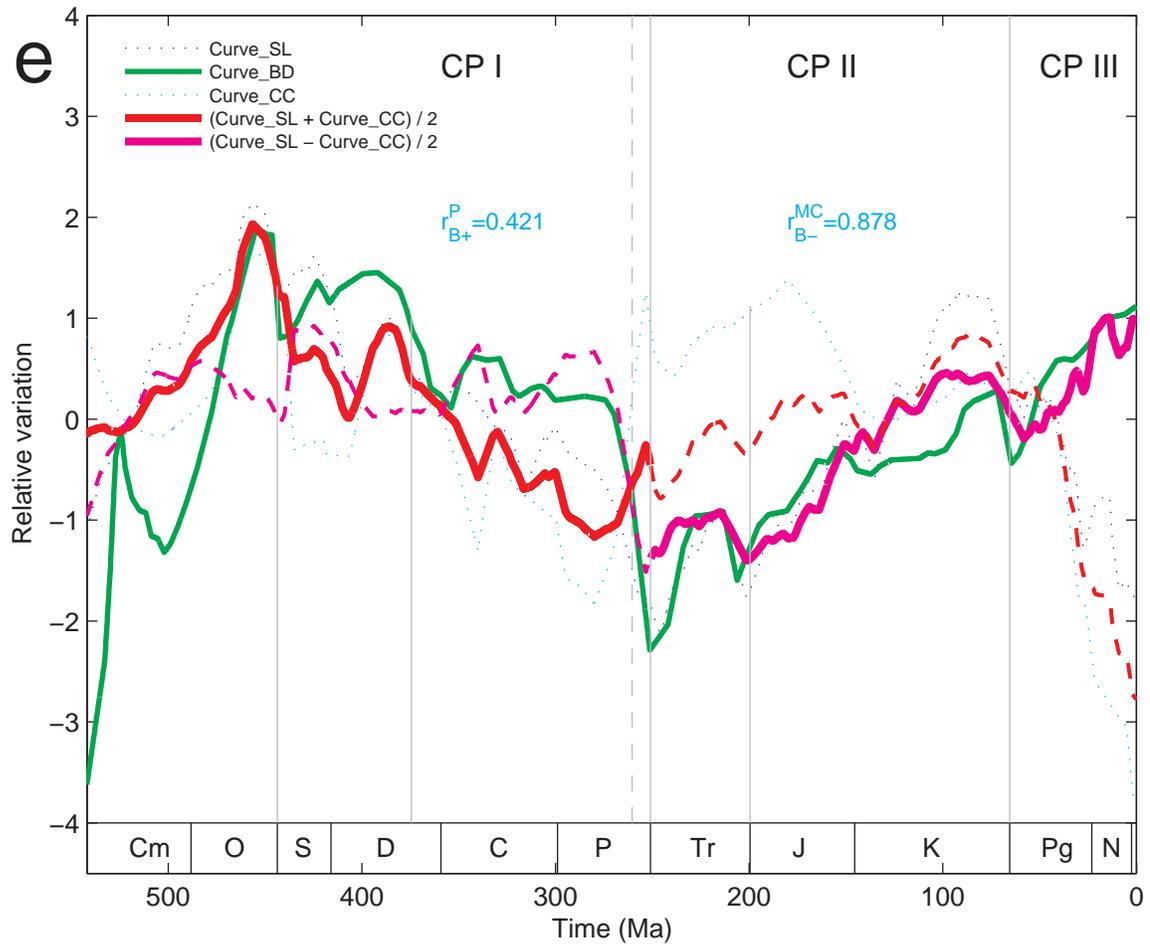}
  \caption{{\bf Fig. S1e}\hspace{0.3cm} The tectonic curve.}\label{1}
\end{figure}

\clearpage
\begin{figure}
  \centering
  \includegraphics[width=18cm]{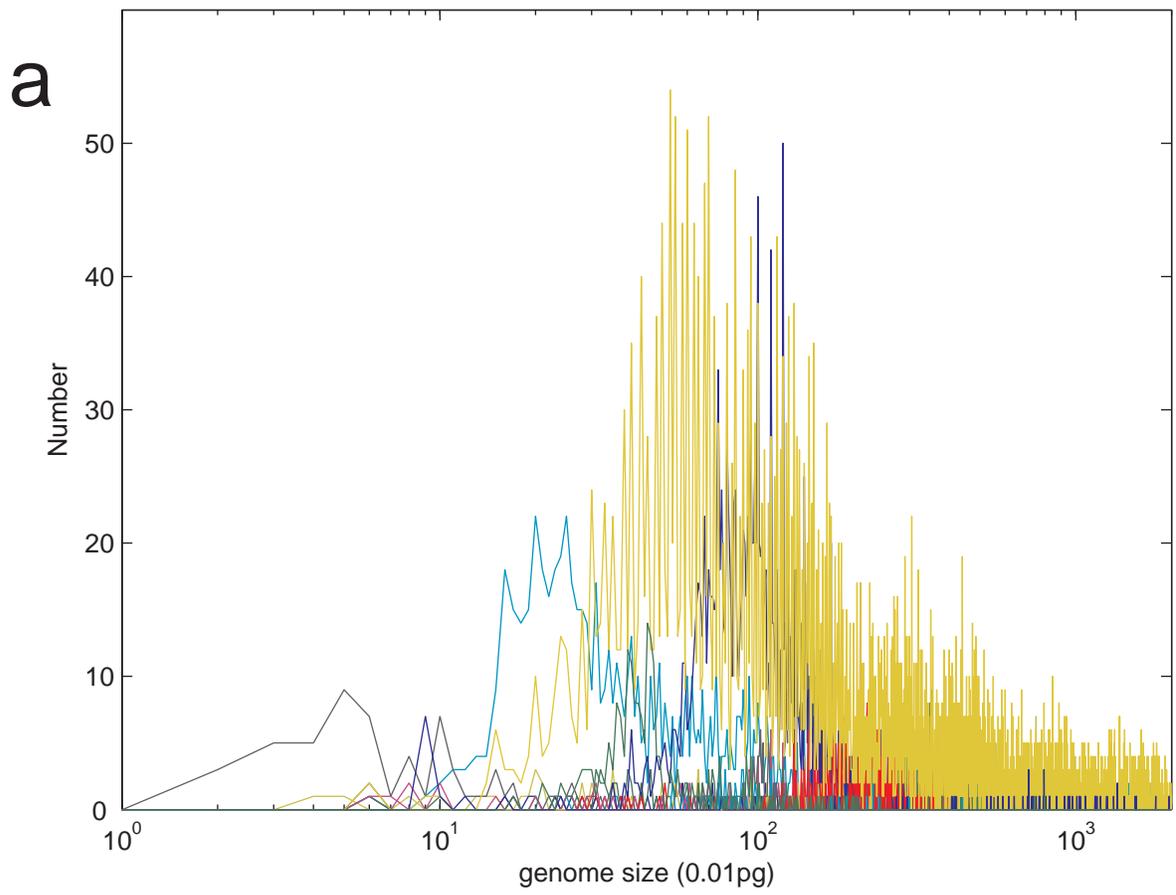}
  \caption{{\bf Fig. S2a}\hspace{0.3cm} Log-normal distributions of genome sizes in taxa.}\label{1}
\end{figure}

\clearpage
\begin{figure}
  \centering
  \includegraphics[width=18cm]{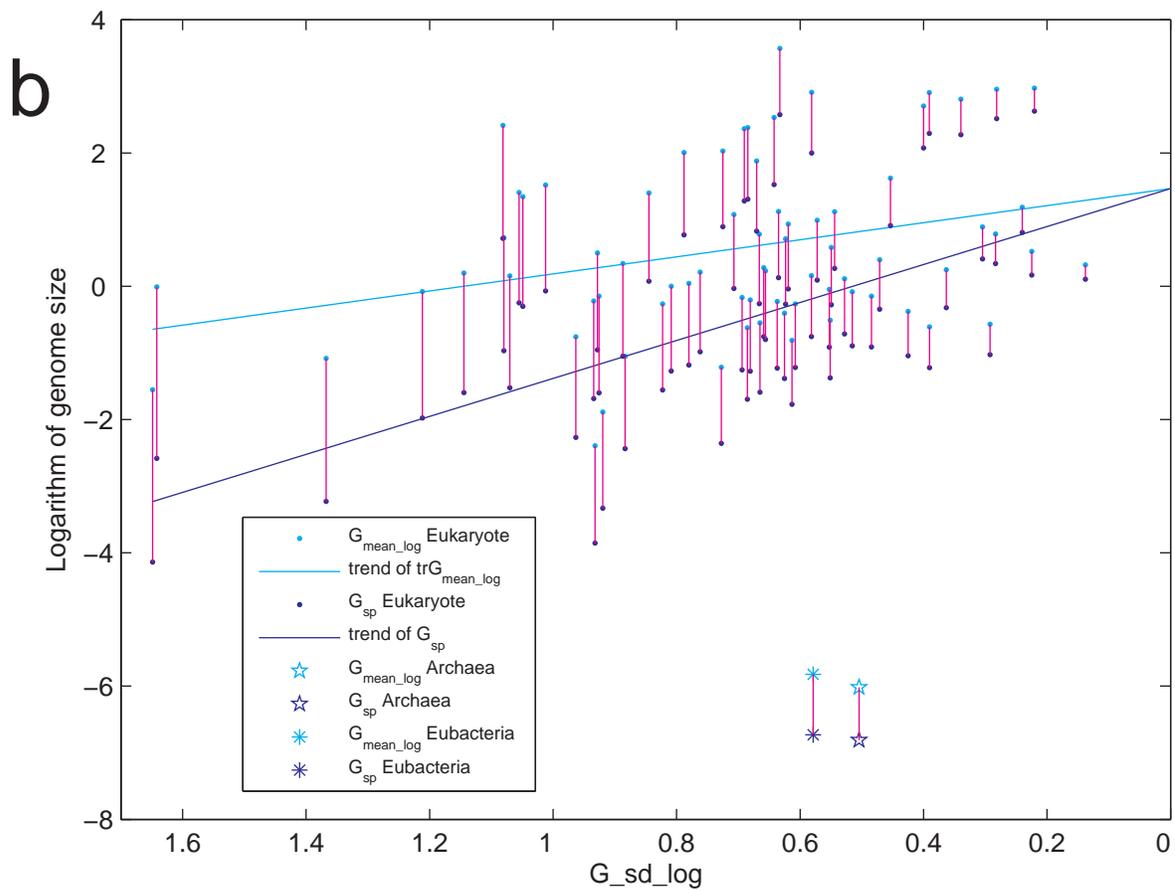}
  \caption{{\bf Fig. S2b}\hspace{0.3cm} $G_{sd\_log}$ tends to decline with respect to $G_{sp}$.}\label{1}
\end{figure}

\clearpage
\begin{figure}
  \centering
  \includegraphics[width=15cm]{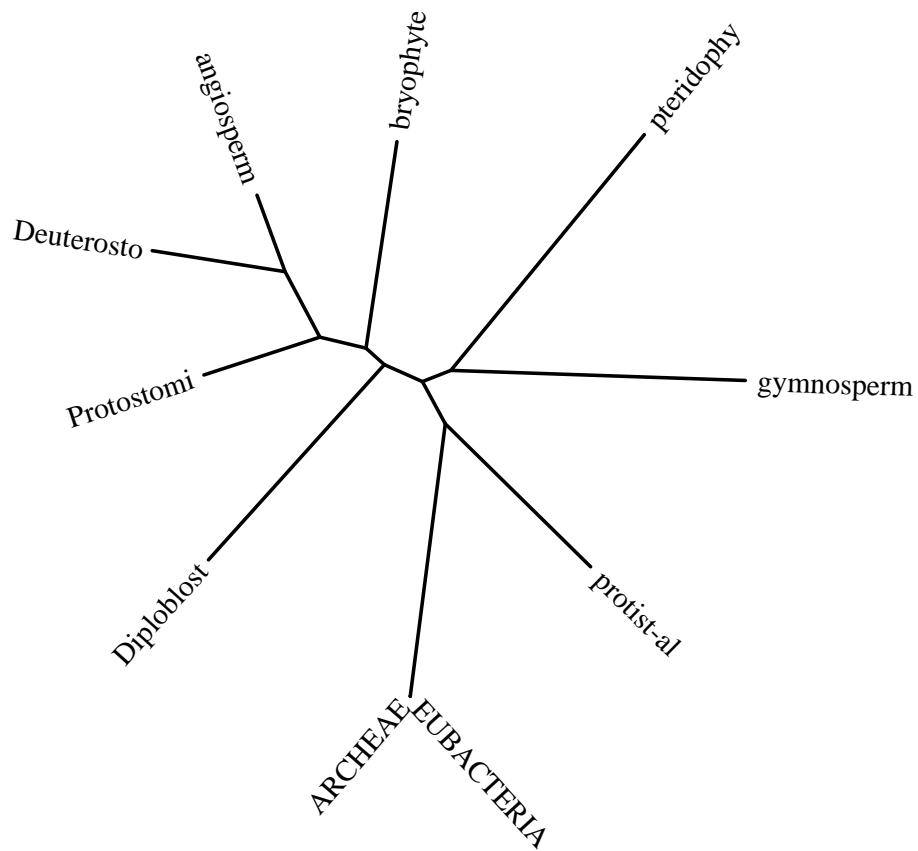}
  \caption{{\bf Fig. S2c}\hspace{0.3cm} The phylogenetic tree based on $M_{gs}^P$.}\label{1}
\end{figure}

\clearpage
\begin{figure}
  \centering
  \includegraphics[width=15cm]{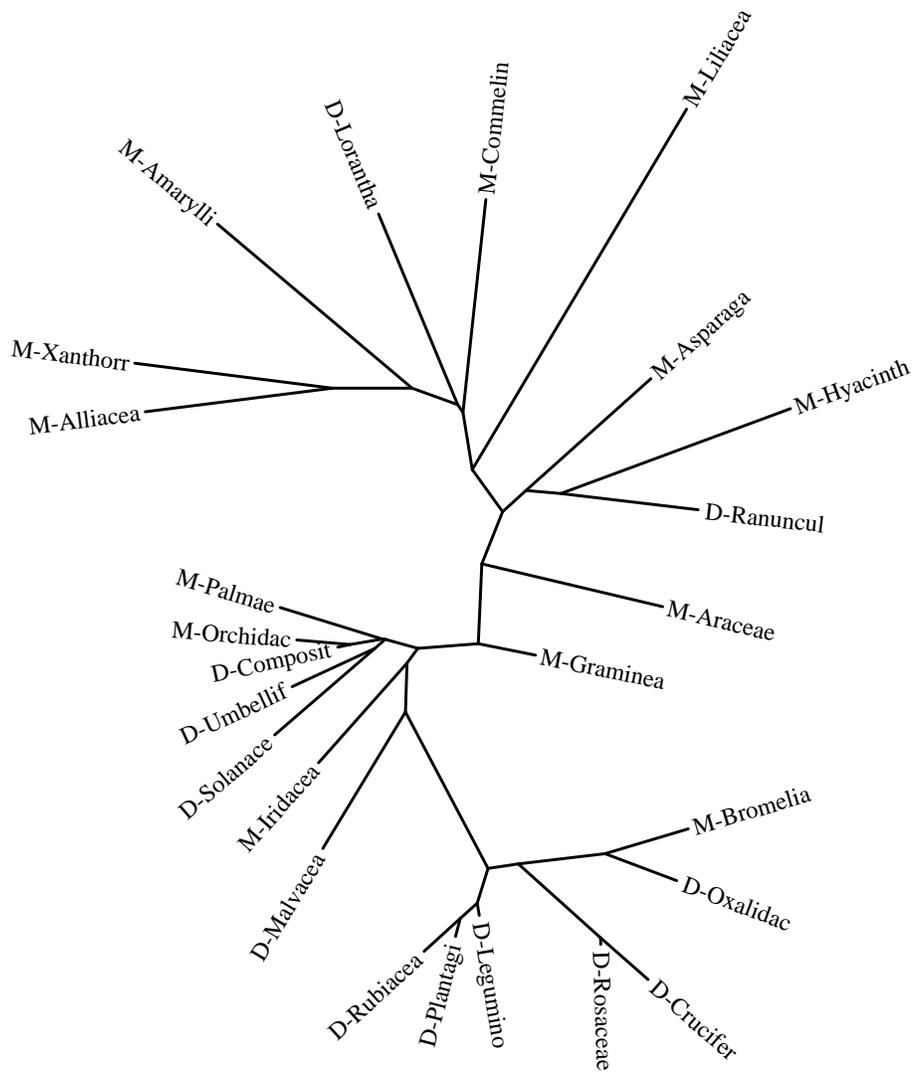}
  \caption{{\bf Fig. S2d}\hspace{0.3cm} The phylogenetic tree based on $M_{gs}^{angiosperm}$.}\label{1}
\end{figure}

\clearpage
\begin{figure}[thbp]
  \centering
  \includegraphics[width=18cm]{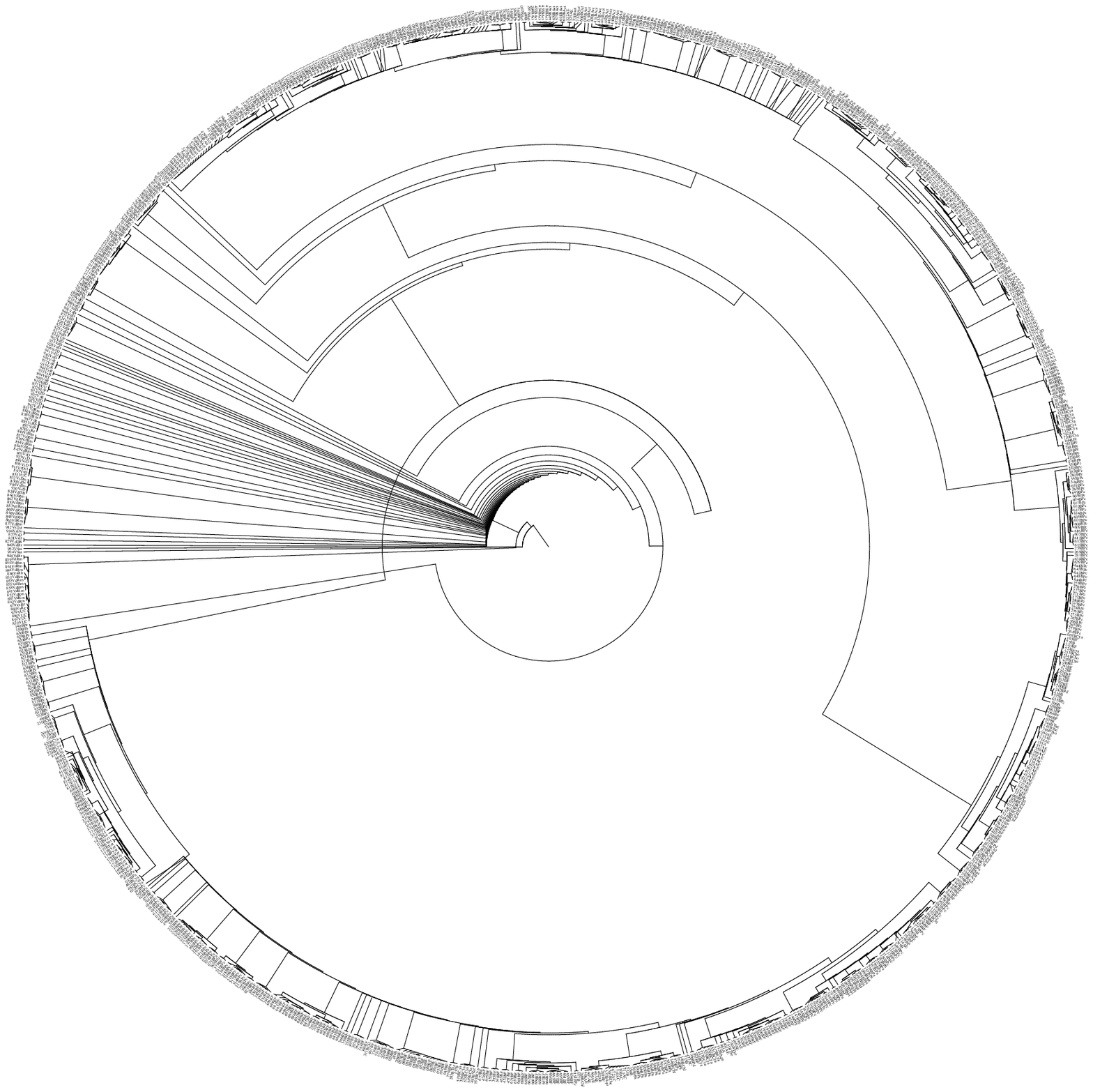}
  \caption{{\bf Fig. S3a}\hspace{0.3cm} The phylogenetic tree based on $M_{ci}^{all}$.}\label{1}
\end{figure}

\clearpage
\begin{figure}
  \centering
  \includegraphics[width=15cm]{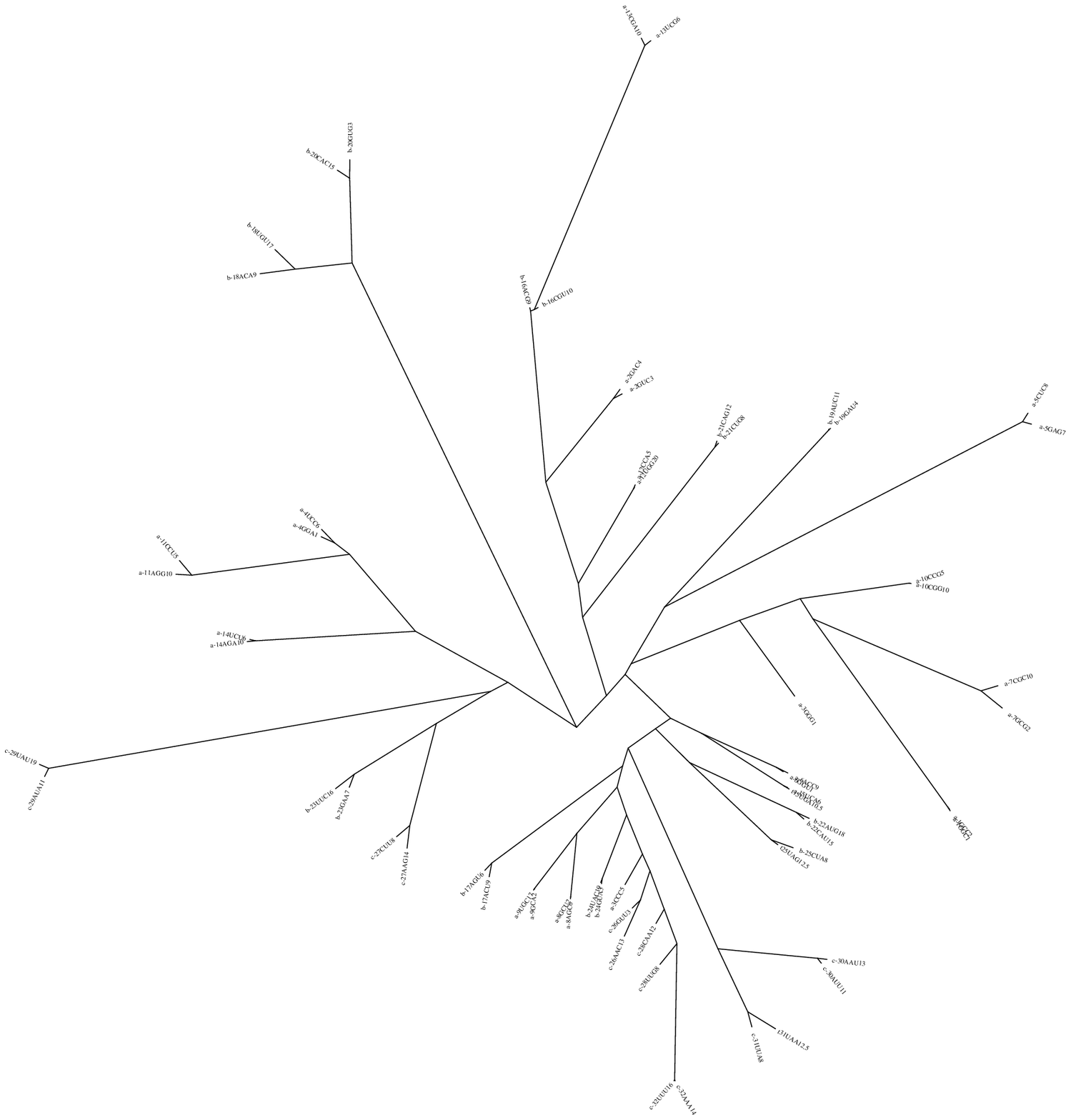}
  \caption{{\bf Fig. S3b}\hspace{0.3cm} Evolutionary tree of codons based on $M_{codon}^{euk}$.}\label{1}
\end{figure}

\clearpage
\begin{figure}
  \centering
  \includegraphics[width=15cm]{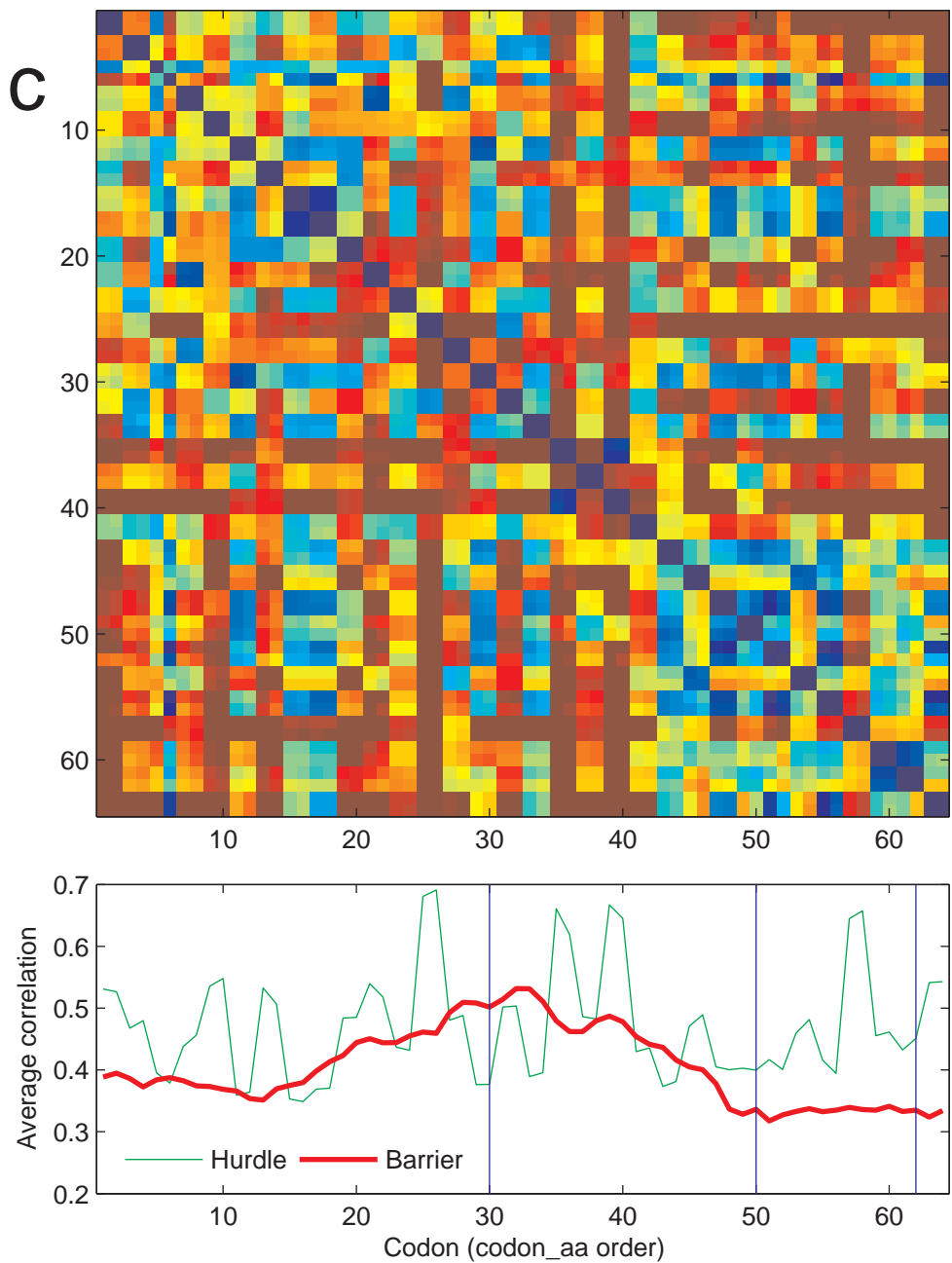}
  \caption{{\bf Fig. S3c}\hspace{0.3cm} The correlation matrix $M_{codon}^{euk}$ and Curve $Barrier$.}\label{1}
\end{figure}

\clearpage

\begin{sidewaystable}
  \centering
  \caption{{\bf Correlation coefficients $r_{\mu\nu}^\rho$} }\label{1}{\scriptsize
\begin{tabular}{|c|c|c|c c c|c c c|c c c c|c c c|c c c|}
  \hline
 n &  & $r_{\mu\nu}^\rho$ & $P$ & $M$ & $C$ & $PMC$ & $PM$ & $MC$ & $P\setminus L$ & $L$ & $L.M.Tr$ & $M\setminus L.M.Tr$ & $R^{+}$ & $R^{-}$ & $\Delta R$ & $Q$ & $Q'$ & $\Delta Q$ \\ \hline \hline
 &  $Curve\_SL$  & $SB$ &{\color{blue} $0.593$} & {\color{blue}$0.905$} & {\color{red}$-0.831$} & {\color{green}$0.564$} & {\color{blue}$0.703$} & {\color{yellow}$0.390$} & {\color{blue}$0.536$} & {\color{blue}$0.923$} & {\color{blue}$0.957$} & {\color{blue}$0.915$} & $ $ & $ $ & $ $ & $ $ & $ $ & \\ \cline{3-13}
1 &  $Curve\_BD$  & $BC$ &{\color{blue} $0.114$} & {\color{red}$-0.431$} & {\color{red}$-0.881$} & {\color{yellow}$-0.287$} & {\color{yellow}$-0.161$} & {\color{red}$-0.752$} & {\color{blue}$0.196$} & {\color{red}$-0.901$} & {\color{blue}$0.205$} & {\color{red}$-0.638$} & $0.516$ & $-0.611$ & $1.126$ & $0.545$ & $0.343$ & $0.202$ \\ \cline{3-13}
 &  $Curve\_CC$  & $CS$ &{\color{blue} $0.494$} & {\color{red}$-0.617$} & {\color{blue}$0.950$} & {\color{yellow}$0.217$} & {\color{yellow}$0.003$} & {\color{yellow}$0.108$} & {\color{blue}$0.658$} & {\color{red}$-0.904$} & {\color{blue}$0.248$} & {\color{red}$-0.726$} & $ $ & $ $ & $ $ & $ $ & $ $ & \\ \hline \hline

 &  $Curve\_SL$  & $SB$ &{\color{blue} $0.593$} & {\color{blue}$0.905$} & {\color{red}$-0.831$} & {\color{green}$0.564$} & {\color{blue}$0.703$} & {\color{yellow}$0.390$} & {\color{blue}$0.536$} & {\color{blue}$0.923$} & {\color{blue}$0.957$} & {\color{blue}$0.915$} & $ $ & $ $ & $ $ & $ $ & $ $ & \\ \cline{3-13}
2 &  $Curve\_BD$  & $BC^1$ &{\color{blue} $-0.233$} & {\color{red}$-0.470$} & {\color{red}$-0.887$} & {\color{yellow}$-0.399$} & {\color{yellow}$-0.339$} & {\color{red}$-0.659$} & {\color{blue}$-0.219$} & {\color{red}$-0.999$} & {\color{blue}$0.980$} & {\color{red}$-0.518$} & $0.308$ & $-0.583$ & $0.891$ & $0.477$ & $0.370$ & $0.106$ \\ \cline{3-13}
 &  $C^1$  & $C^1S$ &{\color{blue} $0.043$} & {\color{red}$-0.502$} & {\color{blue}$0.934$} & {\color{yellow}$-0.006$} & {\color{yellow}$-0.251$} & {\color{yellow}$-0.023$} & {\color{blue}$0.086$} & {\color{red}$-0.929$} & {\color{blue}$0.938$} & {\color{red}$-0.485$} & $ $ & $ $ & $ $ & $ $ & $ $ & \\ \hline \hline

 &  $Curve\_SL$  & $SB$ &{\color{blue} $0.593$} & {\color{blue}$0.905$} & {\color{red}$-0.831$} & {\color{green}$0.564$} & {\color{blue}$0.703$} & {\color{yellow}$0.390$} & {\color{blue}$0.536$} & {\color{blue}$0.923$} & {\color{blue}$0.957$} & {\color{blue}$0.915$} & $ $ & $ $ & $ $ & $ $ & $ $ & \\ \cline{3-13}
3 &  $Curve\_BD$  & $BC^2$ &{\color{blue} $0.335$} & {\color{red}$0.310$} & {\color{red}$-0.804$} & {\color{yellow}$-0.185$} & {\color{yellow}$-0.131$} & {\color{red}$-0.561$} & {\color{blue}$0.572$} & {\color{red}$-0.827$} & {\color{blue}$-0.593$} & {\color{red}$0.148$} & $0.378$ & $-0.038$ & $0.416$ & $0.469$ & $0.372$ & $0.097$ \\ \cline{3-13}
 &  $C^2$  & $C^2S$ &{\color{blue} $-0.246$} & {\color{red}$0.165$} & {\color{blue}$0.891$} & {\color{yellow}$-0.236$} & {\color{yellow}$-0.359$} & {\color{yellow}$0.300$} & {\color{blue}$-0.063$} & {\color{red}$-0.857$} & {\color{blue}$-0.537$} & {\color{red}$-0.009$} & $ $ & $ $ & $ $ & $ $ & $ $ & \\ \hline \hline

 &  $Curve\_SL$  & $SB$ &{\color{blue} $0.593$} & {\color{blue}$0.905$} & {\color{red}$-0.831$} & {\color{green}$0.564$} & {\color{blue}$0.703$} & {\color{yellow}$0.390$} & {\color{blue}$0.536$} & {\color{blue}$0.923$} & {\color{blue}$0.957$} & {\color{blue}$0.915$} & $ $ & $ $ & $ $ & $ $ & $ $ & \\ \cline{3-13}
4 &  $Curve\_BD$  & $BC^3$ &{\color{blue} $-0.031$} & {\color{red}$-0.627$} & {\color{red}$-0.820$} & {\color{yellow}$0.095$} & {\color{yellow}$0.180$} & {\color{red}$-0.821$} & {\color{blue}$-0.098$} & {\color{red}$-1.000$} & {\color{blue}$0.837$} & {\color{red}$-0.805$} & $0.525$ & $-0.748$ & $1.273$ & $0.605$ & $0.458$ & $0.147$ \\ \cline{3-13}
 &  $C^3$  & $C^3S$ &{\color{blue} $0.677$} & {\color{red}$-0.815$} & {\color{blue}$0.940$} & {\color{yellow}$0.564$} & {\color{yellow}$0.549$} & {\color{yellow}$-0.379$} & {\color{blue}$0.669$} & {\color{red}$-0.922$} & {\color{blue}$0.809$} & {\color{red}$-0.883$} & $ $ & $ $ & $ $ & $ $ & $ $ & \\ \hline \hline

 &  $Curve\_SL$  & $SB$ &{\color{blue} $0.593$} & {\color{blue}$0.905$} & {\color{red}$-0.831$} & {\color{green}$0.564$} & {\color{blue}$0.703$} & {\color{yellow}$0.390$} & {\color{blue}$0.536$} & {\color{blue}$0.923$} & {\color{blue}$0.957$} & {\color{blue}$0.915$} & $ $ & $ $ & $ $ & $ $ & $ $ & \\ \cline{3-13}
5 &  $Curve\_BD$  & $BC_{w1}$ &{\color{blue} $-0.027$} & {\color{red}$-0.564$} & {\color{red}$-0.888$} & {\color{yellow}$-0.158$} & {\color{yellow}$-0.011$} & {\color{red}$-0.784$} & {\color{blue}$-0.042$} & {\color{red}$-0.949$} & {\color{blue}$0.965$} & {\color{red}$-0.711$} & $0.509$ & $-0.691$ & $1.200$ & $0.575$ & $0.373$ & $0.202$ \\ \cline{3-13}
 &  $C_{w1}$  & $C_{w1}S$ &{\color{blue} $0.609$} & {\color{red}$-0.700$} & {\color{blue}$0.951$} & {\color{yellow}$0.433$} & {\color{yellow}$0.333$} & {\color{yellow}$0.003$} & {\color{blue}$0.652$} & {\color{red}$-0.930$} & {\color{blue}$0.953$} & {\color{red}$-0.748$} & $ $ & $ $ & $ $ & $ $ & $ $ & \\ \hline \hline

 &  $Curve\_SL$  & $SB$ &{\color{blue} $0.593$} & {\color{blue}$0.905$} & {\color{red}$-0.831$} & {\color{green}$0.564$} & {\color{blue}$0.703$} & {\color{yellow}$0.390$} & {\color{blue}$0.536$} & {\color{blue}$0.923$} & {\color{blue}$0.957$} & {\color{blue}$0.915$} & $ $ & $ $ & $ $ & $ $ & $ $ & \\ \cline{3-13}
6 &  $Curve\_BD$  & $BC_{w2}$ &{\color{blue} $0.027$} & {\color{red}$-0.474$} & {\color{red}$-0.889$} & {\color{yellow}$-0.353$} & {\color{yellow}$-0.262$} & {\color{red}$-0.746$} & {\color{blue}$0.111$} & {\color{red}$-0.915$} & {\color{blue}$0.621$} & {\color{red}$-0.616$} & $0.454$ & $-0.621$ & $1.075$ & $0.506$ & $0.366$ & $0.139$ \\ \cline{3-13}
 &  $C_{w2}$  & $C_{w2}S$ &{\color{blue} $0.344$} & {\color{red}$-0.599$} & {\color{blue}$0.951$} & {\color{yellow}$0.115$} & {\color{yellow}$0.149$} & {\color{yellow}$0.079$} & {\color{blue}$0.503$} & {\color{red}$-0.913$} & {\color{blue}$0.638$} & {\color{red}$-0.652$} & $ $ & $ $ & $ $ & $ $ & $ $ & \\ \hline \hline

 &  $Curve\_SL$  & $SB$ &{\color{blue} $0.593$} & {\color{blue}$0.905$} & {\color{red}$-0.831$} & {\color{green}$0.564$} & {\color{blue}$0.703$} & {\color{yellow}$0.390$} & {\color{blue}$0.536$} & {\color{blue}$0.923$} & {\color{blue}$0.957$} & {\color{blue}$0.915$} & $ $ & $ $ & $ $ & $ $ & $ $ & \\ \cline{3-13}
7 &  $Curve\_BD$  & $BC_w$ &{\color{blue} $-0.006$} & {\color{red}$-0.517$} & {\color{red}$-0.889$} & {\color{yellow}$-0.267$} & {\color{yellow}$-0.135$} & {\color{red}$-0.763$} & {\color{blue}$0.018$} & {\color{red}$-0.931$} & {\color{blue}$0.842$} & {\color{red}$-0.661$} & $0.494$ & $-0.654$ & $1.148$ & $0.545$ & $0.361$ & $0.184$ \\ \cline{3-13}
 &  $C_w$  & $C_wS$ &{\color{blue} $0.529$} & {\color{red}$-0.647$} & {\color{blue}$0.951$} & {\color{yellow}$0.299$} & {\color{yellow}$0.131$} & {\color{yellow}$0.047$} & {\color{blue}$0.617$} & {\color{red}$-0.921$} & {\color{blue}$0.845$} & {\color{red}$-0.697$} & $ $ & $ $ & $ $ & $ $ & $ $ & \\ \hline \hline

 &  $S^1$  & $S^1B$ &{\color{blue} $0.506$} & {\color{blue}$0.901$} & {\color{red}$-0.848$} & {\color{green}$0.481$} & {\color{blue}$0.687$} & {\color{yellow}$0.182$} & {\color{blue}$0.431$} & {\color{blue}$0.998$} & {\color{blue}$0.928$} & {\color{blue}$0.888$} & $ $ & $ $ & $ $ & $ $ & $ $ & \\ \cline{3-13}
8 &  $Curve\_BD$  & $BC$ &{\color{blue} $0.114$} & {\color{red}$-0.431$} & {\color{red}$-0.881$} & {\color{yellow}$-0.287$} & {\color{yellow}$-0.161$} & {\color{red}$-0.752$} & {\color{blue}$0.196$} & {\color{red}$-0.901$} & {\color{blue}$0.205$} & {\color{red}$-0.638$} & $0.471$ & $-0.601$ & $1.072$ & $0.511$ & $0.337$ & $0.174$ \\ \cline{3-13}
 &  $Curve\_CC$  & $CS^1$ &{\color{blue} $0.420$} & {\color{red}$-0.586$} & {\color{blue}$0.915$} & {\color{yellow}$0.182$} & {\color{yellow}$-0.132$} & {\color{yellow}$0.274$} & {\color{blue}$0.584$} & {\color{red}$-0.878$} & {\color{blue}$0.436$} & {\color{red}$-0.709$} & $ $ & $ $ & $ $ & $ $ & $ $ & \\ \hline \hline

 &  $S^2$  & $S^2B$ &{\color{blue} $0.613$} & {\color{blue}$0.894$} & {\color{red}$-0.776$} & {\color{green}$0.530$} & {\color{blue}$0.611$} & {\color{yellow}$0.489$} & {\color{blue}$0.566$} & {\color{blue}$0.039$} & {\color{blue}$0.906$} & {\color{blue}$0.915$} & $ $ & $ $ & $ $ & $ $ & $ $ & \\ \cline{3-13}
9 &  $Curve\_BD$  & $BC$ &{\color{blue} $0.114$} & {\color{red}$-0.431$} & {\color{red}$-0.881$} & {\color{yellow}$-0.287$} & {\color{yellow}$-0.161$} & {\color{red}$-0.752$} & {\color{blue}$0.196$} & {\color{red}$-0.901$} & {\color{blue}$0.205$} & {\color{red}$-0.638$} & $0.521$ & $-0.607$ & $1.128$ & $0.548$ & $0.341$ & $0.207$ \\ \cline{3-13}
 &  $Curve\_CC$  & $CS^2$ &{\color{blue} $0.513$} & {\color{red}$-0.627$} & {\color{blue}$0.909$} & {\color{yellow}$0.207$} & {\color{yellow}$0.131$} & {\color{yellow}$0.001$} & {\color{blue}$0.650$} & {\color{red}$-0.258$} & {\color{blue}$0.030$} & {\color{red}$-0.724$} & $ $ & $ $ & $ $ & $ $ & $ $ & \\ \hline \hline

 &  $S_w$  & $S_wB$ &{\color{blue} $0.594$} & {\color{blue}$0.905$} & {\color{red}$-0.830$} & {\color{green}$0.565$} & {\color{blue}$0.702$} & {\color{yellow}$0.393$} & {\color{blue}$0.538$} & {\color{blue}$0.915$} & {\color{blue}$0.957$} & {\color{blue}$0.915$} & $ $ & $ $ & $ $ & $ $ & $ $ & \\ \cline{3-13}
10 &  $Curve\_BD$  & $BC$ &{\color{blue} $0.114$} & {\color{red}$-0.431$} & {\color{red}$-0.881$} & {\color{yellow}$-0.287$} & {\color{yellow}$-0.161$} & {\color{red}$-0.752$} & {\color{blue}$0.196$} & {\color{red}$-0.901$} & {\color{blue}$0.205$} & {\color{red}$-0.638$} & $0.516$ & $-0.611$ & $1.127$ & $0.545$ & $0.343$ & $0.202$ \\ \cline{3-13}
 &  $Curve\_CC$  & $CS_w$ &{\color{blue} $0.495$} & {\color{red}$-0.618$} & {\color{blue}$0.949$} & {\color{yellow}$0.217$} & {\color{yellow}$0.007$} & {\color{yellow}$0.105$} & {\color{blue}$0.659$} & {\color{red}$-0.902$} & {\color{blue}$0.243$} & {\color{red}$-0.726$} & $ $ & $ $ & $ $ & $ $ & $ $ & \\ \hline

\end{tabular}}
\end{sidewaystable}

\clearpage
\begin{table}[m]
  \centering
  \caption{{\bf Metazoan origination ($G_{sp}=G_{mean\_log}-\chi \cdot G_{sd\_log}$, $\chi=1.5677$)}}\label{1}
{\scriptsize
\begin{tabular}{|c|c|c|c|c|c|c|c|}
  \hline
  No. & Superphylum & Taxon & number of records & $G\_mean\_log$ & $G\_sd\_log$ & $G_{sp}$ & $T_{ori}$ (Ma) \\ \hline
1 & Protostomia & Nematodes  & 66 & -2.394 & 0.93204 & -3.8552 & 572.89 \\ \hline
2 & Deuterostomia & Chordates  & 5 & -1.8885 & 0.91958 & -3.3301 & 566.49 \\ \hline
3 & Diploblostica & Sponges  & 7 & -1.0834 & 1.3675 & -3.2272 & 565.23 \\ \hline
4 & Diploblostica & Ctenophores  & 2 & -0.010305 & 1.6417 & -2.584 & 557.39 \\ \hline
5 & Protostomia & Tardigrades & 21 & -1.2168 & 0.7276 & -2.3574 & 554.63 \\ \hline
6 & Protostomia & Misc\_Inverts  & 57 & -0.75852 & 0.96321 & -2.2686 & 553.54 \\ \hline
7 & Protostomia & Arthropod & 1284 & -0.078413 & 1.2116 & -1.9778 & 550 \\ \hline
8 & Protostomia & Annelid  & 140 & -0.14875 & 0.9258 & -1.6001 & 545.39 \\ \hline
9 & Protostomia & Myriapods  & 15 & -0.54874 & 0.66478 & -1.5909 & 545.28 \\ \hline
10 & Protostomia & Flatworms  & 68 & 0.15556 & 1.0701 & -1.522 & 544.44 \\ \hline
11 & Protostomia & Rotifers  & 9 & -0.51158 & 0.55134 & -1.3759 & 542.66 \\ \hline
12 & Diploblostica & Cnidarians  & 11 & -0.16888 & 0.69379 & -1.2565 & 541.2 \\ \hline
13 & Deuterostomia & Fish & 2045 & 0.23067 & 0.6559 & -0.7976 & 535.6 \\ \hline
14 & Deuterostomia & Echinoderm  & 48 & 0.11223 & 0.52794 & -0.71542 & 534.6 \\ \hline
15 & Protostomia & Molluscs  & 263 & 0.5812 & 0.5493 & -0.27994 & 529.29 \\ \hline
16 & Deuterostomia & Bird  & 474 & 0.32019 & 0.13788 & 0.10403 & 524.61 \\ \hline
17 & Deuterostomia & Reptile  & 418 & 0.78332 & 0.28332 & 0.33916 & 521.74 \\ \hline
18 & Deuterostomia & Amphibian  & 927 & 2.4116 & 1.081 & 0.71691 & 517.13 \\ \hline
19 & Deuterostomia & Mammal  & 657 & 1.1837 & 0.2401 & 0.80727 & 516.03 \\ \hline
\end{tabular}}
\end{table}

\clearpage
\begin{table}[m]
  \centering
  \caption{{\bf Metazoan origination ($G_{sp}'=G_{mean\_log}-\chi_1 \cdot G_{sd\_log}$, $\chi_1=3.1867$)}}\label{1}
{\scriptsize
\begin{tabular}{|c|c|c|c|c|c|c|c|c|}
  \hline
  No. & Superphylum & Taxon & number of records & $G\_mean\_log$ & $G\_sd\_log$ & $G_{sp}$ & $G_{sp}$' ($\chi_1$) & $T_{ori}$ (Ma) \\ \hline
3 & Diploblostica & Sponges  & 7 & -1.0834 & 1.3675 & -3.2272 & -5.4412 & 565.23 \\ \hline
1 & Protostomia & Nematodes  & 66 & -2.394 & 0.93204 & -3.8552 & -5.3641 & 572.89 \\ \hline
4 & Diploblostica & Ctenophores  & 2 & -0.010305 & 1.6417 & -2.584 & -5.2419 & 557.39 \\ \hline
2 & Deuterostomia & Chordates  & 5 & -1.8885 & 0.91958 & -3.3301 & -4.8189 & 566.49 \\ \hline
7 & Protostomia & Arthropod & 1284 & -0.078413 & 1.2116 & -1.9778 & -3.9394 & 550 \\ \hline
6 & Protostomia & Misc\_Inverts  & 57 & -0.75852 & 0.96321 & -2.2686 & -3.828 & 553.54 \\ \hline
5 & Protostomia & Tardigrades & 21 & -1.2168 & 0.7276 & -2.3574 & -3.5354 & 554.63 \\ \hline
10 & Protostomia & Flatworms  & 68 & 0.15556 & 1.0701 & -1.522 & -3.2545 & 544.44 \\ \hline
8 & Protostomia & Annelid  & 140 & -0.14875 & 0.9258 & -1.6001 & -3.099 & 545.39 \\ \hline
9 & Protostomia & Myriapods  & 15 & -0.54874 & 0.66478 & -1.5909 & -2.6672 & 545.28 \\ \hline
12 & Diploblostica & Cnidarians  & 11 & -0.16888 & 0.69379 & -1.2565 & -2.3798 & 541.2 \\ \hline
11 & Protostomia & Rotifers  & 9 & -0.51158 & 0.55134 & -1.3759 & -2.2685 & 542.66 \\ \hline
13 & Deuterostomia & Fish & 2045 & 0.23067 & 0.6559 & -0.7976 & -1.8595 & 535.6 \\ \hline
14 & Deuterostomia & Echinoderm  & 48 & 0.11223 & 0.52794 & -0.7154 & -1.5702 & 534.6 \\ \hline
15 & Protostomia & Molluscs  & 263 & 0.5812 & 0.5493 & -0.2799 & -1.1693 & 529.29 \\ \hline
18 & Deuterostomia & Amphibian  & 927 & 2.4116 & 1.081 & 0.71691 & -1.0332 & 517.13 \\ \hline
17 & Deuterostomia & Reptile  & 418 & 0.78332 & 0.28332 & 0.33916 & -0.1195 & 521.74 \\ \hline
16 & Deuterostomia & Bird  & 474 & 0.32019 & 0.13788 & 0.10403 & -0.1192 & 524.61 \\ \hline
19 & Deuterostomia & Mammal  & 657 & 1.1837 & 0.2401 & 0.80727 & 0.41857 & 516.03 \\ \hline
\end{tabular}}
\end{table}

\clearpage
\begin{table}[m]
  \centering
  \caption{{\bf Metazoan origination ($G_{mean\_log}$)}}\label{1}
{\scriptsize
\begin{tabular}{|c|c|c|c|c|c|c|c|}
  \hline
  No. & Superphylum & Taxon & number of records & $G\_mean\_log$ & $G\_sd\_log$ & $G_{sp}$ & $T_{ori}$ (Ma) \\ \hline
  1 & Protostomia & Nematodes  & 66 & -2.394 & 0.93204 & -3.8552 & 572.89 \\ \hline
2 & Deuterostomia & Chordates  & 5 & -1.8885 & 0.91958 & -3.3301 & 566.49 \\ \hline
5 & Protostomia & Tardigrades & 21 & -1.2168 & 0.7276 & -2.3574 & 554.63 \\ \hline
3 & Diploblostica & Sponges  & 7 & -1.0834 & 1.3675 & -3.2272 & 565.23 \\ \hline
6 & Protostomia & Misc\_Inverts  & 57 & -0.75852 & 0.96321 & -2.2686 & 553.54 \\ \hline
9 & Protostomia & Myriapods  & 15 & -0.54874 & 0.66478 & -1.5909 & 545.28 \\ \hline
11 & Protostomia & Rotifers  & 9 & -0.51158 & 0.55134 & -1.3759 & 542.66 \\ \hline
12 & Diploblostica & Cnidarians  & 11 & -0.16888 & 0.69379 & -1.2565 & 541.2 \\ \hline
8 & Protostomia & Annelid  & 140 & -0.14875 & 0.9258 & -1.6001 & 545.39 \\ \hline
7 & Protostomia & Arthropod & 1284 & -0.078413 & 1.2116 & -1.9778 & 550 \\ \hline
4 & Diploblostica & Ctenophores  & 2 & -0.010305 & 1.6417 & -2.584 & 557.39 \\ \hline
14 & Deuterostomia & Echinoderm  & 48 & 0.11223 & 0.52794 & -0.71542 & 534.6 \\ \hline
10 & Protostomia & Flatworms  & 68 & 0.15556 & 1.0701 & -1.522 & 544.44 \\ \hline
13 & Deuterostomia & Fish & 2045 & 0.23067 & 0.6559 & -0.7976 & 535.6 \\ \hline
16 & Deuterostomia & Bird  & 474 & 0.32019 & 0.13788 & 0.10403 & 524.61 \\ \hline
15 & Protostomia & Molluscs  & 263 & 0.5812 & 0.5493 & -0.27994 & 529.29 \\ \hline
17 & Deuterostomia & Reptile  & 418 & 0.78332 & 0.28332 & 0.33916 & 521.74 \\ \hline
19 & Deuterostomia & Mammal  & 657 & 1.1837 & 0.2401 & 0.80727 & 516.03 \\ \hline
18 & Deuterostomia & Amphibian  & 927 & 2.4116 & 1.081 & 0.71691 & 517.13 \\ \hline
\end{tabular}}
\end{table}

\clearpage
\begin{table}[m]
  \centering
  \caption{{\bf Angiosperm origination ($G_{sp}=G_{mean\_log}-\chi \cdot G_{sd\_log}$, $\chi=1.5677$)}}\label{1}
{\scriptsize
\begin{tabular}{|c|c|c|c|c|c|c|c|}
  \hline
  No. & Superphylum & Taxon & $G\_mean\_log$ & $G\_sd\_log$ & $G_{sp}$ & $T_{ori}$ (Ma) \\ \hline
  1 & Dicotyledoneae & Lentibulariaceae & -1.0532 & 0.88349 & -2.4382 & 177.96 \\ \hline
2 & Monocotyledoneae & Cyperaceae & -0.81211 & 0.61307 & -1.7732 & 169.85 \\ \hline
3 & Dicotyledoneae & Cruciferae & -0.62192 & 0.6855 & -1.6966 & 168.91 \\ \hline
4 & Dicotyledoneae & Rutaceae & -0.22121 & 0.93413 & -1.6856 & 168.78 \\ \hline
5 & Dicotyledoneae & Oxalidaceae & 0.19774 & 1.1445 & -1.5964 & 167.69 \\ \hline
6 & Dicotyledoneae & Crassulaceae & -0.26578 & 0.82268 & -1.5555 & 167.19 \\ \hline
7 & Dicotyledoneae & Rosaceae & -0.40468 & 0.62511 & -1.3847 & 165.11 \\ \hline
8 & Dicotyledoneae & Boraginaceae & -0.20664 & 0.68081 & -1.2739 & 163.76 \\ \hline
9 & Dicotyledoneae & Labiatae & -0.0021905 & 0.80883 & -1.2702 & 163.71 \\ \hline
10 & Monocotyledoneae & Juncaceae & -0.23032 & 0.63698 & -1.2289 & 163.21 \\ \hline
11 & Dicotyledoneae & Vitaceae & -0.60987 & 0.39049 & -1.222 & 163.13 \\ \hline
12 & Dicotyledoneae & Cucurbitaceae & -0.26487 & 0.60779 & -1.2177 & 163.07 \\ \hline
13 & Dicotyledoneae & Onagraceae & 0.040848 & 0.78018 & -1.1822 & 162.64 \\ \hline
14 & Dicotyledoneae & Leguminosae & 0.33968 & 0.88684 & -1.0506 & 161.04 \\ \hline
15 & Dicotyledoneae & Myrtaceae & -0.37801 & 0.42511 & -1.0445 & 160.96 \\ \hline
16 & Monocotyledoneae & Bromeliaceae & -0.56838 & 0.29232 & -1.0266 & 160.74 \\ \hline
17 & Dicotyledoneae & Polygonaceae & 0.20985 & 0.76174 & -0.98433 & 160.23 \\ \hline
18 & Dicotyledoneae & Euphorbiaceae & 0.72687 & 1.0796 & -0.96561 & 160 \\ \hline
19 & Dicotyledoneae & Convolvulaceae & 0.50052 & 0.928 & -0.9543 & 159.86 \\ \hline
20 & Dicotyledoneae & Chenopodiaceae & -0.046809 & 0.5526 & -0.91312 & 159.36 \\ \hline
21 & Dicotyledoneae & Plantaginaceae & -0.15021 & 0.48422 & -0.90932 & 159.31 \\ \hline
22 & Dicotyledoneae & Rubiaceae & -0.084413 & 0.51565 & -0.8928 & 159.11 \\ \hline
23 & Dicotyledoneae & Caryophyllaceae & 0.27683 & 0.65869 & -0.7558 & 157.44 \\ \hline
24 & Dicotyledoneae & Amaranthaceae & 0.15834 & 0.58176 & -0.75369 & 157.42 \\ \hline
25 & Dicotyledoneae & Malvaceae & 0.39517 & 0.47109 & -0.34336 & 152.41 \\ \hline
26 & Monocotyledoneae & Zingiberaceae & 0.24819 & 0.36317 & -0.32115 & 152.14 \\ \hline
27 & Monocotyledoneae & Iridaceae & 1.3429 & 1.0491 & -0.30178 & 151.9 \\ \hline
28 & Dicotyledoneae & Umbelliferae & 0.71003 & 0.6235 & -0.26742 & 151.49 \\ \hline
29 & Dicotyledoneae & Solanaceae & 0.78034 & 0.66585 & -0.26352 & 151.44 \\ \hline
30 & Monocotyledoneae & Orchidaceae & 1.4063 & 1.0551 & -0.24784 & 151.25 \\ \hline
31 & Monocotyledoneae & Araceae & 1.5174 & 1.012 & -0.069152 & 149.07 \\ \hline
32 & Dicotyledoneae & Papaveraceae & 0.93206 & 0.61932 & -0.038854 & 148.7 \\ \hline
33 & Dicotyledoneae & Compositae & 1.0741 & 0.70726 & -0.034657 & 148.65 \\ \hline
34 & Monocotyledoneae & Gramineae & 1.4002 & 0.84476 & 0.075894 & 147.3 \\ \hline
35 & Dicotyledoneae & Cactaceae & 0.98813 & 0.57251 & 0.090608 & 147.12 \\ \hline
36 & Monocotyledoneae & Palmae & 1.1222 & 0.63488 & 0.12691 & 146.68 \\ \hline
37 & Dicotyledoneae & Passifloraceae & 0.52209 & 0.22472 & 0.16979 & 146.15 \\ \hline
38 & Dicotyledoneae & Orobanchaceae & 1.12 & 0.54393 & 0.26726 & 144.97 \\ \hline
39 & Dicotyledoneae & Cistaceae & 0.88905 & 0.30458 & 0.41155 & 143.21 \\ \hline
40 & Monocotyledoneae & Asparagaceae & 2.0053 & 0.78802 & 0.76991 & 138.84 \\ \hline
41 & Dicotyledoneae & Asteraceae & 1.8795 & 0.67031 & 0.82863 & 138.12 \\ \hline
42 & Dicotyledoneae & Ranunculaceae & 2.0285 & 0.72517 & 0.8916 & 137.35 \\ \hline
43 & Monocotyledoneae & Agavaceae & 1.6207 & 0.4537 & 0.90941 & 137.13 \\ \hline
44 & Monocotyledoneae & Hyacinthaceae & 2.3635 & 0.69028 & 1.2814 & 132.6 \\ \hline
45 & Dicotyledoneae & Loranthaceae & 2.3797 & 0.68478 & 1.3062 & 132.3 \\ \hline
46 & Monocotyledoneae & Commelinaceae & 2.5322 & 0.64196 & 1.5258 & 129.62 \\ \hline
47 & Monocotyledoneae & Amaryllidaceae & 2.9085 & 0.5811 & 1.9975 & 123.86 \\ \hline
48 & Monocotyledoneae & Xanthorrhoeaceae & 2.7036 & 0.4 & 2.0765 & 122.9 \\ \hline
49 & Monocotyledoneae & Asphodelaceae & 2.8054 & 0.33968 & 2.2729 & 120.51 \\ \hline
50 & Monocotyledoneae & Alliaceae & 2.9051 & 0.39078 & 2.2924 & 120.27 \\ \hline
51 & Dicotyledoneae & Paeoniaceae & 2.957 & 0.28164 & 2.5155 & 117.55 \\ \hline
52 & Monocotyledoneae & Liliaceae & 3.5678 & 0.63278 & 2.5757 & 116.81 \\ \hline
53 & Monocotyledoneae & Aloaceae & 2.9724 & 0.2203 & 2.627 & 116.19 \\ \hline
\end{tabular}}
\end{table}

\clearpage
\begin{table}[m]
  \centering
  \caption{{\bf Angiosperm origination ($G_{sp}'=G_{mean\_log}-\chi_1 \cdot G_{sd\_log}$, $\chi_1=3.1867$)}}\label{1}
{\scriptsize
\begin{tabular}{|c|c|c|c|c|c|c|c|c|}
  \hline
  No. & Superphylum & Taxon & $G\_mean\_log$ & $G\_sd\_log$ & $G_{sp}$ & $G_{sp}'$ & $T_{ori}$ (Ma) \\ \hline
  1 & Dicotyledoneae & Lentibulariaceae & -1.0532 & 0.88349 & -2.4382 & -3.8686 & 177.96 \\ \hline
5 & Dicotyledoneae & Oxalidaceae & 0.19774 & 1.1445 & -1.5964 & -3.4494 & 167.69 \\ \hline
4 & Dicotyledoneae & Rutaceae & -0.22121 & 0.93413 & -1.6856 & -3.198 & 168.78 \\ \hline
6 & Dicotyledoneae & Crassulaceae & -0.26578 & 0.82268 & -1.5555 & -2.8874 & 167.19 \\ \hline
3 & Dicotyledoneae & Cruciferae & -0.62192 & 0.6855 & -1.6966 & -2.8064 & 168.91 \\ \hline
2 & Monocotyledoneae & Cyperaceae & -0.81211 & 0.61307 & -1.7732 & -2.7658 & 169.85 \\ \hline
18 & Dicotyledoneae & Euphorbiaceae & 0.72687 & 1.0796 & -0.96561 & -2.7135 & 160 \\ \hline
9 & Dicotyledoneae & Labiatae & -0.0021905 & 0.80883 & -1.2702 & -2.5797 & 163.71 \\ \hline
14 & Dicotyledoneae & Leguminosae & 0.33968 & 0.88684 & -1.0506 & -2.4864 & 161.04 \\ \hline
19 & Dicotyledoneae & Convolvulaceae & 0.50052 & 0.928 & -0.9543 & -2.4567 & 159.86 \\ \hline
13 & Dicotyledoneae & Onagraceae & 0.040848 & 0.78018 & -1.1822 & -2.4454 & 162.64 \\ \hline
7 & Dicotyledoneae & Rosaceae & -0.40468 & 0.62511 & -1.3847 & -2.3967 & 165.11 \\ \hline
8 & Dicotyledoneae & Boraginaceae & -0.20664 & 0.68081 & -1.2739 & -2.3762 & 163.76 \\ \hline
10 & Monocotyledoneae & Juncaceae & -0.23032 & 0.63698 & -1.2289 & -2.2602 & 163.21 \\ \hline
17 & Dicotyledoneae & Polygonaceae & 0.20985 & 0.76174 & -0.98433 & -2.2176 & 160.23 \\ \hline
12 & Dicotyledoneae & Cucurbitaceae & -0.26487 & 0.60779 & -1.2177 & -2.2017 & 163.07 \\ \hline
27 & Monocotyledoneae & Iridaceae & 1.3429 & 1.0491 & -0.30178 & -2.0003 & 151.9 \\ \hline
30 & Monocotyledoneae & Orchidaceae & 1.4063 & 1.0551 & -0.24784 & -1.956 & 151.25 \\ \hline
11 & Dicotyledoneae & Vitaceae & -0.60987 & 0.39049 & -1.222 & -1.8542 & 163.13 \\ \hline
23 & Dicotyledoneae & Caryophyllaceae & 0.27683 & 0.65869 & -0.7558 & -1.8222 & 157.44 \\ \hline
20 & Dicotyledoneae & Chenopodiaceae & -0.046809 & 0.5526 & -0.91312 & -1.8078 & 159.36 \\ \hline
15 & Dicotyledoneae & Myrtaceae & -0.37801 & 0.42511 & -1.0445 & -1.7327 & 160.96 \\ \hline
22 & Dicotyledoneae & Rubiaceae & -0.084413 & 0.51565 & -0.8928 & -1.7276 & 159.11 \\ \hline
31 & Monocotyledoneae & Araceae & 1.5174 & 1.012 & -0.069152 & -1.7075 & 149.07 \\ \hline
24 & Dicotyledoneae & Amaranthaceae & 0.15834 & 0.58176 & -0.75369 & -1.6956 & 157.42 \\ \hline
21 & Dicotyledoneae & Plantaginaceae & -0.15021 & 0.48422 & -0.90932 & -1.6933 & 159.31 \\ \hline
16 & Monocotyledoneae & Bromeliaceae & -0.56838 & 0.29232 & -1.0266 & -1.4999 & 160.74 \\ \hline
29 & Dicotyledoneae & Solanaceae & 0.78034 & 0.66585 & -0.26352 & -1.3415 & 151.44 \\ \hline
34 & Monocotyledoneae & Gramineae & 1.4002 & 0.84476 & 0.075894 & -1.2918 & 147.3 \\ \hline
28 & Dicotyledoneae & Umbelliferae & 0.71003 & 0.6235 & -0.26742 & -1.2769 & 151.49 \\ \hline
33 & Dicotyledoneae & Compositae & 1.0741 & 0.70726 & -0.034657 & -1.1797 & 148.65 \\ \hline
25 & Dicotyledoneae & Malvaceae & 0.39517 & 0.47109 & -0.34336 & -1.1061 & 152.41 \\ \hline
32 & Dicotyledoneae & Papaveraceae & 0.93206 & 0.61932 & -0.038854 & -1.0415 & 148.7 \\ \hline
26 & Monocotyledoneae & Zingiberaceae & 0.24819 & 0.36317 & -0.32115 & -0.9091 & 152.14 \\ \hline
36 & Monocotyledoneae & Palmae & 1.1222 & 0.63488 & 0.12691 & -0.901 & 146.68 \\ \hline
35 & Dicotyledoneae & Cactaceae & 0.98813 & 0.57251 & 0.090608 & -0.8363 & 147.12 \\ \hline
38 & Dicotyledoneae & Orobanchaceae & 1.12 & 0.54393 & 0.26726 & -0.6133 & 144.97 \\ \hline
40 & Monocotyledoneae & Asparagaceae & 2.0053 & 0.78802 & 0.76991 & -0.5059 & 138.84 \\ \hline
42 & Dicotyledoneae & Ranunculaceae & 2.0285 & 0.72517 & 0.8916 & -0.2824 & 137.35 \\ \hline
41 & Dicotyledoneae & Asteraceae & 1.8795 & 0.67031 & 0.82863 & -0.2566 & 138.12 \\ \hline
37 & Dicotyledoneae & Passifloraceae & 0.52209 & 0.22472 & 0.16979 & -0.194 & 146.15 \\ \hline
39 & Dicotyledoneae & Cistaceae & 0.88905 & 0.30458 & 0.41155 & -0.0816 & 143.21 \\ \hline
44 & Monocotyledoneae & Hyacinthaceae & 2.3635 & 0.69028 & 1.2814 & 0.16378 & 132.6 \\ \hline
43 & Monocotyledoneae & Agavaceae & 1.6207 & 0.4537 & 0.90941 & 0.17489 & 137.13 \\ \hline
45 & Dicotyledoneae & Loranthaceae & 2.3797 & 0.68478 & 1.3062 & 0.19751 & 132.3 \\ \hline
46 & Monocotyledoneae & Commelinaceae & 2.5322 & 0.64196 & 1.5258 & 0.48647 & 129.62 \\ \hline
47 & Monocotyledoneae & Amaryllidaceae & 2.9085 & 0.5811 & 1.9975 & 1.05671 & 123.86 \\ \hline
48 & Monocotyledoneae & Xanthorrhoeaceae & 2.7036 & 0.4 & 2.0765 & 1.42892 & 122.9 \\ \hline
52 & Monocotyledoneae & Liliaceae & 3.5678 & 0.63278 & 2.5757 & 1.55132 & 116.81 \\ \hline
50 & Monocotyledoneae & Alliaceae & 2.9051 & 0.39078 & 2.2924 & 1.6598 & 120.27 \\ \hline
49 & Monocotyledoneae & Asphodelaceae & 2.8054 & 0.33968 & 2.2729 & 1.72294 & 120.51 \\ \hline
51 & Dicotyledoneae & Paeoniaceae & 2.957 & 0.28164 & 2.5155 & 2.0595 & 117.55 \\ \hline
53 & Monocotyledoneae & Aloaceae & 2.9724 & 0.2203 & 2.627 & 2.27037 & 116.19 \\ \hline
\end{tabular}}
\end{table}

\clearpage
\begin{table}[m]
  \centering
  \caption{{\bf Angiosperm origination ($G_{mean\_log}$)}}\label{1}
{\scriptsize
\begin{tabular}{|c|c|c|c|c|c|c|c|}
  \hline
  No. & Superphylum & Taxon & $G\_mean\_log$ & $G\_sd\_log$ & $G_{sp}$ & $T_{ori}$ (Ma) \\ \hline
  1 & Dicotyledoneae & Lentibulariaceae & -1.0532 & 0.88349 & -2.4382 & 177.96 \\ \hline
2 & Monocotyledoneae & Cyperaceae & -0.81211 & 0.61307 & -1.7732 & 169.85 \\ \hline
3 & Dicotyledoneae & Cruciferae & -0.62192 & 0.6855 & -1.6966 & 168.91 \\ \hline
11 & Dicotyledoneae & Vitaceae & -0.60987 & 0.39049 & -1.222 & 163.13 \\ \hline
16 & Monocotyledoneae & Bromeliaceae & -0.56838 & 0.29232 & -1.0266 & 160.74 \\ \hline
7 & Dicotyledoneae & Rosaceae & -0.40468 & 0.62511 & -1.3847 & 165.11 \\ \hline
15 & Dicotyledoneae & Myrtaceae & -0.37801 & 0.42511 & -1.0445 & 160.96 \\ \hline
6 & Dicotyledoneae & Crassulaceae & -0.26578 & 0.82268 & -1.5555 & 167.19 \\ \hline
12 & Dicotyledoneae & Cucurbitaceae & -0.26487 & 0.60779 & -1.2177 & 163.07 \\ \hline
10 & Monocotyledoneae & Juncaceae & -0.23032 & 0.63698 & -1.2289 & 163.21 \\ \hline
4 & Dicotyledoneae & Rutaceae & -0.22121 & 0.93413 & -1.6856 & 168.78 \\ \hline
8 & Dicotyledoneae & Boraginaceae & -0.20664 & 0.68081 & -1.2739 & 163.76 \\ \hline
21 & Dicotyledoneae & Plantaginaceae & -0.15021 & 0.48422 & -0.90932 & 159.31 \\ \hline
22 & Dicotyledoneae & Rubiaceae & -0.084413 & 0.51565 & -0.8928 & 159.11 \\ \hline
20 & Dicotyledoneae & Chenopodiaceae & -0.046809 & 0.5526 & -0.91312 & 159.36 \\ \hline
9 & Dicotyledoneae & Labiatae & -0.0021905 & 0.80883 & -1.2702 & 163.71 \\ \hline
13 & Dicotyledoneae & Onagraceae & 0.040848 & 0.78018 & -1.1822 & 162.64 \\ \hline
24 & Dicotyledoneae & Amaranthaceae & 0.15834 & 0.58176 & -0.75369 & 157.42 \\ \hline
5 & Dicotyledoneae & Oxalidaceae & 0.19774 & 1.1445 & -1.5964 & 167.69 \\ \hline
17 & Dicotyledoneae & Polygonaceae & 0.20985 & 0.76174 & -0.98433 & 160.23 \\ \hline
26 & Monocotyledoneae & Zingiberaceae & 0.24819 & 0.36317 & -0.32115 & 152.14 \\ \hline
23 & Dicotyledoneae & Caryophyllaceae & 0.27683 & 0.65869 & -0.7558 & 157.44 \\ \hline
14 & Dicotyledoneae & Leguminosae & 0.33968 & 0.88684 & -1.0506 & 161.04 \\ \hline
25 & Dicotyledoneae & Malvaceae & 0.39517 & 0.47109 & -0.34336 & 152.41 \\ \hline
19 & Dicotyledoneae & Convolvulaceae & 0.50052 & 0.928 & -0.9543 & 159.86 \\ \hline
37 & Dicotyledoneae & Passifloraceae & 0.52209 & 0.22472 & 0.16979 & 146.15 \\ \hline
28 & Dicotyledoneae & Umbelliferae & 0.71003 & 0.6235 & -0.26742 & 151.49 \\ \hline
18 & Dicotyledoneae & Euphorbiaceae & 0.72687 & 1.0796 & -0.96561 & 160 \\ \hline
29 & Dicotyledoneae & Solanaceae & 0.78034 & 0.66585 & -0.26352 & 151.44 \\ \hline
39 & Dicotyledoneae & Cistaceae & 0.88905 & 0.30458 & 0.41155 & 143.21 \\ \hline
32 & Dicotyledoneae & Papaveraceae & 0.93206 & 0.61932 & -0.038854 & 148.7 \\ \hline
35 & Dicotyledoneae & Cactaceae & 0.98813 & 0.57251 & 0.090608 & 147.12 \\ \hline
33 & Dicotyledoneae & Compositae & 1.0741 & 0.70726 & -0.034657 & 148.65 \\ \hline
38 & Dicotyledoneae & Orobanchaceae & 1.12 & 0.54393 & 0.26726 & 144.97 \\ \hline
36 & Monocotyledoneae & Palmae & 1.1222 & 0.63488 & 0.12691 & 146.68 \\ \hline
27 & Monocotyledoneae & Iridaceae & 1.3429 & 1.0491 & -0.30178 & 151.9 \\ \hline
34 & Monocotyledoneae & Gramineae & 1.4002 & 0.84476 & 0.075894 & 147.3 \\ \hline
30 & Monocotyledoneae & Orchidaceae & 1.4063 & 1.0551 & -0.24784 & 151.25 \\ \hline
31 & Monocotyledoneae & Araceae & 1.5174 & 1.012 & -0.069152 & 149.07 \\ \hline
43 & Monocotyledoneae & Agavaceae & 1.6207 & 0.4537 & 0.90941 & 137.13 \\ \hline
41 & Dicotyledoneae & Asteraceae & 1.8795 & 0.67031 & 0.82863 & 138.12 \\ \hline
40 & Monocotyledoneae & Asparagaceae & 2.0053 & 0.78802 & 0.76991 & 138.84 \\ \hline
42 & Dicotyledoneae & Ranunculaceae & 2.0285 & 0.72517 & 0.8916 & 137.35 \\ \hline
44 & Monocotyledoneae & Hyacinthaceae & 2.3635 & 0.69028 & 1.2814 & 132.6 \\ \hline
45 & Dicotyledoneae & Loranthaceae & 2.3797 & 0.68478 & 1.3062 & 132.3 \\ \hline
46 & Monocotyledoneae & Commelinaceae & 2.5322 & 0.64196 & 1.5258 & 129.62 \\ \hline
48 & Monocotyledoneae & Xanthorrhoeaceae & 2.7036 & 0.4 & 2.0765 & 122.9 \\ \hline
49 & Monocotyledoneae & Asphodelaceae & 2.8054 & 0.33968 & 2.2729 & 120.51 \\ \hline
50 & Monocotyledoneae & Alliaceae & 2.9051 & 0.39078 & 2.2924 & 120.27 \\ \hline
47 & Monocotyledoneae & Amaryllidaceae & 2.9085 & 0.5811 & 1.9975 & 123.86 \\ \hline
51 & Dicotyledoneae & Paeoniaceae & 2.957 & 0.28164 & 2.5155 & 117.55 \\ \hline
53 & Monocotyledoneae & Aloaceae & 2.9724 & 0.2203 & 2.627 & 116.19 \\ \hline
52 & Monocotyledoneae & Liliaceae & 3.5678 & 0.63278 & 2.5757 & 116.81 \\ \hline
\end{tabular}}
\end{table}

\clearpage
\begin{table}[m]
  \centering
  \caption{{\bf Origination order}}\label{1}
{\scriptsize
\begin{tabular}{|c|c|c|c|c|c|c|}
  \hline
  Superphylum & $T_{ori}$ (Ma) & $G_{mean\_log}$ & $G_{sd\_log}$ & $G_{sp}$ & $G_{max}$ & $G_{min}$ \\ \hline
Diploblostica  & 560 & -0.4731 & 1.095 & -2.1898 & 1.1506 & -2.8134 \\ \hline
Protostomia  & 542 & -0.10229 & 1.2158 & -2.0083 & 4.1685 & -3.912 \\ \hline
Deuterostomia  & 525 & 0.87752 & 1.0869 & -0.82636 & 4.8891 & -2.8134 \\ \hline
bryophyte  & 488.3 & -0.63576 & 0.54685 & -1.4931 & 2.0757 & -1.772 \\ \hline
pteridophyte  & 416 & 1.7359 & 1.6606 & -0.86744 & 4.2861 & -2.4079 \\ \hline
gymnosperm  & 359.2 & 2.8263 & 0.46055 & 2.1043 & 3.5835 & 0.81093 \\ \hline
angiosperm & 145.5 & 0.96878 & 1.2681 & -1.0193 & 5.0252 & -2.8134 \\ \hline
Protist &  & -1.5532 & 1.6488 & -4.1381 & 2.9755 & -7.3475 \\ \hline
Eubacteria &  & -5.8238 & 0.57889 & -6.7313 & -4.5865 & -8.7269 \\ \hline
Archaea &  & -6.02 & 0.50451 & -6.811 & -4.7404 & -6.9616 \\ \hline
\end{tabular}}
\end{table}

\end{document}